\documentstyle[12pt,epsfig]{article}

\def    \be             {\begin{equation}}
\def    \ee             {\end{equation}}

\newcommand{\beq}{\begin{equation}}
\newcommand{\eeq}{\end{equation}}
\newcommand{\beqn}{\begin{eqnarray}}
\newcommand{\eeqn}{\end{eqnarray}}
\newcommand{\beqns}{\begin{eqnarray*}}
\newcommand{\eeqns}{\end{eqnarray*}}

\def\to{\rightarrow}
\def\d{{\rm d}}

\def  \mathrm#1{{\rm #1}}
\def\gapproxeq{{\ \lower 0.6ex \hbox{$\buildrel>\over\sim$}\ }}

\def\etal{\hbox{\it et al.}}    

\def\ww{W^+W^-}
\def\mev{\; {\rm MeV}}
\def\gev{\; {\rm GeV}}

\def\pb{\; {\rm pb}}
\newcommand{\epem}{\mbox{$\mathrm{e^+e^-}$}}
\def\Mw{M_{\mathrm{W}}}
\def\MW{M_{\mathrm{W}}}
\def \MZ     {M_{\mathrm{Z}}}
\def \Mt     {M_{\mathrm{t}}}
\def \eone   {\epsilon_{1}}
\def \etwo   {\epsilon_{2}}
\def \ethree {\epsilon_{3}}

\def \alfasmz {\alpha_{s}(\MZ)}
\newcommand{\Zzero}{\mbox{${\mathrm{Z}^0}$}}
\newcommand{\WW}{\mbox{$\mathrm{W^+W^-}$}}
\newcommand{\qq}{\mbox{$\mathrm{q\overline{q}}$}}

\newcommand{\lnu}{\mbox{$l\nu$}}
\newcommand{\enu}{\mbox{$\mathrm{e}\nu$}}
\newcommand{\mnu}{\mbox{$\mu\nu$}}
\newcommand{\tnu}{\mbox{$\tau\nu$}}

\newcommand{\WWqqqq}{\mbox{\WW$\rightarrow$\qq\qq}}

\newcommand{\Zqq}{\mbox{\mathrm{Z}/$\gamma\rightarrow$\qq}}
\newcommand{\Zgqq}{\mbox{(\Zzero/$\gamma)^{*}\rightarrow$\qq}}
\newcommand{\ZZ}{\mbox{(\Zzero/$\gamma)^{*}$(\Zzero/$\gamma)^{*}$}}
\newcommand{\WWqqlnu}{\mbox{\WW$\rightarrow$\qq\lnu}}
\newcommand{\WWqqenu}{\mbox{\WW$\rightarrow$\qq\enu}}
\newcommand{\WWqqmnu}{\mbox{\WW$\rightarrow$\qq\mnu}}
\newcommand{\WWqqtnu}{\mbox{\WW$\rightarrow$\qq\tnu}}
\newcommand{\WWlnulnu}{\mbox{\WW$\rightarrow$\lnu\lnu}}
\newcommand{\Ecm}{\mbox{$E_{\mathrm{cm}}$}}

\def \Mz       {\mbox{$M_{\mathrm{Z}}$} }
\def \Mh       {\mbox{$M_{\mathrm{H}}$} }

\newcommand{\Zee}{\mbox{\Zzero\ ee}}

\newcommand{\Lepone}{\mbox{LEP1}}
\newcommand{\Ebeam}{\mbox{$E_{\mathrm{beam}}$}}

\def    \be             {\begin{equation}}
\def    \ee             {\end{equation}}
\newcommand{\GW}{\Gamma_{\mathrm{W}}}
\newcommand{\GZ}{\Gamma_{\mathrm{Z}}}
\renewcommand{\d}{\mathrm{d\,}}
\newcommand{\dd }{\mathrm{d^2\,}}

\newcommand{\MWIindx}[1]{\mbox{\scriptsize #1}}
\newcommand{\MWIbar}[1]{\overline{\mbox{#1}}}
\newcommand{\MWImw}{\mbox{$M_{\mbox{\scriptsize W}}$}}
\newcommand{\MWIgw}{\mbox{$\Gamma_{\mbox{\scriptsize W}}$}}
\newcommand{\MWIq}{\mbox{q}}
\newcommand{\MWIqbar}{\overline{\mbox{q}}}
\newcommand{\MWIe}{\mbox{e}}
\newcommand{\MWIw}{\mbox{W}}
{\end{list}}
\newcounter{MWIenumct}

\begin{document}

%%%%%%%%%%%%%%%maintext%%%%%%%%%%%
\begin{center}
{\large \bf DETERMINATION  OF THE MASS OF THE  W BOSON}
\end{center}
\begin{center}
{\it Conveners}: Z.~Kunszt and W.~J.~Stirling
\end{center}
%\begin{center}
{\it Working group}: 
A.~Ballestrero,
S.~Banerjee,
A.~Blondel,
M.~Campanelli,
F.~Cavallari,
D.~G.~Charlton, 
H.~S.~Chen,
D.~v.~Dierendonck,
A.~Gaidot,
Ll.~Garrido,
D.~Gel\'e,
M.~W.~Gr\"unewald,
G.~Gustafson,
% Hakkinen,
C.~Hartmann, 
F.~Jegerlehner,
A.~Juste,
S.~Katsanevas, 
V.~A.~Khoze, 
N.~J.~Kj{\ae}r, 
L.~L\"onnblad, 
E.~Maina, 
M.~Martinez,
R.~M{\o}ller,
G.~J.~van~Oldenborgh, 
J.~P.~Pansart,
P.~Perez, 
P.~B.~Renton, 
T.~Riemann, 
M.~Sassowsky,
J.~Schwindling, 
T.~G.~Shears,
T.~Sj\"ostrand, 
\v{S}.~Todorova, 
A.~Trabelsi,
A.~Valassi, 
C.~P.~Ward,
D.~R.~Ward, 
M.~F.~Watson,
N.~K.~Watson,
A.~Weber,
G.~W.~Wilson
%\end{center}
\vspace*{2.0cm}
%\vspace*{1.0cm}
%
\tableofcontents
\newpage

%\input s1_new.tex
% this version: Monday 8th January, final corrections... 
% this version: Monday 11th December, minor updates, Fred's new table... 
% this version: Monday 4th December, mssm figure now included 
% this version: Friday 24th November 3.30pm 
%
%%%%%%%%%%%%%%%%%%%%%%%%%%%%%%%%%%%%%%%%%%%%%%%%%%%%%%%%%%%%%%%%%%%%%%%%%%
%   please note the following conventions and macros:                    %
%                                                                        %
%   Fig.~\ref{}, Table~\ref{}, Section~\ref{} and Ref.~\cite{}.          %
%                                                                        %
%   for masses please use $\Mw$, $\Mh$, $\Mt$, etc.,                     %
%   and if you wish, units are provided by e.g. $25\mev$, $161\gev$,     %
%   $100\pb^{-1}$ which puts the units in roman and adds a space after   %
%   the number.                                                          %
%                                                                        %
%%%%%%%%%%%%%%%%%%%%%%%%%%%%%%%%%%%%%%%%%%%%%%%%%%%%%%%%%%%%%%%%%%%%%%%%%%

\section{Introduction and Overview\protect\footnotemark[1]}
\footnotetext[1]{prepared by F.~Jegerlehner, Z.~Kunszt, G.-J.~van Oldenborgh,
P.B.~Renton, T.~Riemann, W.J.~Stirling}

\label{sec:mw-intro}
Previous studies \cite{aachen} of the physics potential
of LEP2 indicated that  with the design
luminosity of $500\pb^{-1}$ one may get a direct measurement
of the  W  mass with a precision  in the range
$30-50\mev$.
 This report presents an updated  evaluation of the estimated
error  on $\Mw$ based on  recent simulation work and
improved theoretical input. The most efficient experimental
methods which will be used are also described.

\subsection{Machine parameters}
\label{sec:lep2wmachine}

The LEP2 machine parameters are by now  largely determined.
Collider energy values and time-scales for the various runs,
expected luminosities and errors on the   beam energy and  luminosity 
are discussed and summarized elsewhere in this report
\cite{lep2wmachine,machine}.
Here we note that (i) collider energies in the range $161-192\gev$ 
will be available,
and (ii)  the  total luminosity is expected to be
approximately $500\pb^{-1}$ per experiment.
It is likely that the bulk of the luminosity will be delivered at
high energy ($\sqrt{s}\gapproxeq 175\gev$). The
 beam energy will be known to within an uncertainty of $12\mev$,
%up to $\sqrt{s}=195\gev$, 
and the luminosity is expected to be
measured  with a precision better than 1\%.

\subsection{Present status of $\Mw$ measurements}

%----------------------------------------------------------------------
\begin{figure}[htb]
\vspace*{0.1cm}
\centerline{
\epsfig{figure=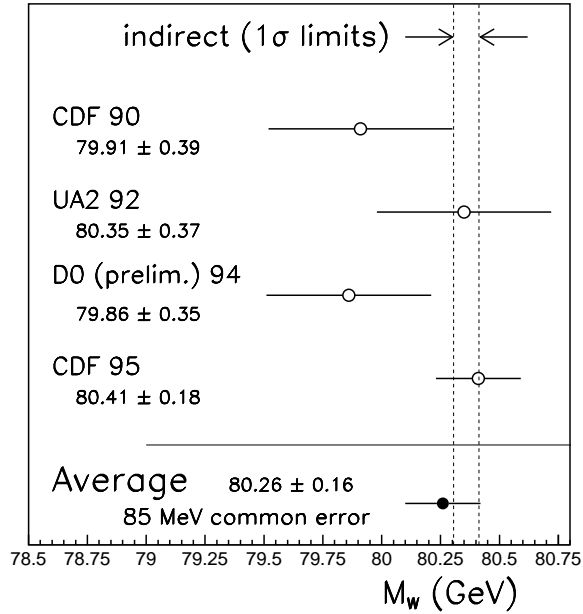,height=9cm,angle=0}
}
\caption{\it
Direct measurements of $\Mw$ together
 with their average. The dashed lines correspond
to the indirect limits ($\pm$ 1 $\sigma$) from a global fit in
the  Standard Model.}
\label{fig-MW}
\end{figure}

 Precise measurements of the masses of the heavy gauge W and Z bosons
are of fundamental physical importance.
 The current precision from direct measurements 
is $\Delta\MZ$ = $\pm$ 2.2 MeV
and  $\Delta$M$_{W}$ = $\pm$ 160 MeV \cite{PBR}. So far, $\Mw$ has been
measured at the CERN \cite{UA2MW} and Fermilab Tevatron
\cite{CDFMW1,CDFMW2,D0MW} $\mathrm{p}\bar\mathrm{p}$ colliders.
The present measurements are summarized in
Fig.~\ref{fig-MW}.
In calculating the world average, a common
systematic error  of $\pm 85\mev$
arising from uncertainties in the  parton distributions
functions is taken into account.
The {\it current world average value} is
\beq\label{mwworldav}
 \Mw = 80.26\, \pm\, 0.16\, \gev\,.
\eeq
An indirect determination of $\Mw$
 from a global Standard Model (SM) fit to electroweak data from LEP1 and
 SLC \cite{PBR}  gives the  more   accurate value
\beq
\Mw = 80.359\, \pm\, 0.051\; ^{+0.013}_{-0.024} \gev\,.
\label{eq:mwfit}
\eeq
In Fig.~\ref{fig-MW} this range is indicated by dashed
vertical lines.  Note that the central value in (\ref{eq:mwfit})
 corresponds to
$\Mh = 300\gev$ and the second error indicates the change
in $\Mw$ when $\Mh$ is varied between $60\gev$ and $1000\gev$ -- 
increasing $\Mh$ decreases $\Mw$.
%There is a correlation between the allowed range
%of  $\Mw$ and the value of the top quark mass $\Mt$, as
%shown in Fig.~\ref{MWMTN}.
%----------------------------------------------------------------------
\begin{figure}[htb]
\vspace*{0.1cm}
\centerline{
\epsfig{figure=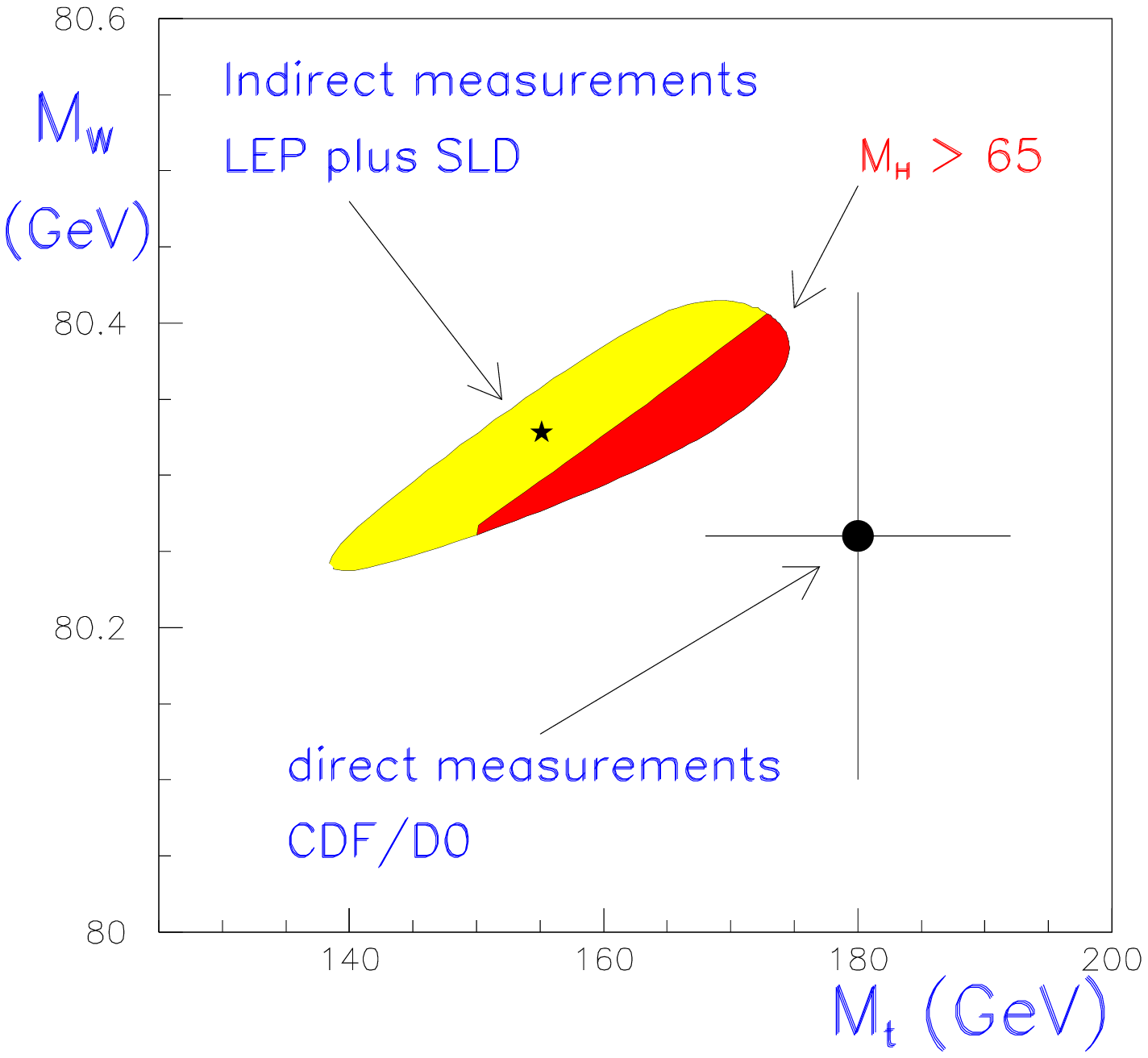,height=9cm,angle=0}
}
\caption{\it
A comparison of direct versus indirect determinations of $\Mt$ and $\Mw$.
The contour for the indirect determination corresponds to a 70\% confidence
level. The dark shaded region within the contour is that compatible with
the direct Higgs search limit, $\Mh$ $ > 65\gev$.
}
\label{MWMTN}
\end{figure}
The direct measurement of $\Mw$ becomes particularly 
interesting if its error can be made
 comparable to, or smaller than, the error
of the indirect measurement, i.e. $ < O(50)\mev$.
In particular,  a precise value of $\Mw$ obtained
from direct measurement could contradict the value determined
indirectly from the global fit, thus
 indicating a breakdown of the Standard Model. 
An improvement
in the precision of the $\Mw$ measurement can be used
to further constrain the allowed values of the Higgs boson mass
in the Standard Model, or the  parameter space of the
Minimal Supersymmetric Standard Model (MSSM) .

 Standard Model fits to electroweak data determine values for $\Mw$
(or $\Mh$), $\Mt$, and $\alfasmz$.
The direct determinations of the top quark mass
\cite{CDFTOP,D0TOP} give an
average value of $\Mt = 180 \pm 12 \gev$.
Fig.~\ref{MWMTN} compares the direct
determinations of $\Mt$ and $\Mw$ with
the indirect determinations obtained from fits to
electroweak data \cite{PBR}.
Note the correlation between the two masses in the latter.
Within the current accuracy, the
direct and indirect measurements are in approximate agreement. 
The central values of $\Mw$ and their errors, determined in several
ways from indirect electroweak fits, are
given in Table~\ref{MWVALS}.
%----------------------------------------------------------------------------
\begin{table}[hbt]
\begin{center}
\begin{tabular}{|c|c|c|c|}
\hline
\rule[-1.2ex]{0mm}{4ex}   &   all data & R$_{b}$ and R$_{c}$ excluded & R$_{b}$, R$_{c}$ 
and A$_{LR}$ excluded  \\
\hline
\rule[-1.2ex]{0mm}{4ex}$\Mt$ (GeV) & 155$^{+11}_{-11}$ & 162$^{+21}_{-12}$  & 167 $^{+45}_{-20}$ \\
\rule[-1.2ex]{0mm}{4ex} $\Mh$ (GeV) & 32$^{+49}_{-17}$  & 44$^{+207}_{-27}$   
& 144$^{+856}_{-111}$   \\
\rule[-1.2ex]{0mm}{4ex}$\alfasmz$ & 0.1221$^{+0.0037}_{-0.0037}$ & 0.1217$^{+0.0037}_{-0.0037}$ &
0.1233$^{+0.0084}_{-0.0040}$ \\
\hline
\rule[-1.2ex]{0mm}{4ex}$\Mw$ (GeV)  & 80.329$^{+0.022}_{-0.042}$ & 80.358$^{+0.032}_{-0.103}$ &
80.319$^{+0.076}_{-0.148}$ \\
\hline
\end{tabular}
\end{center}
\caption{\it Results of Standard Model fits to $\Mt$, $\Mh$ and $\alfasmz$.
The fits are to LEP1 and SLC data. An upper limit $\Mh = 1000\gev$ has
been imposed. The electromagnetic
coupling constant $\alpha$(M$_{Z}$) is also determined in these fits. 
Also given is 
the result of the fit given
in terms of $\Mw$. The results are given for all data, as well as for
excluding R$_{b}$ and R$_{c}$ and also A$_{LR}$.
}
\label{MWVALS}
\end{table}
%----------------------------------------------------------------------------
The results are evidently somewhat sensitive to
the inclusion (or not) of data on  the Z partial width
ratios R$_{b}$ and R$_{c}$ and the SLD/SLC measurement of A$_{LR}$, all of
which differ by 2.5 standard deviations or more from the Standard Model values.
However, the conclusion on the agreement of the direct and indirect 
determinations is unchanged. 
%It is again clear from
% Fig.~\ref{MWMTN} that if the directly measured central values
%of $\Mw$ and $\Mt$ were to remain the
%same, and if the errors were to be  reduced significantly,
%then an inconsistency in the Standard Model could be established.
As we shall see in the following sections, a
 significant reduction in the error on $\Mw$ is  expected from both
LEP2 and the Tevatron.

\subsection{Improved precision on $\Mw$ from the Tevatron}

 The Tevatron data so far analysed, and shown in Fig.~\ref{fig-MW},
come from the 1992/3 data-taking (Run 1a). The results from 
CDF \cite{CDFMW2} are based on approximately $19\pb^{-1}$ and are final,
whereas those from D0 \cite{D0MW} are based on
approximately $13\pb^{-1}$ and
are still preliminary. It is to be expected that
the final result will have a smaller error. In addition, there will be a
significantly larger data sample from the 1994/6 data-taking (Run 1b). 
This should amount to more than $100\pb^{-1}$ of useful data for each
experiment. When these data are analysed it is envisaged that the
total combined error on $\Mw$ will be  reduced to
about $70\mev$.
% or maybe better depending on progress in 
%reducing the systematic errors. 
In particular, the combined CDF/D0 result
will depend on the common systematic error arising from uncertainties
in the parton distribution functions.
Thus when the first $\Mw$ measurements emerge from LEP2
one may assume that the world average error will have
approximately this value.
For more details see Ref.~\cite{einsweiler}.

After 1996 there will be a significant
break in the Tevatron programme. Data-taking will start
 again in 1999 with a much
higher luminosity (due to the main injector and other improvements).
Estimating the error on $\Mw$ which will ultimately be achievable
(with several fb$^{-1}$ of total luminosity) is clearly more difficult.
If one assumes that an increase in the size of the  data sample
leads to a steady reduction in the systematic errors, one might
optimistically envisage that
the combined precision from the Tevatron experiments
 will eventually be in the
 $\Delta \Mw = 30 - 40\mev$ range, assuming a  common systematic
error of about $25\mev$ \cite{demarteau}. However it is important
to remember that these improved values will be obtained
{\it after} the LEP2 measurements.

\subsection{Impact of a precision measurement of $\Mw$}

 Within the Standard Model, the value of $\Mw$ is sensitive to both $\Mt$
and $\Mh$. For example, for a fixed value of $\Mh = 300 \gev$, a precision
of $\Delta\Mw = \pm 25\mev$ translates to a precision on $\Mt$
of $\pm 3.9\gev$.
 The impact of a precise measurement of $\Mw$ on the indirect
determination of $\Mh$  is shown in Table~\ref{MHVALS}.

%----------------------------------------------------------------------------
\begin{table}[hbtp]
\begin{center}
\begin{tabular}{|c|c|c|c|c|}
\hline
\rule[-1.2ex]{0mm}{4ex} & \multicolumn {2}{c|}{ $\Mt = 180\gev$ (fixed)} &
\multicolumn {2} {c|}{ $\Mt = 180 \pm 5\gev$} \\
\hline
\rule[-1.2ex]{0mm}{4ex} $\Mh$ (GeV) & $\Delta\Mw = 25\mev$ & $\Delta\Mw= 50\mev$
& $\Delta\Mw = 25\mev$ & $\Delta\Mw = 50\mev$ \\
\hline
\rule[-1.2ex]{0mm}{4ex}  $100$  & $+48$, $-36$ & $+112$, $-63$ & $+86$, $-54$ & $+140$, $-72$ \\
\hline
\rule[-1.2ex]{0mm}{4ex}  $300$  & $+111$, $-84$ & $+259$, $-148$ & $+196$, $-126$ & $+323$,
$-168$ \\
\hline
\rule[-1.2ex]{0mm}{4ex} $800$  & $+297$, $-212$ & $+740$, $-367$ & $+538$, $-310$ & $+958$,
$-413$ \\
\hline
\end{tabular}
\end{center}
\caption{\it Estimated error on $\Mh$ (in GeV) for several values
of $\Mh$ and for $\Delta\Mw = 25$ and $50\mev$. The estimates
are given for both $\Mt = 180\gev$ (fixed) and for $\Mt = 180 \pm 5\gev$.
All other Standard Model parameters are fixed.}
\label{MHVALS}
\end{table}

\begin{figure}[htb]
\vspace*{0.1cm}
\centerline{
\epsfig{figure=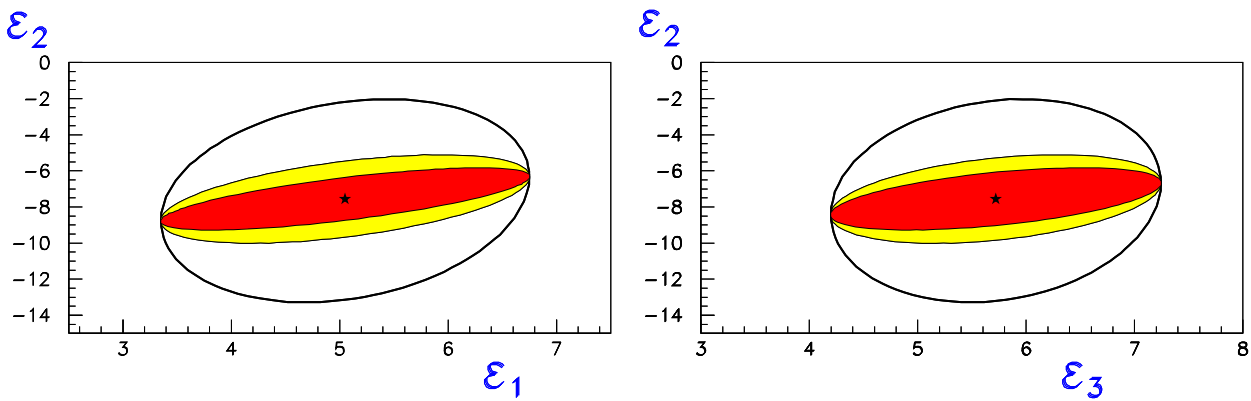,height=6cm,angle=0}
}
\caption{\it 70\% confidence level contour plots
for fits to $\epsilon$'s in units of $10^{-3}$. The outer contour is for
a fit with the current error, $\Delta\Mw = \pm 160\mev$, whereas the
inner contours are for  $\Delta\Mw = \pm 50\mev$
and $\Delta\Mw = \pm 25\mev$ respectively.
 }
\label{fig-eps}
\end{figure}

 In order to assess the impact of a precise measurement of $\Mw$ it is
necessary to make an estimate of the improvements which will be made on
the electroweak data from LEP1 and SLC. Details of the improvements 
which are assumed here are discussed in \cite{PBR1}. The importance of
a precise measurement of $\Mw$ can perhaps best be appreciated by
considering the (almost) model independent $\epsilon$ parameters \cite{Altar}.
The parameter $\eone$ ($= \Delta\rho$)  is sensitive mainly to the
Z partial and total widths. The parameter $\ethree$ depends linearly on both
$\Delta\rho$  and $\Delta\kappa$, where $\Delta\kappa$ is determined from
$\sin^{2}\theta_{eff}$. The parameter $\etwo$ depends linearly on
$\Delta\rho$, $\Delta\kappa$ and $\Delta$r$_{W}$. This latter quantity 
is determined essentially from $\Mw$, and so improvements in the precision
of $\etwo$ depend directly on improving the error on $\Mw$.
This is illustrated
in Fig.~\ref{fig-eps}, which shows the 70\% confidence level contours
for fits to projected global electroweak data. The different contours
correspond to different values of $\Delta\Mw$. In these fits all
electroweak data measurements have been set to correspond to the Standard
Model values $\Mt = 180\gev$, $\Mh = 100\gev$
and $\alpha_{s}(\MZ) = 0.123$. The $\epsilon$ variables are
constructed to be sensitive to vector boson propagator effects,
from both physics within the Standard Model and beyond.

%----------------------------------------------------------------------

 Numerically, the projected data  give a precision 
\begin{equation}
  \Delta\eone = \pm 1.1 \times 10^{-3} \hspace*{1.5cm}
  \Delta\ethree = \pm 1.0 \times 10^{-3} \, .
\end{equation}
For $\Delta\Mw = 160\mev$, the error $\Delta\etwo = \pm 3.7
\times 10^{-3}$
is obtained, whereas for the projected errors on $\Mw$ one obtains
\begin{equation}
  \Delta\etwo = \pm 1.0 \, (\pm 1.6)\, \times 10^{-3} \hspace*{1.0cm}
\mbox{for} \hspace*{0.6cm} \Delta\Mw = 25 \, (50)\mev\, .
\end{equation}
 The smaller the  volume in $\epsilon$
space allowed by the precision electroweak measurements, the greater the
constraint on physics beyond the Standard Model.

\begin{figure}[htb]
%\vspace*{0.1cm}
\centerline{
\epsfig{figure=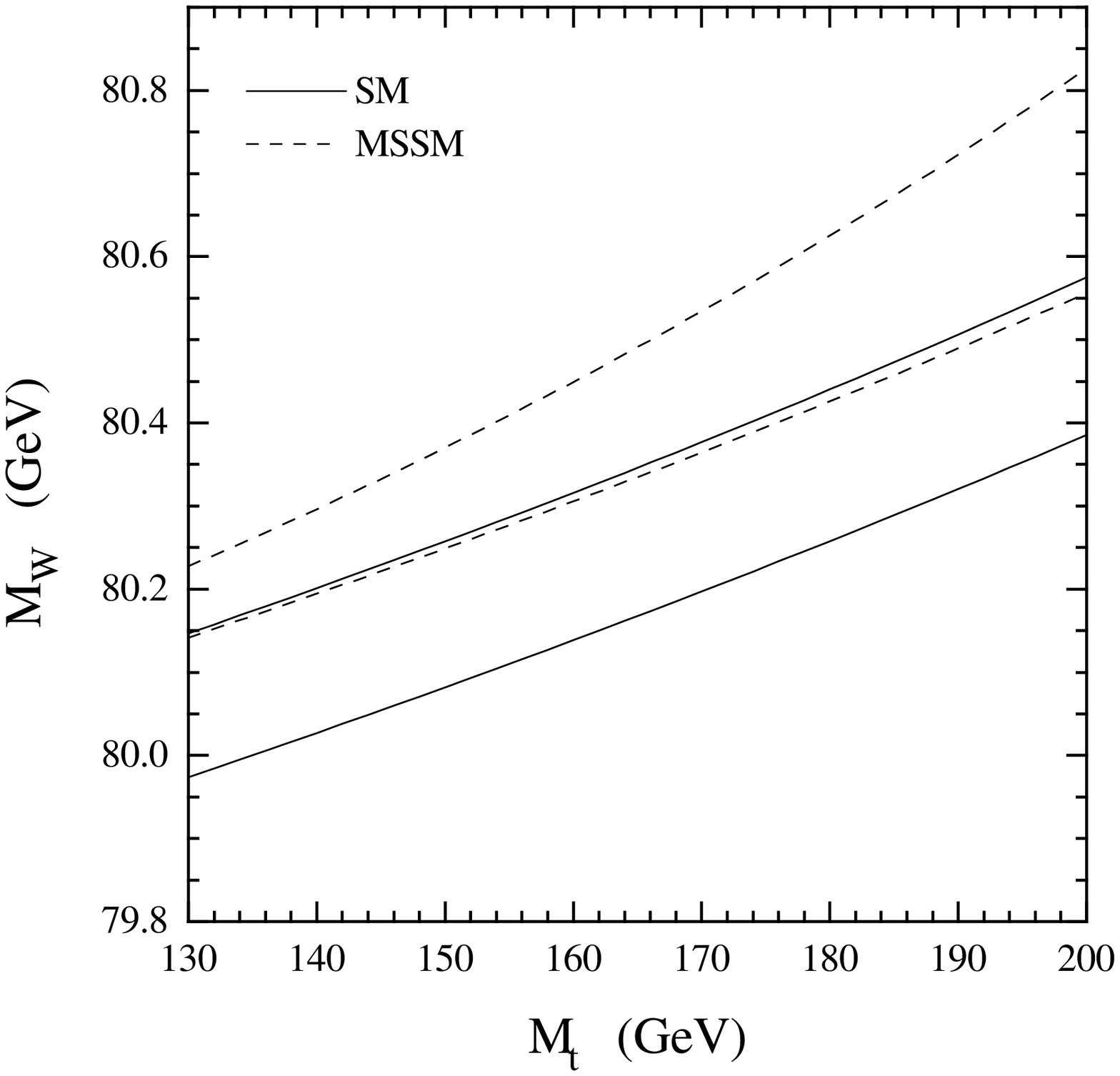,height=8cm,angle=0}
}
\caption{\it Predictions for $\Mw$ as a function of $\Mt$ in the 
SM (solid lines) and in 
the MSSM (dashed lines), from Ref.~\protect\cite{holliketal}.
 In each case the prediction is a band of values,
corresponding to a variation of the model parameters 
 consistent with current measurements and limits.
An additional constraint of `no SUSY particles at LEP2' is imposed in
the MSSM calculation.
 }
\label{mw:mssm}
\end{figure}

The MSSM is arguably the most promising new-physics candidate.
It is therefore especially important to consider the MSSM prediction
for  $\Mw$. Figure~\ref{mw:mssm} \cite{holliketal} shows
$\Mw$ as a function of $\Mt$ in the SM (solid lines) and in 
the MSSM (dashed lines). In each case the prediction is a band of values,
corresponding to a variation of the model parameters (dominantly
$\Mh$ in the SM case, with $90\gev  <\Mh < 1000\gev$ chosen here)
 consistent with current measurements and limits.
An additional constraint of `no SUSY particles at LEP2' is imposed in
the MSSM calculation.
%----------------------------------------------------------------------

\subsection{Methods for measuring $\Mw$}
\label{sec:wmass_int_methods}

Precise measurements of  $\Mw$ can in principle be obtained using
the enhanced statistical power of
 the rapidly varying total cross-section at threshold, 
 the sharp (Breit-Wigner) peaking  behaviour of the invariant-mass distribution
of the W$^\pm$ decay products  and  the sharp end-point spectrum
of the lepton energy in W$^\pm$ decay.
One can obtain a rough idea of the relative power of these
methods by estimating their  statistical precision assuming
100\% efficiency, perfect detectors and no background.
More complete discussions  are given in Sections~\ref{sec:wmass_threshold}
and \ref{sec:mwdir}.
%%%%%%%%%%%%%%%%%%%%%%%
\begin{itemize}
\item[{\bf A)}\ ] {\bf Threshold cross-section measurement }
of the process $\epem \to\WW$. 
The statistical power
of this method, assuming 100\% signal  efficiency and no background, is 
\be\label{delmwA}
\Delta \Mw \ge 91\mev \sqrt{{100\pb^{-1}\over {\cal L}}}\, ,
\ee
where the minimum value is attained 
 at $\sqrt{s} \simeq  161\gev$. Here ${\cal L}$ denotes the total integrated
luminosity. 
\item[{\bf B)}\ ] {\bf Direct reconstruction} methods, which reconstruct
the Breit-Wigner resonant shape from the W$^\pm$ final states 
 using  kinematic fitting
  techniques to improve the mass resolution.
The statistical power
of this method, again  assuming 100\% efficiency, perfect detector
resolution  and no background, can be estimated as  
\be
\label{delmwB}
%\Delta \Mw \ge \frac{\Gamma_{\rm W}}{\sqrt{N}}
\Delta \Mw \sim \frac{\Gamma_{\rm W}}{\sqrt{N}}
\approx  50 \mev\sqrt{{100\pb^{-1}\over {\cal L}}}\, ,
\ee
 approximately  independent of the collider energy.
This order of magnitude estimate 
 is confirmed by more detailed studies, see below.
\item[{\bf C)}\ ] Determination of $\Mw$
from the {\bf lepton end-point energy}. The 
end-points
of the lepton spectrum in $\WW \to  \qq l  \nu$ depend 
quite sensitively on the  W mass. For on-shell W bosons at
leading order:
\begin{equation}
E_-  \leq E_{l}  \leq E_+\; , \qquad
 E_\pm =  {\sqrt{s}\over 4}\; \left( 1 \pm \sqrt{1-4\Mw^2/s} \right) \; .
\end{equation} 
In this case the statistical error on $\Mw$ is determined
by the statistical error on the measurement of the 
lepton end-point energy, 
\be
\Delta \Mw = \frac{\sqrt{s-4\Mw^2}}{\Mw}\; \Delta E_\pm   \, .
\label{eq:endpoint}
\ee
In practice, however, the end-points of the distribution are considerably
smeared by finite width effects and by initial state radiation, and only 
a fraction of 
events close to the end-points are sensitive to $\Mw$. This 
significantly weakens the statistical power of this method from what
the naive estimate (\ref{eq:endpoint}) would predict.
%Again assuming ideal conditions, the statistical error on $\Mw$ is
%\be
%\label{delmwC}
%\Delta \Mw \ge \frac{E_+\sqrt{s/4\Mw^2-1}}{\sqrt{N}}
%\approx  550 \mev\sqrt{{100\pb^{-1}\over {\cal L}}}
%\quad \mbox{\it to be corrected!}
%\ee
%at $\sqrt{s}=176\gev$, and about twice as large at
%$\sqrt{s}=192\gev$.
\end{itemize}
%%%%%%%%%%%%%%%%%%%%%%%%%%
The  detailed studies  described in the following sections
show that  the errors which can realistically be achieved in practice
are somewhat larger than the above estimates for Methods A and B. 
%although the {\it relative}
%precision of the two methods remains roughly the same.
The statistical precisions of the two methods
 are in fact  more comparable (for the 
same integrated luminosity) than the factor 2 difference suggested
by the naive estimates (\ref{delmwA}) and (\ref{delmwB}).
The overall statistical error for Method C has been estimated 
at ${\cal O}(300\mev)$ \cite{aachen} for $ {\cal L}=500\pb^{-1}$\,,
 significantly larger than that 
of the other two methods.
It will not therefore  be considered further here, although 
it is still a   valid measurement for cross-checking the other  results.

It is envisaged that  most of the LEP2 data will be collected at energies
well above threshold,  and so the statistically 
most precise determination of $\Mw$  will come from
Method B. However with a relatively modest amount of luminosity
spent at the \WW\ threshold (for example $50\pb^{-1}$ per experiment), 
Method A can provide a statistical error of order $100\mev$, not 
significantly worse than Method B and with very different systematics.
The two methods can therefore be regarded as complementary tools,
and both should be used to provide an internal cross-check 
on the measurement of the W mass at LEP2.
This constitutes the main motivation
for spending some luminosity in the threshold region.

The threshold cross-section method   is also of interest 
because it appears to fit very well into 
the expected schedule for LEP2 operation in 1996.
It is anticipated that 
the maximum beam energy at LEP2
will increase in steps,
with the progressive installation of
more superconducting RF cavities,
in such a way that a centre-of-mass energy of  161~GeV
will indeed be achievable during the first running period of 1996.
This would then be the ideal time to perform such a threshold measurement.
The achievable statistical error on $\Mw$ depends of course critically on the 
available luminosity at the threshold energy. 
In Section~\ref{sec:wmass_threshold} we present quantitative
estimates based on integrated  luminosities of 25, 50 and 100~pb$^{-1}$
per experiment.

\subsection{Theoretical input information}
\label{sec:theoryinput}
\subsubsection{Cross-sections for the \WW\ signal and backgrounds}

Methods (A) and (B) 
for measuring $\Mw$ described above require rather different
theoretical input. The threshold method relies on the comparison of 
an absolute cross-section measurement with a theoretical calculation
which has $\Mw$ as a  free parameter. The smallness of the cross-section
near threshold is compensated by the enhanced sensitivity to $\Mw$  
in this region. In contrast, the direct reconstruction method makes use of
the large \WW\ statistics at the higher LEP2 energies, $\sqrt{s} \gapproxeq
175$~GeV. Here the more important issue is the  accurate modeling 
of the W$^\pm$ {\it line-shape}, i.e. the distribution in the invariant mass
of the W$^\pm$ decay products.

In this section we describe some of the important features of the theoretical
cross-sections which are relevant for the $\Mw$ measurements. A more
complete discussion can be found in the contribution of the WW and Event 
Generators Working Group to this Report \cite{WWGROUP}.

We begin by writing the cross-section for
$\epem \to 4\mathrm{f}\; (+ \gamma,\mathrm{g},\ldots)$,
 schematically, as
\begin{eqnarray}
\sigma &= & \sigma_{\rm WW} \; +\; \sigma_{\rm bkd} \; , \nonumber \\
\sigma_{\rm WW} &= & \sigma_0^{\rm WW} \; (1 + \delta_{\rm EW}
+ \delta_{\rm QCD}) \; ,
\label{eq:decomposition}
\end{eqnarray} 
We note
that this decomposition of the cross-section into
 `signal' and `background' contributions is practical rather than
theoretically rigorous, since neither contribution is separately 
exactly gauge
invariant nor experimentally distinguishable in general. 
The various terms in (\ref{eq:decomposition}) correspond to 
\begin{itemize}
\item[{(i)}] $\sigma_0^{\rm WW}$: the Born contribution from the 3 
`CC03' leading-order
diagrams for $\epem \to \WW$ involving $t$-channel $\nu$ exchange
and $s$-channel $\gamma$ and Z$^0$ exchange, calculated using off-shell W
propagators.
\item[{(ii)}]
 $\delta_{\rm EW}$: higher-order electroweak radiative corrections,
including loop corrections, real photon emission, etc.
\item[{(iii)}] $\delta_{\rm QCD}$: higher-order QCD corrections
to \WW\ final states containing \qq\ pairs. For the threshold measurement,
where in principle only  the {\it total} cross-section is of primary interest,
the  effect of these is to generate small
corrections  to the hadronic branching ratios which are entirely
straightforward to calculate and take into account.
More generally, such QCD corrections can lead to additional jets in the final
 state, e.g. $\WW\to\qq\qq\mathrm{g}$ from one hard gluon emission.
This affects the direct reconstruction method, insofar as the kinematic fits
to reconstruct $\Mw$  assume a four-jet final state, 
and both methods insofar as cuts have to be imposed in order to suppress
the QCD background
(see Sections~\ref{sec:wmass_threshold},\ref{sec:mwdir} below). 
Perturbative QCD corrections, real gluon emission to $O(\alpha_s^2)$
and real plus virtual emission  to $O(\alpha_s)$, have been recently 
discussed in Refs.~\cite{MAINA,MAINABIS} respectively, 
together with their impact on the measurement of $\Mw$.
\item[{(iv)}] $\sigma_{\rm bkd}$: `background' contributions, for example from 
non-resonant diagrams (e.g. $e^+e^-\to \mu^+\nu_\mu \mathrm{W}^-$) 
and QCD contributions
$\epem\to \qq \mathrm{gg}(\gamma),\; \qq\qq(\gamma)$ to the four-jet final state.
All of the important backgrounds  have been calculated, see
Table~\ref{tab-MCXsect} below. At threshold, the QCD four-jet background
is particularly large in comparison to the signal. 
\end{itemize}
In what follows we consider (i) and (ii) in some detail. Background
contributions and how to suppress them are considered in later sections.

\subsubsection{The \WW\ off-shell cross-section}
\noindent
The leading-order cross-section for off-shell \WW\
production was first presented in Ref.~\cite{MUTA}:
\begin{equation}
\sigma(s) = \int_0^s d s_1 \int_0^{(\sqrt{s}-\sqrt{s_1})^2} ds_2\;
\rho(s_1)\rho(s_2)\; \sigma_0(s,s_1,s_2) \; ,
\label{sig}
\end{equation}
where
\begin{equation}
\rho(s) = {1\over \pi} {\GW\over \Mw}\; {s\over 
(s-\Mw^2)^2 + s^2 \GW^2/\Mw^2 }  \; .
\label{rho}
\end{equation}

The cross-section $\sigma_0$ can be written in terms of the $\nu$, $\gamma$ and
Z exchange contributions and their interferences:
 \begin{equation}
\sigma_0(s,s_1,s_2) = {g^4\over 256 \pi s^2 s_1 s_2} \;
\left[ a_{\gamma\gamma} +  a_{ZZ} + a_{\gamma Z} +
a_{\nu\nu} +  a_{\nu Z} + a_{\nu\gamma} \right] \; ,
\label{siglo}
\end{equation}
where 
$g^4=e^4/\sin^4\theta_W  $. Explicit  expressions for the various contributions
can be found in Ref.~\cite{MUTA} for example.
The stable (on-shell) \WW\ cross-section is simply
\begin{equation}
\sigma^{\rm on}(s) =  \sigma_0(s,\Mw^2,\Mw^2)\; .
\label{sigon}
\end{equation}
A theoretical ansatz of this kind  will be the basis of any
experimental determination of the mass and width of the W boson.
The reason for this is the large effect of the virtuality of the W
bosons produced around the nominal threshold.
An immediate conclusion from Eq.~(\ref{sig}) may be drawn: 
the W mass influences the cross sections exclusively through the
off-shell $W$ propagators; all the other parts are independent of 
$\Mw$ and $\GW$ (neglecting for the moment the relatively minor
dependence due to radiative corrections).
It will be an important factor in the discussion which follows 
that near threshold the (unpolarized) cross-section is completely dominated
by the $t$-channel neutrino exchange diagram. This leads to an $S$-wave
threshold behaviour $\sigma_{t} \sim \beta$, whereas the $s$-channel
vector boson exchange diagrams give the characteristic
  $P$-wave behaviour $\sigma_{s} \sim \beta^3$.

By tradition (for example at LEP1 with the Z boson), the virtual W
propagator in Eq.~(\ref{rho}) uses an $s$-dependent width,
%----------------
\begin{equation}
\GW(s) = {s \over \Mw^2} \; \GW\; ,
 \label{ggg}
\end{equation}
%--------------
where $\GW \equiv \GW(\Mw^2)$. Another choice, equally well justified
from a theoretical point of view,
 would be to use a constant width in the W propagator (for a discussion see
Ref.~\cite{fred}):
%----------------
\begin{eqnarray}
{\overline \rho} = \frac{1}{\pi} \frac{ {\overline M}_{\mathrm{W}} {\overline
    \Gamma}_{\mathrm{W}}}{\left(s- {\overline M}_{\mathrm{W}}^2\right)^2
  + {\overline M}_{\mathrm{W}}^2{\overline \Gamma}_{\mathrm{W}}^2} \; .
\label{barrho}
\end{eqnarray}
%--------------
The numerical values of the width and mass in the two expressions are
related~\cite{wmasstrafo}:
%----------------
\begin{eqnarray}
{\overline M}_{\mathrm{W}} &=& \Mw -\frac{1}{2} \frac{\GW^2}{\Mw}=\Mw - 26.9\mev\; ,
\\
{\overline \Gamma}_{\mathrm{W}} &=& \GW - \frac{1}{2} \frac{\GW^3}{\Mw^2} =
\GW - 0.7\mev\; .
%\label{}
\end{eqnarray}
%--------------
These relations may be derived from the following identity: 
$
(s-{\overline M}_{\mathrm{W}}^2 +i {\overline M}_{\mathrm{W}} 
 {\overline \Gamma}_{\mathrm{W}}) 
= 
(s-\Mw^2+is\GW / \Mw) / (1+i \GW / \Mw)
$. Numerically, the consequences are below the anticipated experimental
accuracy.   
  
\subsubsection{Higher-order electroweak corrections }

 The complete set of $O(\alpha)$ next-to-leading order corrections to
\WW\ production has been calculated by several groups 
\cite{Bielefeld,Leiden-Wuerzburg}, for 
the {\it on-shell} case only, see for example 
 Refs.~\cite{BeenDenn,WWGROUP} and references
therein. There has been some progress 
 with the off-shell (i.e. four
fermion production) corrections but the calculation is not yet complete.
However using the on-shell calculations as a guide, it is already possible to
predict some of the largest effects.
For example,
it has been shown that close to threshold the dominant contribution
comes from the Coulomb correction, i.e. the long-range electromagnetic
interaction between almost stationary heavy particles.
 Also important is the emission
of photons collinear with the initial state e$^\pm$
(`initial state radiation') which gives rise
to logarithmic corrections $\sim \alpha \ln(s/m_e^2)$. 
These leading logarithms
can be resummed to all orders, and incorporated for example
 using a `structure function'
formalism. In this case, the generalization from on-shell to off-shell
$W$'s appears to be straightforward. For the Coulomb corrections, however,
one has to be much more careful, since in this case the  inclusion of
 the finite $W$ decay width has a dramatic effect. Finally, one can
incorporate certain
 important higher-order fermion and boson loop corrections by a judicious
 choice of electroweak coupling constant. Each of these effects will be
discussed  in turn below.

In summary, certain $O(\alpha)$ corrections are already known
to be large because their coefficients involve large factors
like $\ln(s/m_e^2)$, $\sqrt{\Mw/\GW}$, $\Mt^2/\Mw^2$,  etc.
Once these are taken into account, one can expect that the remaining
corrections are no larger than ${\cal O}(\alpha )$.
When estimating the theoretical systematic uncertainty on the
W mass in Section~\ref{sec:wmass_threshold} below, we shall therefore assume a
conservative overall uncertainty on the cross-section
 of $\pm 2\%$ from the as yet uncalculated $O(\alpha)$ and higher-order
  corrections.

\subsubsection{Coulomb corrections}

\noindent The Coulomb corrections for on-shell
and off-shell \WW\ production have been discussed in detail
in Refs.~\cite{FADIN1,FADIN3,BARDIN1},
 where a complete set of  references to earlier studies
can also be found.

The result for on-shell \WW\ production is well-known \cite{SOMMERFELD} 
---  the $O(\alpha)$ correction diverges as $\alpha \pi /v_0$ 
as the relative velocity
 $v_0 = 2 [ 1-4 \Mw^2/ s]^{1/2} $ of the
$W$ bosons approaches zero at threshold. 
Note that $\sigma_0 \sim v_0$ near threshold and so the 
Coulomb-corrected cross-section is formally 
non-vanishing  when $\sqrt{s} = 2 \Mw$.
 
For unstable \WW\ production the finite decay width $\GW$
 {\it screens} the Coulomb singularity \cite{FADIN1}, so that very close to
 threshold the perturbative expansion in $\alpha\pi/v_0$ is effectively 
 replaced  by an expansion in $\alpha \pi\sqrt{\Mw/\GW}$ \cite{FADIN3}. 
In the calculations which follow we use the expressions for the
$O(\alpha)$ correction given in 
Ref.~\cite{FADIN3}.
 The net effect
 is a correction which reaches a maximum of approximately
 $+6\%$ in the threshold region.
Although this does not appear to be large, we will see below that it
changes  the threshold cross-section by an amount equivalent 
to a shift in $\Mw$  of order $100$~MeV.
In Ref.~\cite{FADIN3} the  $O(\alpha)$
 result is generalized to all orders.
However the contributions from second order and above change
the cross-section by $ \ll 1\%$ in the threshold region 
\cite{KMS} (see also \cite{BARDIN1}) and can therefore be safely neglected.

Note also that the Coulomb correction to the off-shell \WW\ cross-section
\begin{equation}
\sigma(s) = \int_0^s d s_1 \int_0^{(\sqrt{s}-\sqrt{s_1})^2} ds_2\;
\rho(s_1)\rho(s_2)\; \sigma_0(s,s_1,s_2) [ 1+ \delta_C(s,s_1,s_2)]
\label{eq:coul}
\end{equation}
provides an example of a (QED) interconnection effect between the two
W bosons: the exchange of a soft photon distorts the line shape 
($d\sigma/d\sqrt{s_1}$)
of the W$^\pm$ and therefore, at least in principle, affects the
direct reconstruction method \cite{MWIKSt,MWIKSj,MWIMY}.
In Ref.~\cite{MWIKSt}, for example, it is shown that 
 the Coulomb interaction between the W bosons
causes a downwards shift in the average reconstructed mass of $O(\pi\alpha
\GW)\sim O(20\mev)$. Selecting events close to the Breit-Wigner peak
reduces the effect somewhat.
 However the calculations 
are  not yet complete, in that QED interactions between the decay
products of the two W bosons are not yet fully included.

\subsubsection{Initial state radiation}
\label{sec:theory_isr}

\noindent Another  important class of electroweak radiative corrections
comes from the emission of photons from the incoming e$^+$ and e$^-$.
In particular, the emission of virtual and soft real photons with
energy $E < \omega$ gives rise to doubly logarithmic contributions
$\sim \alpha \ln(s/m_e^2) \ln(s/\omega^2)$ at each order in
perturbation  theory.  The infra-red ($\ln\omega$) logarithms cancel
when hard photon contributions are added, and the remaining
collinear ($\ln(s/m_e^2)$) logarithms can be resummed and incorporated in
the cross-section using a `flux function' or a `structure function' 
 \cite{FADIN4} (see
%also Refs.~\cite{BERENDS1,CACCIARI,BERENDS2,BARDIN2,zeuthen3,WWGROUP}).
also Refs.~\cite{BERENDS1,CACCIARI,BARDIN2,zeuthen3,WWGROUP}).

The ISR corrected cross section in the flux function ({\tt FF}) approach is
\begin{eqnarray}
\frac{d\sigma_{\mathrm{ISR}}(s)}{ds_1ds_2}
&=&
\int
\limits_{s_{\min}}^s \frac{ds'}{s} 
\Biggl\{ 
F(x,s) \Bigl[ \sigma_{\tt CCn}(s',s_1,s_2) 
+~\delta_C \, \sigma_{\tt CC3}(s',s_1,s_2) \Bigr] 
%\nonumber \\ &&
+~ \sigma_{\mathrm{ISR}}^{\mathrm{non-univ}}\Biggr\},
\label{sigff}
\end{eqnarray}
%---
where $x=1-s'/s$ and 
%---
\begin{eqnarray}
F(x,s)
&=&
t x^{t-1} (1+S) + H(s',s)\; ,
\end{eqnarray}
%---
with
%---
\begin{eqnarray}
\label{beta.e}
t&=&\frac{2\alpha}{\pi}\left[\ln\left(\frac{s}{m_e^2}\right)-1\right]\; .
\end{eqnarray}
The ${\cal S}$ term comes from soft and virtual photon
emission, the ${\cal H}$ term comes from hard photon
emission,   $\delta_C$ contains the Coulomb correction~(\ref{eq:coul}),
$\sigma_{\tt CC3}$
is the doubly resonating Born cross section, and  
$\sigma_{\tt CCn}$ the background contributions. 
Explicit expressions can be found in the above  references.
The additional term $\sigma_{\mathrm{ISR}}^{\mathrm{non-univ}}$ 
is discussed in Refs.~\cite{BARDIN2} and~\cite{WWGROUP}.
The {\it invariant mass lost} to photon radiation may be calculated as
%---------------------------
\begin{eqnarray}
\langle m_{\gamma} \rangle
&=&
\frac{1}{\sigma} 
\int ds_1 ds_2 \, 
\int \frac{ds'}{s} \frac{\sqrt{s}}{2}\left(1-\frac{s'}{s}\right)
\frac{d\sigma}{ds_1ds_2ds'},
\label{eq:22}
\end{eqnarray}
%-----------------------------------------------
where $d\sigma/ds_1ds_2ds'$ is the contents of the curly brackets
in~Eq.~(\ref{sigff}). 

Alternatively, the
structure function ({\tt SF}) approach may be used:
%-----------------------------------------
\begin{eqnarray}
\frac{d\sigma_{\mathrm{QED}}(s)}{ds_1ds_2} 
&=& 
\int
 \limits_{x_1^{\min}}^1 dx_1 
\int
 \limits_{x_2^{\min}}^1 dx_2 
\, D(x_1,s) D(x_2,s)
\Biggl\{
\sigma_{\tt CCn}(x_1x_2s,s_1,s_2) 
%\nonumber \\ && 
+~
\delta_C~ \sigma_{\tt CC3}
%(x_1x_2s,s_1,s_2)
\Biggr\}\;  
, 
\label{sigsf}
\end{eqnarray}
where 
%---
\begin{eqnarray}
D(x,s)
&=&\frac{t}{2}
(1-x)^{t/2-1}  (1+{\overline S}) + {\overline H}(x,s)\; .
\label{dxs}
\end{eqnarray}
%-------------------
Here, the {\it invariant mass loss} is
%---------------------------
\begin{eqnarray}
\langle m_{\gamma} \rangle
&=&
\frac{1}{\sigma} \int ds_1 ds_2
\int dx_1dx_2 \, D(x_1,s) D(x_2,s) \frac{\sqrt{s}}{2}(1-x_1x_2)
\frac{d\sigma(x_1x_2s,s_1,s_2)}{ds_1 ds_2dx_1dx_2}.
\label{eq:25}
\end{eqnarray}
In addition, the radiative {\it energy loss} may be determined,
%------------------------
\begin{eqnarray}
\langle E_{\gamma} \rangle
&=&
\frac{1}{\sigma} 
\int ds_1 ds_2
\int dx_1dx_2 \, D(x_1,s) D(x_2,s) \frac{\sqrt{s}}{2}(2-x_1-x_2)
\frac{d\sigma}{ds_1 ds_2dx_1dx_2} \; .
\nonumber \\
\label{eq:26}
\end{eqnarray}
%-------------------

Initial state radiation affects the W mass measurement in two ways.
Close to threshold the cross-section is smeared out, thus reducing
the sensitivity  to $\Mw$ 
(see Fig.~\ref{fig-sigww} below). 
For the direct reconstruction method, the relatively large
average energy carried away by the radiated photons leads to
a large positive mass-shift if it is not 
 taken into account in the rescaling of the final-state momenta
to the beam energy (see Section~\ref{sec:mwdir} below). By rescaling
to the nominal beam energy 
we obtain for the mass-shift $\Delta \Mw$\,$=$\,
$\langle E_\gamma \rangle \Mw /\sqrt{s}$. Note however that a  fit to the
mass distribution gives more weight to the peak, and  therefore in practice
the effective value of $\langle E_\gamma \rangle$ or
$\langle m_\gamma \rangle$ is less than that given by 
Eqs.~(\ref{eq:22},\ref{eq:25},\ref{eq:26})
(see Section~\ref{sec:sys_error}).  
Table~\ref{ISRlost} shows  the influence of the various cross-section 
contributions on the average energy and invariant mass losses.
The invariant mass loss may be calculated both in the {\tt SF} and
{\tt FF} approaches. 
A comparison shows that the predictions in both schemes differ only
slightly, which allows us to use the numerically faster 
{\tt FF} approach for the numerical estimates.
At the lower LEP2 collider energies,  the energy  and invariant mass
losses are nearly equal, while at higher energies their difference 
is non-negligible.  
Note also that the inclusion of the non-universal ISR corrections and 
 background terms is of minor influence. 
The latter has been studied only for
{\tt CC11} processes; for reactions of the {\tt CC20} type  the background
is larger and the numerical estimates are not yet available.
The Coulomb correction is  numerically important and cannot be
neglected \cite{BARDIN1}.   
The dependence of the predictions on  the details of the treatment of QED is
discussed in detail in~Ref.~\cite{WWGROUP} and will not be repeated here.  

%ttttttttttttttttttttttttttttttttttttttttttttttttttttttt
% new table supplied by Fred 10 Dec
%\begin{table}[bhtp]
\begin{table}[htbp]
\begin{center}
{
\begin{tabular}[]{|c|c|c|c|c|}
\hline %---------------------------------------------------
%   &&&&
%\\
%  & \hspace{.5cm}  $ \hspace{.5cm} \sqrt{s}$ \hspace{.5cm}  &
  & \hspace{.5cm}  $  \sqrt{s}$ (GeV) \hspace{.5cm}  &
  \hspace{.5cm}   175 \hspace{.5cm}  & \hspace{.5cm}  192
  \hspace{.5cm}  & \hspace{.5cm}  205 \hspace{.5cm}
\\ \hline %---------------------------------------------------
{\tt SF}, {\tt CC3}  & $\sigma$ (pb)
 & 13.182  & 16.488 & 17.077
\\     &
$\langle E_\gamma \rangle$
&  1115 &  2280 &  3202
\\     & $\langle m_\gamma \rangle$ &  1112 &  2271 &  3185
\\     & $\langle E_\gamma \rangle$--$\langle m_\gamma \rangle$ &
      3               &  9      &  17
\\ \hline %-------------------------------------------------
change to {\tt FF}     & $\Delta \langle m_\gamma \rangle$
&0.5  &--0.8  &--4.2
\\ \hline %-------------------------------------------------
add $\delta_C$     & $\Delta \langle m_\gamma \rangle$
&11.8  &16.3  &20.8
\\ \hline %-------------------------------------------------
add $\sigma_{\rm{ISR}}^{non-univ}$     & $\Delta \langle m_\gamma \rangle$
&0.2  &0.3  & 0.3
\\ \hline %--------------------------------------------------
add $\sigma_{\tt CCn}$     & $\Delta \langle m_\gamma \rangle$
&$\leq$0.1  &0.2  &0.5
\\ \hline %-------------------------------------------------
\end{tabular}
}
\caption{\it 
Influence of different cross-section treatments on the average energy
loss
$\langle E_\gamma \rangle$ and invariant mass loss $\langle m_\gamma \rangle$
in MeV.
\label{ISRlost}
}
\end{center}
\end{table}
%ttttttttttttttttttttttttttttttttttttttttttttttttttttttt

\begin{table}[htb]
\centering
\begin{tabular}{|c|c|}
\hline
\rule[-1.2ex]{0mm}{4ex} parameter &  value \\
\hline
\rule[-1.2ex]{0mm}{4ex} $\MZ$ &  91.1888 \\
\rule[-1.2ex]{0mm}{4ex} $\Mw$ &  80.23 \\
\rule[-1.2ex]{0mm}{4ex} $\GZ$ & 2.4974  \\
\rule[-1.2ex]{0mm}{4ex} $\GW$ &  2.078 \\
\rule[-1.2ex]{0mm}{4ex} $\alpha^{-1}$ &  137.0359895 \\
\rule[-1.2ex]{0mm}{4ex} $G_\mu$ & $1.16639 \times 10^{-5}$~GeV$^{-2}$  \\
\rule[-1.2ex]{0mm}{4ex}
$\sin^2\theta_W \equiv \sin^2\theta_W^{(\ell)\rm eff}$ &  0.2320 \\
\rule[-1.2ex]{0mm}{4ex} $m_e$ & $5.1099906\times 10^{-4}$ \\
\rule[-1.2ex]{0mm}{4ex} $(\hbar c)^2$ & $3.8937966 \times 10^{8}$~pb~GeV$^{2}$ \\
\hline
\end{tabular}                                                                  \caption{\it
Parameter values used in the numerical calculations in 
Table~\protect\ref{tab:decomp}
and Fig.~\protect\ref{fig-sigww}. Masses and widths
are given in GeV.}
\label{tab-inputpar}
\end{table}
%%%%%%%%%%%%%%%%

\subsubsection{Improved Born approximation}

\noindent In the Standard Model, three parameters are 
sufficient to parametrise
the electroweak interactions, and the conventional choice is
$\{\alpha,G_\mu,\MZ\}$ since these are the three which
 are measured most accurately.
In this case the value of $\Mw$ is a {\it prediction} of the model.
Radiative corrections to the expression
for $\Mw$ in terms of these parameters
 introduce non-trivial dependences on $\Mt$ and $\Mh$, and
so a measurement of $\Mw$ provides a constraint on these masses.
However the choice  $\{\alpha,G_\mu,\MZ\}$ does not appear to be
well suited to \WW\ production. The reason is that a variation
of the parameter $\Mw$, which appears explicitly in the phase space
and in the matrix element,
has to be accompanied by an adjustment of the charged and neutral
weak couplings. Beyond leading order this is a complicated procedure.

It has been argued \cite{JEGERLEHNER} that a more appropriate choice
of parameters for LEP2 is the set $\{\Mw,G_\mu,\MZ\}$
(the so-called $G_\mu$--scheme), since in this case
the quantity of prime interest is one
of the parameters of the model. Using the tree-level relation
\begin{equation}
g^2 = e^2/\sin^2\theta_W = 4 \sqrt{2} G_\mu  \Mw^2
\label{gmuscheme}
\end{equation}
we see that the dominant $t$-channel neutrino
exchange amplitude, and hence the corresponding
contribution to the cross-section,
depends only on the parameters $\Mw$ and $G_\mu$.
It has also been shown \cite{JEGERLEHNER} that  in the
$G_\mu$--scheme there are no large next-to-leading order contributions to
the cross-section which depend on $\Mt$, either quadratically
or logarithmically. One can go further and choose the couplings
which appear in the other terms in the Born cross-section
such that all  large corrections at
next-to-leading order are absorbed, see for example Ref.~\cite{DITTMAIER}.
However for the  threshold
cross-section, which is dominated by the 
$t$-channel 
exchange amplitude, one can   simply use combinations of
 $e^2$ and $g^2$ defined  by Eq.~(\ref{gmuscheme})
for the neutral and charged weak
couplings which appear in the Born cross-section, Eq.~(\ref{siglo}).

In summary, the most model-independent approach when defining
the parameters for computing the $\epem\to\WW$ cross-section
appears to be the $G_\mu$--scheme, in which $\Mw$ appears explicitly
as a parameter of the model. Although this makes a non-negligible
 difference when calculating the Born cross-section, compared to using
$\alpha$ and $\sin^2\theta_W$
to define the weak couplings (see Table~\ref{tab:decomp} below),
a full next-to-leading-order
calculation will remove much of this scheme dependence~\cite{JEGERLEHNER}.
%%%%%%%%%%%%%%%%%%%%%%%%%
\begin{table}[htb]
\centering
\begin{tabular}{|l|c|c|}
\hline
\rule[-1.2ex]{0mm}{4ex} $\sigma_{WW}$ & $\sqrt{s} = 161$~GeV &
 $\sqrt{s} = 175$~GeV \\
\hline
\rule[-1.2ex]{0mm}{4ex} $\sigma_0$ (on-shell, $\alpha$) & 3.813  & 15.092   \\
\rule[-1.2ex]{0mm}{4ex} $\sigma_0$ (on-shell, $G_\mu$) & 4.402  & 17.425   \\
\rule[-1.2ex]{0mm}{4ex} $\sigma_0$ (off-shell with
$\GW(\Mw^2)$, $G_\mu$) & 4.747  & 15.873   \\
\rule[-1.2ex]{0mm}{4ex} $\sigma_0$ (off-shell with
$\GW(s)$, $G_\mu$) & 4.823  &   15.882 \\
\rule[-1.2ex]{0mm}{4ex} $\ldots\ +$ $O(\alpha)$ Coulomb & 5.122  & 16.311   \\
\rule[-1.2ex]{0mm}{4ex} $\ldots\ +$ $O(\alpha)$ Coulomb $+$ ISR&3.666&13.620\\
\hline
\end{tabular}
%%%%%
\caption{\it 
Decomposition of the theoretical $e^+e^-\to W^+W^-$ cross-section (in
picobarns)
as defined and discussed in the text, at two LEP2 collider energies.}
\label{tab:decomp}
\end{table}

\subsubsection{Numerical evaluation of the cross-section}

\begin{figure}[htb]
\vspace*{0.1cm}
\centerline{
\epsfig{figure=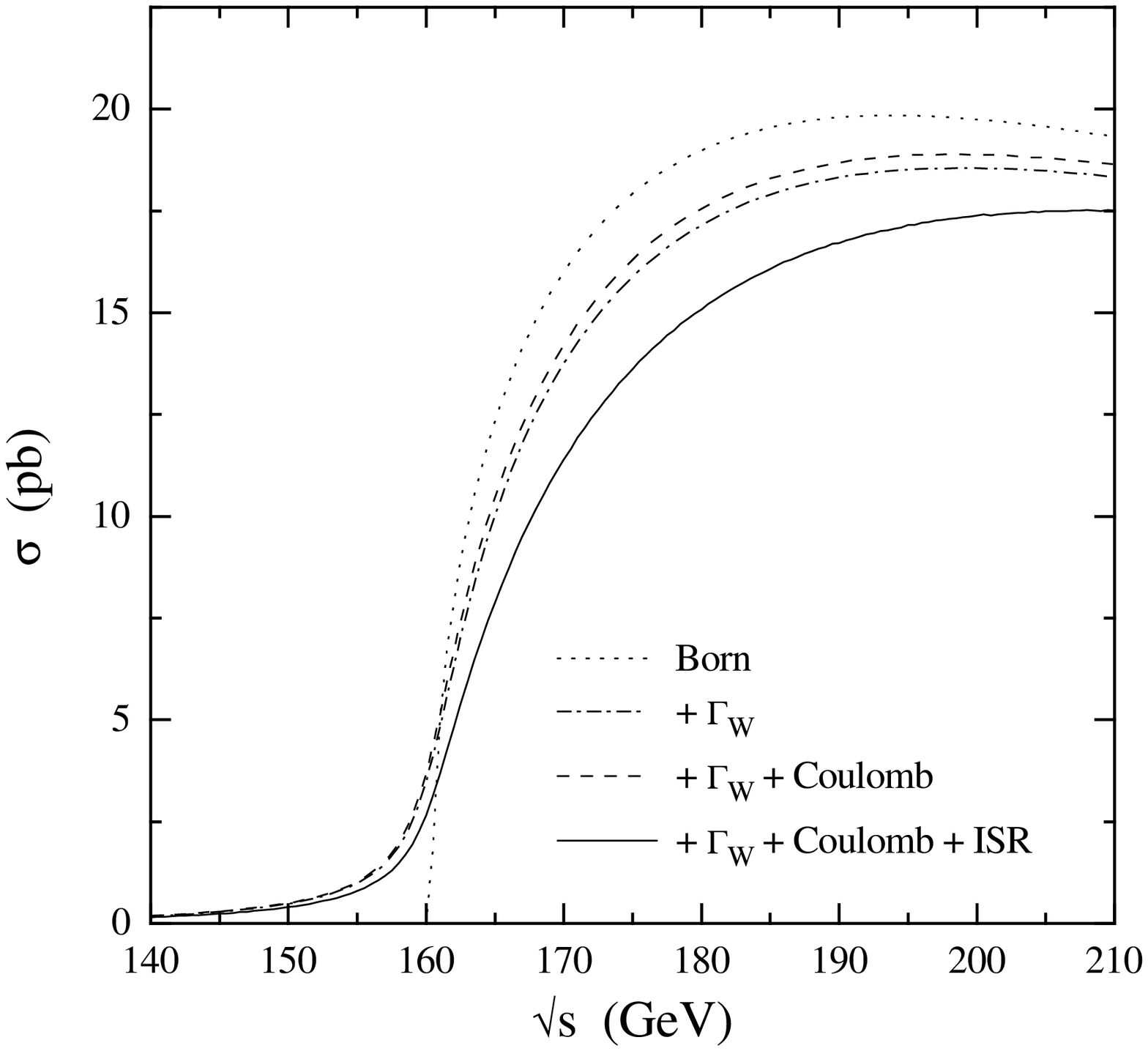,height=11cm,angle=0}
}
%\vspace*{-2cm}
\caption{\it
The cross-section for $\epem\to\ww $ in various approximations:
(i) Born (on-shell) cross-section, (ii) Born (off-shell)
cross-section, (iii) with first order Coulomb corrections,
and (iv) with initial state radiation. The parameter values are
listed in Table~\protect\ref{tab-inputpar}.
}
\label{fig-sigww}
\end{figure}

\noindent  
Figure~\ref{fig-sigww}  shows the $\epem\to\WW$ cross-section at LEP2 energies.
The different curves correspond to the sequential inclusion of the different
effects discussed above. The parameters used in 
the calculation are listed in Table~\ref{tab-inputpar}.
Note that both the initial state radiation and the finite width smear
the sharp threshold behaviour at $\sqrt{s} = 2 \Mw$ of the on-shell
cross-section.
%----------------------------------------------------------------------
 The different contributions
are quantified in Table~\ref{tab:decomp}, which
 gives the values of the cross-section
in different approximations just above threshold ($\sqrt{s} = 161$~GeV)
and at the standard LEP2 energy of $\sqrt{s} = 175$~GeV.
At threshold we see that the effects of initial state radiation
and the finite  W width are large and comparable in magnitude.
For the threshold method, the primary interest is the dependence 
of the cross-section on $\Mw$. 
%%%%%%%%%%%%%%%%%%%%%%%%%%%%%%%%%%%
% From: MX%"Miriam.Watson@cern.ch" 22-NOV-1995 09:23:49.27
\begin{table}[!htbp]
\begin{center}
\begin{tabular}{|l|c|c|c|}
\hline
%Reaction & Cross-section/pb & Cross-section/pb & Cross-section/pb \\
 Reaction & \multicolumn {3}{c|}{Cross-section (pb)}  \\
          &    at 161 GeV    &    at 175 GeV    &    at 192 GeV    \\
\hline
$\epem\rightarrow\WW\rightarrow\mathrm{all}$  &  3.64  & 13.77 & 17.10 \\
$\epem\rightarrow\WWqqqq$                     &  1.67  &  6.30 &  7.85 \\
$\epem\rightarrow\WWqqlnu$                    &  1.59  &  6.02 &  7.46 \\
$\epem\rightarrow\WWlnulnu$                   &  0.38  &  1.44 &  1.79 \\
\hline
$\epem\rightarrow\mathrm{Z}/\gamma\rightarrow\mathrm{all}$               
                                              & 221.  & 172. & 135. \\
$\epem\rightarrow\mathrm{Z}/\gamma\rightarrow\qq$ & 151.  & 116. &  91. \\
$\epem\rightarrow\mathrm{Z}/\gamma\rightarrow l^+l^-,\nu\overline{\nu}$
                                              &  70.  &  56. &  44. \\
\hline
$\epem\rightarrow(\mathrm{Z}/\gamma)(\mathrm{Z}/\gamma)
\rightarrow\mathrm{all}$   
                                              & 0.46   & 0.44  & 1.12 \\
$\epem\rightarrow\mathrm{Zee}\rightarrow\mathrm{all}$     
                                              & 2.53   & 2.70  & 2.85  \\
$\epem\rightarrow\mathrm{We}\nu\rightarrow\mathrm{all}$     
                                              & 0.37   & 0.51  & 0.72 \\
\hline
\hline
$\epem\rightarrow\epem$, $|\cos\theta|<0.95$
                                              & 351.  & 297. & 247. \\
\hline
\end{tabular}
\end{center}
\vspace{-4mm}
\caption{\it Cross-sections (in picobarns) for signal and background
  channels as given by PYTHIA.  $\WW$,
  $(\mathrm{Z}/\gamma)(\mathrm{Z}/\gamma)$, $\mathrm{We}\nu$ and
  $\mathrm{Zee}$ refer to four-fermion production with intermediate
  formation of either one or two vector bosons; in the case of Z pair
  production, the generator includes the photon contribution.
%The interactions $\mathrm{Zee}$
%  and $\mathrm{We}\nu$ imply single boson production, in which photons
%  radiated by one incoming beam particle interact with the other beam
%  particle to produce a Z or W boson.  
  The process $\mathrm{Z}/\gamma\rightarrow\mathrm{all}$ refers to the
  production of a fermion pair via a Z boson or photon.  All
  interactions include initial state radiation, and therefore may
  contain additional photons in the final state.  The final row lists
  cross-sections generated for the Bhabha scattering process
  (including t-channel exchange) using the BABAMC program.}
\label{tab-MCXsect}
\end{table}
%%%%%%%%%%%%%%%%%%%%%%%%%%%%%%%%%%%
This will be quantified in Section~\ref{sec:wmass_threshold}
below. For both methods, the size of the background cross-sections
is important. For completeness, therefore, we list in   
Table~\ref{tab-MCXsect}  some relevant cross-section
values  obtained using  the PYTHIA Monte Carlo. This includes
finite-width effects, 
initial state radiation and Coulomb corrections. Notice that the 
values for $\sigma_{WW}$  agree to within about 
 1\% accuracy with those given in the last row of
 Table~\protect\ref{tab:decomp}.

%%%%%%%%%%%%%%%%%%%%%%%%%%%%%%%%%%%%%%%%%%%

%\input s2_new.tex
% this version: Monday 8th January 
% this version: Monday 4th December 5.00pm 
% this version: Thursday 23rd November 1.20pm 
%  -- small changes by WJS
%
% this version: Wednesday 22nd November 8.50am 
% From: MX%"Miriam.Watson@cern.ch" 22-NOV-1995 09:23:49.27

%%%%%%%%%%%%%%%%%%%%%%%%%%%%%%%%%%%%%%%%%%%%%%%%%%%%%%%%%%%%%%%%%%%%%%%%%%
%   please note the following conventions and macros:                    %
%                                                                        %
%   Fig.~\ref{}, Table~\ref{}, Section~\ref{} and Ref.~\cite{}.          %
%                                                                        %
%   for masses please use $\Mw$, $\Mh$, $\Mt$, etc.,                     %
%   and if you wish, units are provided by e.g. $25\mev$, $161\gev$,     %
%   $100\pb^{-1}$ which puts the units in roman and adds a space after   %
%   the number.                                                          %
%                                                                        %
%%%%%%%%%%%%%%%%%%%%%%%%%%%%%%%%%%%%%%%%%%%%%%%%%%%%%%%%%%%%%%%%%%%%%%%%%%

\section{Measurement of $\Mw$ from the \WW\ Threshold
 Cross-Section\protect\footnotemark[2]}
\footnotetext[2]{prepared by D.~Gel\'e, T.G.~Shears, W.J.~Stirling, A.~Valassi,
 M.F.~Watson}

\label{sec:wmass_threshold}

%\subsection{Introduction}
%\label{sec:wmass_threshold_introduction}
As discussed in Section~\ref{sec:wmass_int_methods}, one can  exploit 
the rapid increase  of the \WW\ production cross-section
at $\sqrt{s} \sim 2\Mw$ to measure the W mass.
In the following,
we briefly discuss the basic features of this method,
suggest an optimal collider strategy for data-taking, and estimate
the statistical and systematic errors.
The intrinsic statistical limit to the resolution on $\Mw$
is shown to be energy-dependent:
in particular,
arguments are presented in favour of
a single cross-section measurement
at a fixed energy
$\sqrt{s} \sim 161\;\mathrm{GeV}$.

%The  analysis presented here shows that 
%this method is expected to 
%attain a statistical resolution on $\Mw$
%equivalent to that from direct W mass reconstruction,
%but with very different systematics.
%The two methods can therefore be regarded as complementary tools,
%and both should be used to provide an internal cross-check 
%on the measurement of the W mass at LEP2.
%This constitutes the main motivation
%for spending some luminosity in the threshold region
%(in contrast to the direct reconstruction method, 
%which  requires data-taking at higher energies,
%where the cross-section is larger).

%In addition, this measurement is also of interest 
%because it appears to fit very well into 
%the expected schedule for LEP2 operation in 1996.
%It is anticipated that 
%the maximum beam energy at LEP2
%will increase in steps,
%with the progressive installation of
%more superconducting RF cavities,
%in such a way that a centre-of-mass energy of  161~GeV
%will indeed be achievable during the first running period of 1996.
%This would then be the ideal time to perform such a measurement.
%The achievable statistical error on $\Mw$ depends of course on the 
%available luminosity at the threshold energy. We present quantitative
%estimates based on integrated  luminosities of 25, 50 and 100~pb$^{-1}$
%per experiment.

\subsection{Collider strategy}
\label{sec:wmass_threshold_strategy}
\begin{figure}[htb]
  \centerline{\epsfysize12cm\epsffile{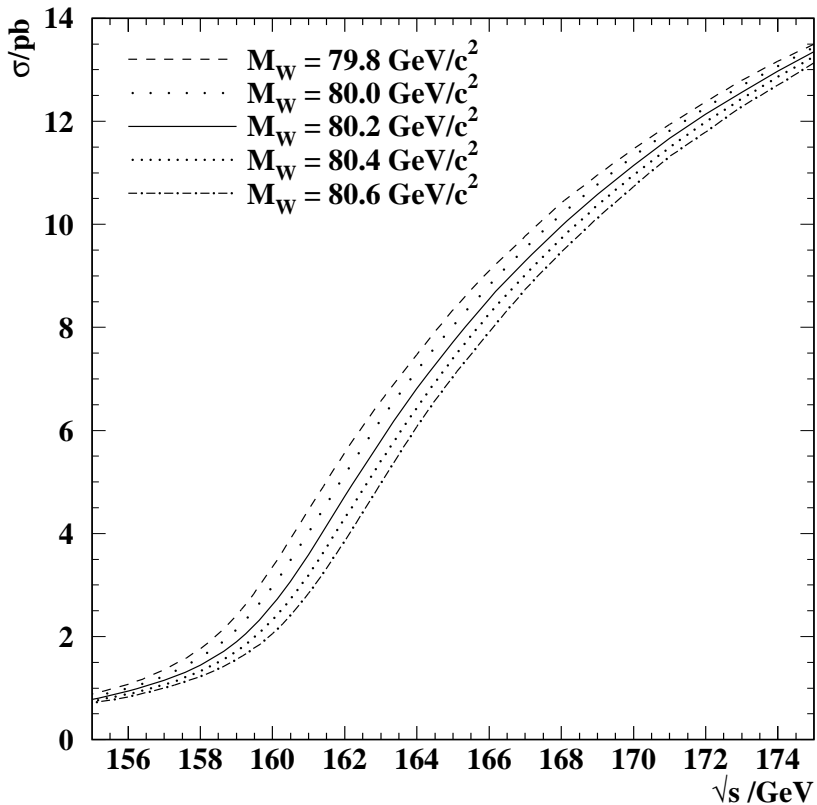}}
  \caption{\it 
      The cross-section for \WW\ production as a function of
     $\protect\sqrt{s}$ \ 
 in the threshold region,
      for various values of $\Mw$.
      Finite-width effects,
      the QED Coulomb correction and Initial State Radiation are included.
          }
  \label{fig:xsec_xsec}
\end{figure}

The cross-section for \WW\ production
increases very rapidly 
near the nominal kinematic threshold $\sqrt{s} = 2\Mw$,
although the finite W width and ISR
smear out the abrupt rise of the Born on-shell cross-section.
This means that
for a given $\sqrt{s}$ near threshold,
the value of the cross-section
is very sensitive to $\Mw$. This is illustrated 
in Fig.~\ref{fig:xsec_xsec},
where the \WW\ excitation curve is plotted
for various values  of the W mass. The calculation is
the same as that discussed in Section~\ref{sec:theoryinput},
 and includes finite W width
effects, ISR and QED Coulomb corrections.
A measurement of the cross-section 
in this region therefore  directly yields  a measurement of $\Mw$.

\begin{figure}[htb]
\begin{center}
    \mbox{\epsfysize10cm\epsffile{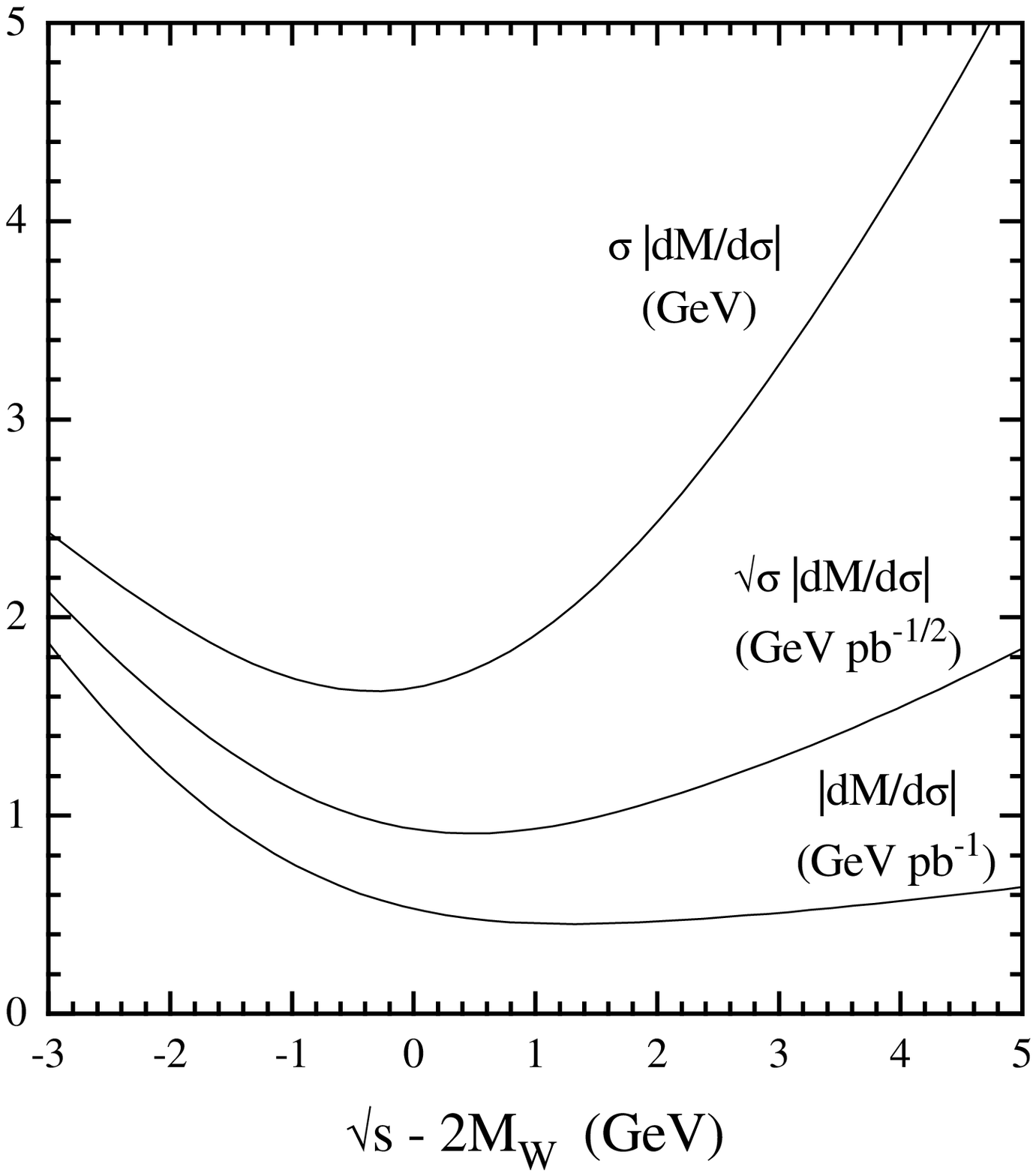}}
\end{center}
  \caption{\it 
      The sensitivity of the \WW\ cross-section to the W mass,
      plotted as a function of $\protect\sqrt{s}-2\Mw$.
      The significance of the three curves to the 
      W mass measurement
      is discussed in the text.
      A value of $\Mw=80.26\;\mathrm{GeV}$ has been used in the calculations.
          }
  \label{fig:xsec_dxsec}
\end{figure}

For an integrated luminosity $\cal{L}$
and an overall signal efficiency
$\epsilon_{WW}=\sum_i\epsilon_i\mathrm{BR}_i$
(where the sum extends over the various channels selected,
with branching ratios BR$_i$ and efficiencies $\epsilon_i$),
the error on the \WW\ cross-section
due to signal statistics is given by
\begin{equation}
\Delta\sigma_{WW}={\sigma_{WW}\over \sqrt{N}} = 
{\sqrt{\sigma_{WW}} \over \sqrt{\epsilon_{WW}{\cal L}}},
\end{equation}
where $N=\epsilon_{WW}\sigma_{WW}{\cal L}$ is the number of selected
signal events.
The corresponding error on the W mass is
\begin{equation}
\Delta \Mw = 
 \sqrt{\sigma_{WW}}\left\vert\frac{d\Mw}{d\sigma_{WW}}\right\vert
 \frac{1}{\sqrt{\epsilon_{WW}\cal{L}}}.
\end{equation}
The  sensitivity factor $\sqrt{\sigma_{WW}}|{d\Mw}/d\sigma_{WW}|$ is
plotted in Fig.~\ref{fig:xsec_dxsec}
as a function of $\sqrt{s}-2\Mw$. There is a  minimum at 
%has a minimum of approximately 0.89 $\mathrm{GeV}/\sqrt{\mathrm{pb}}$ at 
\begin{equation}
%(\sqrt{s})^{\,\mathrm{opt}} - 2\Mw \simeq 0.55\;\mathrm{GeV},
 (\sqrt{s})^{\,\mathrm{opt}}\simeq  2\Mw + 0.5\;\mathrm{GeV},
\label{eq:min}
\end{equation}
corresponding to a minimum value of approximately
$0.91\mathrm{\gev\pb}^{-1/2}$.
Note that the $0.5\;\mathrm{GeV}$ offset of the minimum of the sensitivity
above the nominal threshold is insensitive to the actual value of $\Mw$,
since in the threshold region the cross-section is to a first approximation
a function of $\sqrt{s} - 2 \Mw$ only.

As discussed below,
the statistical uncertainty is expected to be
the most important source of error
for the threshold measurement of $\Mw$:
the optimal strategy for data-taking
consists therefore in operating at the collider energy
$(\sqrt{s})^{\,\mathrm{opt}}$
in order to minimize the statistical error on $\Mw$.
The statistical sensitivity factor is essentially flat within
$(\sqrt{s})^{\,\mathrm{opt}} \pm 0.65\;\mathrm{GeV}$,
where it increases at most to $0.95\mathrm{\gev\pb}^{-1/2}$ (+4\%);
bearing in mind that the present uncertainty on $\Mw$ from direct measurements
is 160~MeV (and is expected to decrease further in the coming years), 
this corresponds 
to $\pm 2\sigma$ on the current world average $\Mw$ value.
%Bearing in mind that the present uncertainty on $\Mw$ from direct measurements
% is 160~MeV (and is expected to decrease further in the coming year), the
% range in collider energy over which the statistical sensitivity 
% is flat is significantly larger than the uncertainty in the position
% of the threshold from the present $\Mw$ measurement.
In other words,
$\Mw$ is already known to a level of precision
good enough to choose, {\em a priori}, 
{\em one} optimal energy 
for the measurement of the \WW\ cross-section
at the threshold.
Using the latest world average value    
$\Mw = 80.26 \pm 0.16\;\mathrm{GeV}$ (see Eq.~(\ref{mwworldav}))
gives an optimal collider energy of
 $(\sqrt{s})^{\,\mathrm{opt}} \simeq 161.0\;\mathrm{GeV}$.

\subsection{Event selection and statistical errors}
\label{wmass_threshold_selection}

The error on the W mass due to the statistics of \WW\ events collected
has been given in the previous section.
Background contamination
with an effective cross-section 
$\sigma_{\mathrm{bkg}}$ 
introduces an additional statistical error.
The overall effect
is that the statistical error on $\Mw$ is modified according to
\begin{equation}
\Delta \Mw \rightarrow \Delta \Mw 
 \sqrt{1+\frac{\sigma_{\mathrm{bkg}}}{\epsilon_{WW}\sigma_{WW}}}.
\end{equation}
In the following subsections we present estimates of this statistical
error for realistic event selections, for an integrated luminosity of
$100\;\mathrm{pb}^{-1}$ at $\sqrt{s} = 161\;\mathrm{GeV}$.  Tight
selection cuts are required to reduce the background contamination
while retaining a high efficiency for the signal, especially as the
signal cross-section is a factor of 4--5 lower at threshold than at
higher centre-of-mass energies.

The studies are based on samples of signal and background events
generated by means of Monte Carlo programs 
(mainly PYTHIA \cite{jetset74}) tuned to
LEP1 data. These events were run through the complete simulation
program giving a realistic detector response, and passed through the
full reconstruction code for the pattern recognition.

%\paragraph
\subsubsection{Fully hadronic channel, \WWqqqq.}

\noindent The pure four quark \WW\ decay mode benefits from a substantial
branching ratio (46\%) corresponding to a cross-section
$\sigma(\WWqqqq ; \sqrt{s} = 161\; \mathrm{GeV})= 1.67$~pb.
Obviously, the typical topology of such events consists of four or
more energetic jets in the final state.  Due to its large
cross-section (see Table~\ref {tab-MCXsect}), the main natural
background to this four-jet topology comes from $\epem \to \mathrm{Z}
(\to \qq (+n\mathrm{g}))\gamma$ events which can be separated into
two classes depending on the virtuality of the Z: (i) the production
of an on-mass-shell Z accompanied by a radiative photon of nearly
55~GeV (at $\sqrt{s} = 161\;\mathrm{GeV}$), which is experimentally
characterised by  missing momentum carried by the photon escaping
inside the beam pipe (typically 70\% of the time), and (ii) events
with a soft ISR $\gamma$ and a large total visible energy, which
potentially constitute the most dangerous QCD background contribution.

Note that a semi-analytical calculation of the genuine four-fermion
background cross-section  \cite{fourferm} for a wide
range of four-fermion final states (with non-identical fermions) shows that in
the threshold region $\sigma_{\mathrm{bkgd}}(4\mathrm{f})/\sigma_{WW}\ll 1\%$, 
and therefore the
effect on the $\Mw$ determination from these final states is
negligible.

\begin{figure}[htb]
\vspace{0.1cm}
\centerline{
 \epsfig{figure=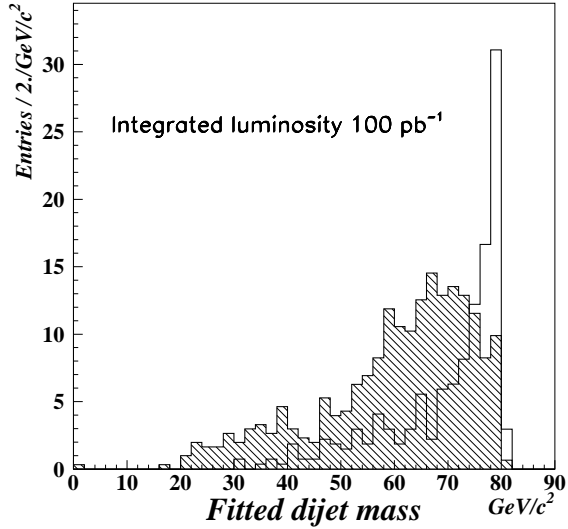,height=8cm,angle=0}
}
\caption{\it Distributions of the fitted dijet mass using a 
constrained kinematic fit as described in the text. The solid histogram
shows the $\WW \to \qq\qq $ signal
 and the hatched histogram represents the
$\epem \to q\bar q\gamma$ background events.  }
\label{threshold_qqqq}
\end{figure}

Although the effective four-jet-like event selection
 depends somewhat
  on the specific detector under consideration, a
general and realistic guideline selection can  be described.
The most relevant conditions to be fulfilled by the selected events can be
summarised as follows:
\begin{itemize}
\item A minimum number of reconstructed tracks of charged and neutral
particles is required.
A typical value is 15. This cut removes nearly all
low multiplicity reactions, such as dilepton production
($\epem\to l\bar l, l=\mathrm{e},\mu,\tau$) and two-photon processes.
\item A veto criterion against hard ISR photons from $\qq\gamma$ events
in the
 detector acceptance can be implemented by rejecting events with an
isolated cluster with significant electromagnetic energy
(larger than 10~GeV for example).
\item A large visible energy,  estimated using the information from
 tracks of charged particles and from the
electromagnetic and hadron calorimeters. For example, a minimum energy cut value
of 130~GeV reduces by a factor of 2 the number of $\qq\gamma$ events
 with a photon  collinear to the beam axis.
\item A minimum number (typically 5) of reconstructed tracks per jet.
This criterion acts on the low multiplicity jets from $\tau$
decays  as well
as from $\gamma$ conversions
 or $\gamma$ interactions with the detector material.
\item A minimum jet polar angle. The actual cut value
depends on the
detector setup, but is likely to be around $10^\circ-15^\circ$. This cut is
 mainly needed to eliminate  poorly measured jets in the very forward
 region, where the experiments are generally less well instrumented.
\end{itemize}
These selection criteria almost completely
remove the harmless background sources
($\epem \to \mathrm{ZZ,Zee}$ and $\mathrm{We}\nu$), but there is still an
unacceptable  level of $\epem\to \mathrm{Z}(\to \qq(+n\mathrm{g}))\gamma$
 contamination (about
three times  higher than the signal).
The second step is  to suppress  the remaining QCD
background (from $\epem\to \qq \mathrm{gg}\gamma_{soft}$ events)
 by performing a
W$^{\pm}$ mass reconstruction based on a constrained kinematic  fit.
The following additional  criteria can then be imposed:
\begin{itemize}
\item A $\chi^2$ probability cut associated with a minimum
constrained dijet mass requirement -- a typical choice of standard values
 of 1\% and 70~GeV respectively is used here.
 This procedure appears to be an
efficient tool to improve the mass resolution and therefore to reduce
the final background, see Fig.~\ref{threshold_qqqq}.
\end{itemize}

In summary, a reasonable signal detection efficiency in excess of 50\%
is achievable for \WWqqqq\ events.  Although the final rejection
factor of QCD events is approximately 500, a substantial residual
four-jet background still remains, giving a purity of around
70\%.  The contributions from other backgrounds ($\mathrm{ZZ}$,
$\mathrm{Zee}$, $\mathrm{We}\nu$ and two-photon events) are negligible
for the four-jet analysis.

The signal and background efficiencies for
the typical event selection described
above are given in Table~\ref{tab:stats}, assuming a total integrated
luminosity of 100~pb$^{-1}$ (i.e. 25 pb$^{-1}$ per interaction point).
%the corresponding number of events are also listed.

% Results table
%\begin{table}[hbtp]
%\begin{center}
%\begin{tabular}{|c|c|c|c|c|}
%\hline
% & efficiency (\%)    & $\sigma_{selected}$ & $N_{evt}$  & $\Delta \Mw(stat)$
% (MeV)
% \\ \hline
% Signal     &  &  &  & \\
%$e^+e^-\to W^+W^- \to qqqq$ & 55 & 0.94 &  $94 \pm 3$ & 180  \\
%\hline
%Background &  &  &  & \\
%$e^+e^- \to qq\gamma$ &0.25 & 0.39 & $39 \pm 3 $  & 106 \\ \hline
%\end{tabular}
%\end{center}
%\caption{Typical selection for the $W^+W^- \to qqqq$ channel.
% $\sigma_{selected}$ is the selected cross-section. $N_{evt}$ is the number of
%expected events assuming a total 100 pb$^{-1}$ luminosity and the errors on
%$N_{evt}$ are statistical.   }
%\label{wwqqqqsel}
%\end{table}

%
% Table of selection efficiencies etc. at 161 GeV
%
\begin{table}[htbp]
\begin{center}
\begin{tabular}{|l|c|c|c|}
\hline
 & \WWqqqq  &\WWqqlnu  & \WWlnulnu \\ \hline
Signal cross-section            & 0.94~pb & 0.76~pb & 0.23~pb \\
Signal efficiency               &   55\% &   47\% &   60\% \\ 
$\Delta \Mw$ (stat.) for signal     &  180~MeV &  197~MeV &  354~MeV \\ \hline
Background cross-section    & 0.39~pb & 0.03~pb & 0.01~pb \\
$\Delta \Mw$ (stat.) for background &  106~MeV &   37~MeV &   74~MeV \\ \hline
Total $\Delta \Mw$ (stat.)      &  209~MeV &  200~MeV &  360~MeV \\
\hline
\end{tabular}
\end{center}
\vspace{-4mm}
\caption{\it Typical accepted cross-sections
         at $\protect\sqrt{s} = 161$~GeV and corresponding statistical
         uncertainty on $\Mw$, assuming an integrated
         luminosity of 100~pb$^{-1}$. Note that $l=\mathrm{e},\mu,\tau$.}
\label{tab:stats}
\end{table}
%
%The estimated overall
%statistical precision on $\Mw$ with the threshold method
%using a realistic four jets selection and 100 pb$^{-1}$ integrated
%luminosity is
%\begin{equation}
%\Delta \Mw(stat) = 180\  \mbox{(signal,stat.)}\
%\oplus \ 106 \ \mbox{(bkgd,stat.)} = 209 \;\mathrm{MeV} ,
%\end{equation}
%where the first uncertainty is for the signal and the second one for
%the residual background to be subtracted.

\subsubsection{Semileptonic channel, \WWqqlnu .}

\begin{figure}[htb]
\vspace{0.1cm}
\centerline{
  \epsfig{figure=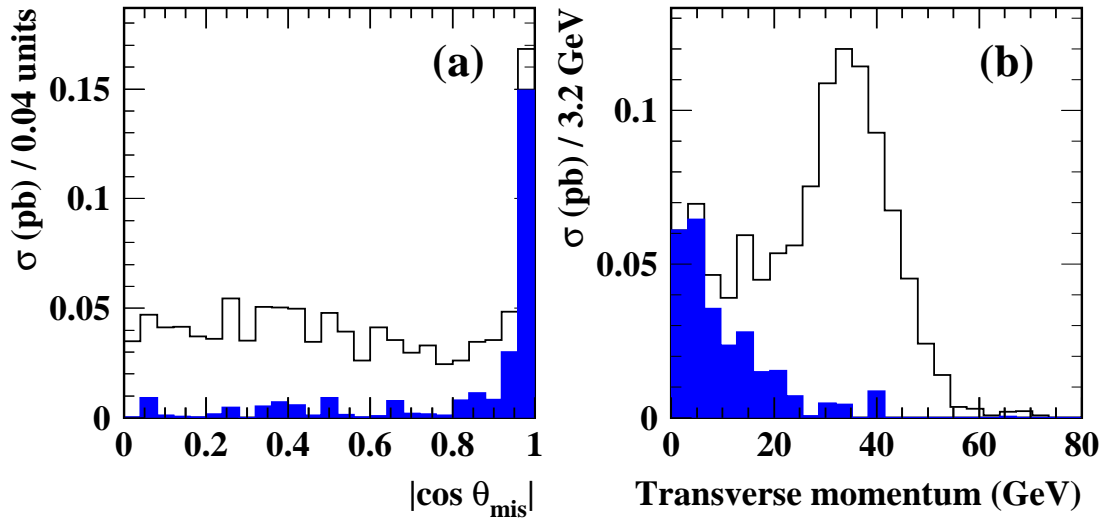,height=7.7cm,angle=0}
}
\caption{\it  Examples of the momentum imbalance in candidate events
  for the \WWqqlnu\ channel: (a) direction of the
  missing momentum, (b) total momentum transverse to the beam
  direction.  The open histogram indicates events which contain an
  isolated, identified lepton and pass all other selection cuts; the
  background from the $\mathrm{Z}/\gamma$, $\mathrm{ZZ}$,
$\mathrm{Zee}$ and $\mathrm{We}\nu$ channels is
  shaded.  All samples include detector simulation.}
\label{fig:threshold_qqlv}
\end{figure}

\noindent The decay channel \WWqqlnu\ is characterised by the presence of
two or more hadronic jets, an isolated, energetic lepton or a narrow
jet in the case of hadronic $\tau$ decays, and missing energy and
momentum due to the undetected neutrino.  Since the \WW\ cross-section
is small at threshold, processes such as
$\epem\rightarrow\mathrm{Z}/\gamma$ and $\epem\rightarrow\mathrm{Zee}$
constitute important backgrounds.  Their cross-sections at 161~GeV are
listed in Table~\ref{tab-MCXsect}.  Typical experimental selection cuts
for the \WWqqenu\ and \WWqqmnu\ channels include:
\begin{itemize}

\item A multiplicity cut to reject purely leptonic events.  By
  requiring at least 6 charged tracks in the event, backgrounds from
  the \WWlnulnu, $\mathrm{Z}/\gamma\rightarrow l^+ l^-$ and
  leptonic two-photon channels can be removed. 

\item Identification of an electron or muon using standard
  experimental cuts.  The lepton is required
  to have a high momentum and to be isolated from the hadronic jets.
  This isolation can be achieved by restricting the energy in a cone
  around the lepton candidate, or by requiring a minimum
  transverse momentum relative to the nearest jet.  An example of such
  a cut is to require less than 2.5~GeV of energy (charged and
  neutral) inside a 200~mrad cone around the track.  This suppresses
  hadronic background, much of which originates from the semi-leptonic
  decays of heavy quarks inside jets.

\item Cuts on missing momentum and its direction.  The neutrino
  carries away momentum, leaving the event with a net momentum
  imbalance.  In order to distinguish the signal events from
  $\mathrm{Z^0}/\gamma$ or two photon backgrounds, in which the
  missing momentum tends to be along the beam direction, the event is
  required either to have a large transverse momentum, or to have
  missing momentum pointing away from the beam direction.  These
  quantities are illustrated in Fig.~\ref{fig:threshold_qqlv}(a) and
  (b) and show clear differences between signal and background events.
  Effective experimental cuts are $|\cos\theta_{\mathrm{mis}}|<0.95$ and
  $p_{\mathrm{mis}}/\Ecm >0.07$.

\item Cuts on the corrected, visible energy in the event.  Requiring
  $E_{\mathrm{vis}}/\Ecm <1$ removes some residual background from
  $\mathrm{Z}/\gamma\rightarrow\mathrm{hadrons}$.  Similar results can
  be achieved by cutting on the energy or mass of the hadronic system. 

\end{itemize}
%
% This is a sample Figure
%

Additional improvements to the selection may be obtained by
applying a kinematic fit, in which energy-momentum constraints are
applied in conjunction with assumptions about the W masses.  The
background events tend to give poor probabilities in the fit.

Selection cuts based on these quantities give efficiencies of 70--75\%
for the \qq\enu\ and \qq\mnu\ channels, but only about 5\% for the
\qq\tnu\ decays.  The purity of the selected sample is 90--95\%, with
typical accepted cross-sections as listed in Table~\ref{tab:stats}.
The corresponding statistical uncertainty on $\Mw$ is approximately
200~MeV for 100~pb$^{-1}$.  More work will be needed to improve the
efficiency for selecting $\tau$ decays, and hence to enhance the
statistical precision of the cross-section measurement.  Hadronic
$\tau$ decays can be identified as a third jet with low multiplicity
and a small opening angle.

\subsubsection{Fully leptonic channel, \WWlnulnu.}

\noindent The fully leptonic channel
is not used by the direct reconstruction method
due to the lack of sufficient constraints 
on the kinematics of the event,
but it {\it can}  be used for event-counting
in the threshold measurement of $\Mw$.
It has, however,
a branching ratio of only 11\% ($l=\mathrm{e},\mu,\tau$),
a factor 4 lower than the other two channels. This
 is  reflected in the relative weight
of this channel to the overall precision on $\Mw$ using  the threshold
method.

\begin{figure}[tbh]
  \centerline{
%    \vspace{10truecm}
      \mbox{\epsfysize8cm\epsffile{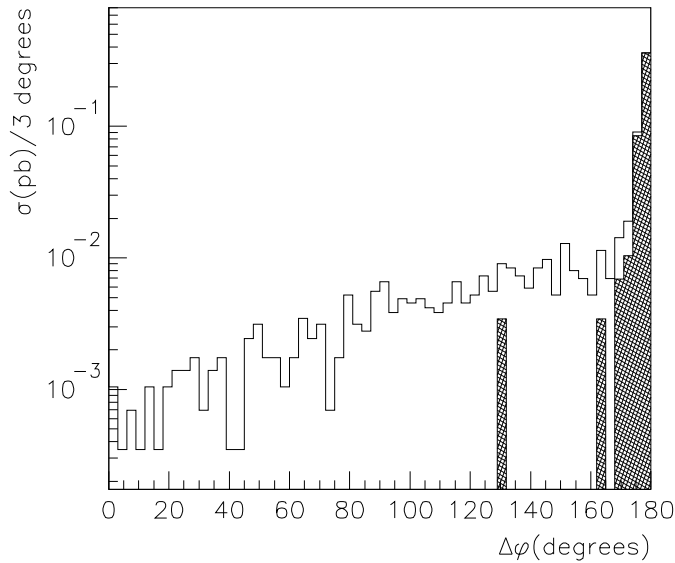}}
             }
  \caption{\it 
      Distribution
      of the jet acoplanarity $\Delta\phi$,
      in the selection of \WWlnulnu\ events at threshold.
      The solid histogram includes the events selected 
      by all cuts (except that on $\Delta\phi$).
      The shaded area corresponds to the background 
      from the $\mathrm{Z}/\gamma$ and $\gamma\gamma$ dilepton events.
          }
  \label{fig:threshold_lvlv}
\end{figure}

The \WWlnulnu\  channel is characterised
by two acoplanar, energetic leptons
and large missing momentum.
In 4/9 (1/9) of the cases, however,
one (both) of these leptons is a $\tau$,
which typically decays to a narrow hadronic jet.
Typical experimental selections for 
all \WWlnulnu\ channels require:
\begin{itemize}
\item
Low charged multiplicity (typically not greater than 6),
which allows the rejection of most of the hadronic backgrounds.
The most important background to this channel 
is then given by dilepton ($\mathrm{Z}/\gamma$, two-photon or Bhabha) events.
At least two good charged tracks (with a typical minimum energy of 1 GeV)
are required in any case.
\item
One identified electron or muon
(requiring two would lose most of the events with one
$\mathrm{W}\rightarrow\tau\nu_{\tau}$ decay);
this lepton is required to have an angle of typically at least $25^{\circ}$
to the beam axis, 
as both Bhabhas and two-photon events are concentrated at low angles.
Since the leptons from \WW\ decays at threshold
have approximately half the beam energy,
an energy window may be imposed on the lepton,
centered around this value
(such as [28--55]~GeV);
this is effective both against 
low-energy two-photon events
(which can be further reduced by requiring a large total visible mass),
and against di-muons and Bhabhas,
where the energy of each lepton is typically equal to the beam energy.
\item
Explicit tagging of events with
isolated, energetic photons or luminosity clusters
allows one to reject radiative Z events
with hard, detected ISR.
\item
Cuts on the missing momentum and its direction can also be used:
large transverse missing momentum (typically more than 4~GeV) is required
to discriminate against two-photon events,
while combined cuts on the angle between the jets 
and on the missing momentum out of the beam-thrust plane
are very effective against di-tau $\mathrm{Z}/\gamma$ events.
\item
To reconstruct the two original lepton directions, 
the event is forced into two jets,
which must be of low invariant mass
and well contained in the detector.
A cut on the acoplanarity $\Delta\phi$ between these two jets 
(such as $\Delta\phi<174^{\circ}$)
is effective against all residual backgrounds.
The distribution of this variable for signal and background events
is shown in Fig.~\ref{fig:threshold_lvlv}.
\end{itemize}

Selection cuts of this kind give 
overall efficiencies of approximately  60\%
(60\% to 70\% in all individual channels
except the $\tau\nu_{\tau}\tau\nu_{\tau}$ channel,
where it is only of the order of 30\%),
with a purity close to 95\%.
Typical accepted cross-sections
are listed in Table~\ref{tab:stats}.
The corresponding statistical error
on $\Mw$ is approximately 360~MeV
(354~MeV from signal statistics
plus 74~MeV from background statistics)
for a total integrated luminosity of 100 $\mathrm{pb}^{-1}$.

\subsection{Systematic errors}

%\subsection{Systematic errors}
%\label{wmass_threshold_sys}

Uncertainties in the \WW\ production cross-section translate
directly into systematic errors on the W mass. The uncertainties fall
into three categories: 
multiplicative uncertainties in the cross-section;
%additive uncertainties in the cross-section;
an additive uncertainty due to background subtraction;
other sources, for example beam energy and W width.
 
\subsubsection{Luminosity, higher-order corrections and selection 
efficiency}
%\noindent{\bf (a) Multiplicative uncertainties}
%\subsubsection{Multiplicative uncertainties}
%\label{wmass_threshold_sys_mult}
 
\noindent Quantities which enter multiplicatively in the
calculation or measurement of the cross-section 
%(e.g. luminosity,
%higher-order corrections and signal  
%efficiency), 
contribute an error
to the W mass which can be expressed as
\begin{equation}
\Delta \Mw = \left\vert \frac{d \sigma}{d\Mw} \right\vert^{-1}
\sigma \mbox{\hspace{0.2cm}} \frac{\Delta \mathrm C}{\mathrm C} ,
\label{threshold_mult_1}
\end{equation}
where $\Delta$C is the uncertainty on the multiplicative quantity C,
and the cross-section $\sigma$ includes contributions from signal and
background weighted by their relative efficiencies.
The quantity 
%$ \sigma_{WW} \vert d \sigma_{WW}/d\Mw \vert^{-1}$
$ \sigma \vert d \sigma/d\Mw \vert^{-1}$
is shown in Fig.~\ref{fig:xsec_dxsec} as a function
of $\sqrt{s}$. Like the statistical
sensitivity factor, it also has a minimum near the nominal
threshold and has the value of approximately
 1.7~GeV at $(\sqrt{s})^{\,\mathrm{opt}} \simeq 161$~GeV.
 
The three most important uncertainties of this type are:
\begin{itemize}
\item The luminosity, which
is expected to be known to about 
$0.5\%$, including  both the known theoretical error on the Bhabha
cross-section and the expected experimental error, thus contributing
 about 8~MeV to $\Delta \Mw$.
\item Unknown higher-order corrections to the theoretical cross-section (see
Section 1.1.6 and Ref.~\cite{WWGROUP}), 
which if we conservatively assume a value of $\pm 2\%$,
contribute about 34~MeV to  $\Delta \Mw$.
\item Uncertainties in the signal efficiency, described in
Section~\ref{wmass_threshold_sys_exp} below, which depend
 on the particular decay channel under consideration.
\end{itemize}

\subsubsection{Background subtraction}
%\noindent {\bf (b) Additive uncertainties}
%\subsubsection{Additive uncertainties}
%\label{wmass_threshold_sys_add}
 
\noindent An uncertainty 
$\Delta \sigma_{\mathrm{bkgd}}$
on the residual background cross-section predicted by Monte Carlo
propagates as an additive uncertainty
to the measured \WW\ cross-section, 
from which the background has to be subtracted.
It contributes an error 
\begin{equation}
\Delta \Mw = \left\vert \frac{d \sigma}{d \Mw} \right\vert^{-1}
\frac{\Delta \sigma_{\mathrm{bkgd}}}{\epsilon_{WW}} 
\label{threshold_add_1}
\end{equation}
where $\epsilon_{WW}$ is the signal efficiency,
which is found by multiplying the selection efficiency for a given channel
by the appropriate branching ratio.
The quantity  
%$\vert d \sigma_{WW}/d \Mw\vert^{-1}$ 
$\vert d \sigma /d \Mw\vert^{-1}$ 
is shown in
Fig.~\ref{fig:xsec_dxsec}  as a function of
$\sqrt{s}$, and is approximately 470~MeV~pb$^{-1}$ at
 $(\sqrt{s})^{\,\mathrm{opt}} \simeq 161$~GeV\@.
Experimental methods for determining the uncertainty in the background
are described in Section~\ref{wmass_threshold_sys_exp}  below.
%\ref{wmass_threshold_sys_exp}. 
A systematic contribution to $\Delta \Mw$ of about 59 MeV (32 MeV) in
the \WWqqqq\ (\WWqqlnu, \WWlnulnu) channels is expected.

\subsubsection{Beam energy and W width}
%\subsubsection{Beam energy and W width}
%\label{wmass_threshold_sys_bewwidth}

\noindent The error introduced by an uncertainty in the beam energy 
$\Ebeam$
to $\Mw$ is
\begin{equation}
\Delta \Mw = \left\vert \frac{d \sigma}{d \Mw} \right\vert^{-1} \left\vert
\frac{d \sigma}{d \Ebeam} \right\vert
\Delta \Ebeam. 
\label{threshold_be_1}
\end{equation}
In the threshold region the cross-section $\sigma_{WW}$
is essentially a function of the single variable $\sqrt{s} - 2 \Mw$
only (see Fig.~\ref{fig:xsec_xsec}), and hence the ratio of derivatives
in (\ref{threshold_be_1}) is approximately unity, i.e.
$\Delta \Mw \simeq 1.0\; \Delta \Ebeam$.
It is estimated that the beam energy
will be known to an accuracy of 12~MeV.
 
The error on $\Mw$ introduced by an uncertainty in the W width
$\GW$  is
\begin{equation}
\Delta \Mw = \left\vert \frac{d \sigma}{d \Mw} \right\vert^{-1}
\left\vert \frac{d
\sigma}{d \GW} \right\vert \Delta \GW
\simeq 0.16\; \Delta \GW ,
\label{threshold_wwidth_1}
\end{equation}
where the value of the ratio of the derivatives corresponds
to $\sqrt{s} = 161~$GeV\@.
In the Standard Model,
$\GW$ is proportional to the third power of $\Mw$,
\begin{equation}
\GW = 2.080 \; \left( {\Mw \over 80.26\;\mathrm{GeV}}\right)^3 .
\end{equation}
If the current world average value of $\Mw = 
80.26 \pm 0.16$~GeV
is used then $\Delta \GW = 12$~MeV\@.
In contrast, a (combined) measurement of $\GW$ by the CDF and D0
collaborations at the Tevatron
$\mathrm{p} \bar\mathrm{p}$ collider \cite{directgammaw} 
%gives $\GW = 2.067 \pm 0.082$~GeV, 
gives $\GW = 2.051  \pm 0.061 \mbox{(expt.)}\pm 0.013 \mbox{(input)}$~GeV, 
consistent with the Standard Model calculation, but with
a much larger error. 
%(Note that the direct measurement error
%is expected to decrease significantly in the near future.)
If we use the  Standard Model width, then the contribution (2~MeV)
to $\Delta \Mw$ is negligible. See Ref.~\cite{stirling} for
a further discussion.
 
\subsubsection{Experimental determination of systematic uncertainties}
%\subsubsection{Experimental determination of systematic uncertainties}
\label{wmass_threshold_sys_exp}
 
\noindent Two methods have been proposed to determine the uncertainty in the
signal efficiency and background cross-section.
 The first method examines the sensitivity
of the signal efficiencies and background cross-section 
to uncertainties in fragmentation. Some preliminary
studies have been performed in the \WWqqqq\
channel by varying the fragmentation
parameters Q$_0$, $\sigma_q$, $b$ 
and $\Lambda_{\mathrm{QCD}}$~\cite{jetset74} within
one standard error bounds, and noting the effect on 
the signal efficiency and background cross-section
 in PYTHIA~\cite{jetset74}
generated events.
 
The second method uses data and Monte Carlo event samples from LEP1
to determine the uncertainty in the background cross-section.  The
selections described in Section~\ref{wmass_threshold_selection} have
been scaled to the LEP1 centre-of-mass energy and applied to both
real and simulated data.  The difference between data and Monte Carlo
gives the error on the background cross-section at $\sqrt{s}\simeq \MZ$.
The results are then rescaled to $\sqrt{s}=161$~GeV\@.  It is assumed
that the fractional uncertainty on the selection efficiencies for the
background is the same at the two energies.

The uncertainty on the signal efficiency appears small
from the results of the fragmentation test,
within the low
statistics of the event samples tested. 
Further studies are needed to quantify any effects. Therefore a
conservative estimate of the signal efficiency error of 2\%
is used at present,  contributing an error of 34~MeV to $\Mw$.
The uncertainty on the background cross-section 
in the \WWqqqq\ channel (8\%)
is 
estimated to contribute about 59~MeV to the error on $\Mw$.
The uncertainty in the \WWqqlnu\ and \WWlnulnu\ channels,
estimated to be 50\% and 100\% respectively, both give an error
of about 32~MeV\@.
These errors are expected to decrease with further study.

%\noindent{\bf (e) Conclusions}
\subsubsection{Conclusions}
\label{wmass_threshold_sys_concl}
 
\noindent Table~\ref{threshold_sys_total} summarizes the estimated systematic
errors for the \WWqqqq, \WWqqlnu\ and \WWlnulnu\ channels.
\begin{table}[h]
\begin{center}
\begin{tabular}{|l|c|c|c|} \hline
Source & \WWqqqq  & \WWqqlnu  & \WWlnulnu  \\ \hline
Luminosity (*) & 8 & 8 & 8 \\ 
HO corrections (*) & 34 & 34 & 34 \\
Beam energy (*) & 12 & 12 & 12 \\
W width (*) & 2 & 2 & 2 \\ \hline
Signal efficiency & 34 & 34 & 34 \\
Background cross-section & 59 & 32 & 32 \\ \hline
Total  (MeV) & 77 & 60 & 60 \\ \hline
\end{tabular}
\end{center}
\caption{\it Summary of systematic error contributions, in MeV, 
to the threshold
measurement in the \WWqqqq, \WWqqlnu\ and \WWlnulnu\ channels.
The quantities denoted (*) are common to all channels.
The total for each channel is found by combining the sources in quadrature.}
\label{threshold_sys_total}
\end{table}

\subsection{Summary}

Table~\ref{threshold_errors_summary} summarizes the results presented
above for the estimated statistical, systematic and total  
errors on $\Mw$ (for all decay channels combined) 
using the threshold method, i.e. by
measuring the \WW\ cross-section at the optimal collider energy
of 161~GeV. Our estimates for some of the systematic errors, for example
the unknown higher-order theoretical corrections, are probably too
conservative, and others, for example the uncertainty in the estimates
of the various background cross-sections, will almost certainly decrease
with more study. Nevertheless, for the  amount of luminosity likely to
be available for the threshold measurement the overall error is
dominated by statistics.
\begin{table}[h]
\begin{center}
\begin{tabular}{|l|c|c|c|} \hline
Total luminosity &  $ \Delta \Mw$ (stat)  & $ \Delta \Mw$ (stat$+$sys$_1$)
& $ \Delta \Mw$ (total)  \\ \hline
$4 \times 25\pb^{-1} = 100\pb^{-1}$  & 134  & 139 & 144 \\
$4 \times 50\pb^{-1} = 200\pb^{-1}$  & 95  & 101 & 108 \\ 
$4 \times 100\pb^{-1} = 400\pb^{-1}$  & 67  & 76 & 84 \\ \hline
\end{tabular}
\end{center}
\caption{\it 
Summary of statistical and systematic errors (in MeV) on $\Mw$ from a
cross-section measurement (all channels) at $\protect\sqrt{s}=
161$~GeV. The total luminosity refers to four 
experiments combined. The third column includes the channel-dependent systematic
errors only (see Table~\protect\ref{threshold_sys_total}), 
and the fourth column includes
the overall common systematic error.}
\label{threshold_errors_summary}
\end{table}

%\input s3_new.tex

% this version: Monday 4th December 5.00pm (slightly tidied up)
% date 1995 nov 22 
%Dear Zoltan and James,
%                      Below is the updated section3.tex file. I'm
% not sure who I'm supposed to be sending it to, so I'm sending it
% to both of you. I will also send new versions of some of the
% figures - 5 more files.
%
% Please can I request that the title of the section be changed
% from "W mass" to "Determination of the mass of the W boson"
%
%                                           Regards, Pat
%%%%%%%%%%%%%%%%%%%%%%%%%%%%%%%%%%%%%%%%%%%%%%%%%%%%%%%%%%%%%%%%%%%%%%%%%%
%   please note the following conventions and macros:                    %
%                                                                        %
%   Fig.~\ref{}, Table~\ref{}, Section~\ref{} and Ref.~\cite{}.          %
%                                                                        %
%   for masses please use $\Mw$, $\Mh$, $\Mt$, etc.,                     %
%   and if you wish, units are provided by e.g. $25\mev$, $161\gev$,     %
%   $100\pb^{-1}$ which puts the units in roman and adds a space after   %
%   the number.                                                          %
%                                                                        %
%%%%%%%%%%%%%%%%%%%%%%%%%%%%%%%%%%%%%%%%%%%%%%%%%%%%%%%%%%%%%%%%%%%%%%%%%%

\section{Direct Reconstruction of $\Mw$\protect\footnotemark[3]}
\footnotetext[3]{prepared by M.~Gr\"unewald, N.~J.~Kj{\ae}r, Z.~Kunszt,
 P.~Perez, C.~P.~Ward}

\label{sec:mwdir}

%\subsection{Introduction}
In this section, we discuss the measurement of the W mass by kinematic
reconstruction of the invariant mass of the W decay products. The
statistical precision of this method which could be obtained by combining
four experiments each with 500~pb$^{-1}$ at $\sqrt{s} = 175\gev$, assuming
100\% efficiency and perfect detector resolution, is about $10\mev$, limited 
by the finite width of the W. In practice, this ideal precision will be
degraded, partly through loss of statistics, but mainly because detector
effects limit the resolution on the reconstructed mass. This has been
studied in detail by the four experiments, using Monte Carlo events
with full detector simulation. We discuss methods of improving the
mass resolution over that obtained by simple calculation of invariant
masses. In particular, a kinematic fit using the constraints of energy
and momentum conservation, together with the equality of the two W
masses in an event, proves to be a very powerful technique for improving the
mass resolution, and also turns out to be a useful background rejection
criterion. For this reason, we concentrate on the channels \WWqqqq\ and 
\WWqqlnu\ where the lepton is an electron or muon, for which constrained 
fits are most useful.  

We start by discussing the basic selection of W-pair events in these
two channels, and the reconstruction of jets. We discuss the techniques of
the constrained fit in some detail, followed by the determination of
$\Mw$ from the distribution of reconstructed masses, indicating the 
statistical error which may be expected. Finally we describe the
main sources of systematic error pertaining to this measurement.

%%%%%% Event Selection Subsection for Direct Reconstruction %%%%%

\subsection{Event selection and jet reconstruction}
\label{sec:mwdir_evsel}  
The criteria used to select W-pair events are essentially the same
as those described in Section~\ref{sec:wmass_threshold}, but at the higher
energies used for direct reconstruction the background from
\Zqq\ is lower, so looser cuts can be used.

\subsubsection{\WWqqqq}

\begin{figure}[htbp]
\vspace{0.1cm}
\begin{center}
\epsfig{figure=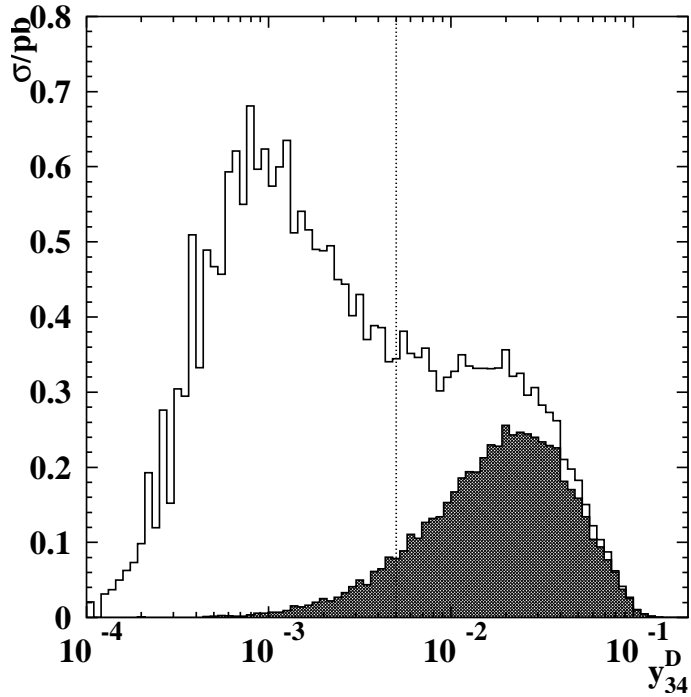,height=10cm,angle=0}
\end{center}
\caption{\it Distribution of $y_{34}^{D}$, the value of $y_{cut}$ in the 
      Durham (k$_{\mathrm{T}}$) jet-finding scheme at which events change 
      from three jets to four jets, after cuts on multiplicity, visible
      energy and missing momentum have been applied to events at
      175~GeV. The solid histogram shows \WWqqqq\ events, the open
      histogram background, mainly non-radiative \Zgqq\ events. The
      dotted line indicates a typical cut value.}
\label{fig:mwdir_y34}
\end{figure}

\WWqqqq\ events are characterised by high multiplicity (about twice
that of a \Zqq\ event at \Lepone), high visible energy, and exhibit
a four-jet structure. The main background comes from \Zgqq\ events, for which
the cross-section is much higher, as shown in Table~\ref{tab-MCXsect}.
Typical selection cuts for \WWqqqq\ events include:
\begin{itemize}
\item high multiplicity of tracks and calorimeter clusters to remove
      purely leptonic events; a significant fraction of \WWqqlnu\ and
      \Zgqq\ events can also be removed.
\item high visible energy and low missing momentum, to remove \WWqqlnu\
      and radiative \Zqq\ events.
\item explicit removal of events with an isolated electromagnetic cluster
      consistent with being an initial state photon to remove 
      radiative \Zqq\ events.
\item event shape variables to separate the four-jet \WWqqqq\ events
      from the remaining background, almost entirely composed of
      non-radiative \Zqq\ events. For example, Fig.~\ref{fig:mwdir_y34}
      shows the distribution of $y_{34}^{D}$, the value of $y_{cut}$ in the 
      Durham (k$_{\mathrm{T}}$) jet-finding scheme at which events change 
      from three jets to four jets, after cuts on the above quantities
      have been applied. The \WWqqqq\ events tend to have larger
      values of this variable than the background. Other event shape
      variables, such as jet broadening measures can also be used.
\end{itemize}

In Table~\ref{tab:mwdir_eff1} we show values of efficiency, purity,
accepted cross-sections and numbers of events produced by typical cuts on
these variables. The efficiency of selection
cuts tends to fall slightly with energy because as $\sqrt{s}$ is increased
the W's are more boosted and event shape variables have less
separating power. The purity can be further enhanced by using kinematic fits, 
as described below, though at a cost in efficiency.

\begin{table}[htbp]
\begin{center}
\begin{tabular}{|l|c|c||c|c|}
\hline
 &\multicolumn{2}{|c||}{$\sqrt{s}$ = 175~GeV} 
 &\multicolumn{2}{|c|}{$\sqrt{s}$ = 192~GeV} \\ \hline
 &Accepted  &Events  &Accepted &Events \\
 & cross-section (pb) &for 500pb$^{-1}$ & cross-section (pb) 
 &for 500pb$^{-1}$ \\ \hline
\multicolumn{5}{|l|}{After basic selection cuts:} \\ \hline 
\WWqqqq &5.3 &2650 &6.2 &3100 \\
\Zgqq   &3.1 &1550 &2.1 &1050 \\
\ZZ     &0.1 &50   &0.4 &200  \\ \hline
efficiency &\multicolumn{2}{|c||}{83\%} &\multicolumn{2}{|c|}{79\%} \\
purity &\multicolumn{2}{|c||}{62\%} &\multicolumn{2}{|c|}{62\%} \\ \hline
\multicolumn{5}{|l|}{After kinematic fit:} \\ \hline 
\WWqqqq &4.5 &2250 &5.0 &2500 \\
\Zgqq   &1.8 &900  &1.2 &600  \\
\ZZ     &0.08 &40   &0.3 &150  \\ \hline
efficiency &\multicolumn{2}{|c||}{71\%} &\multicolumn{2}{|c|}{64\%} \\
purity &\multicolumn{2}{|c||}{71\%} &\multicolumn{2}{|c|}{77\%} \\ \hline
$\Delta(\Mw)$ (stat) &\multicolumn{2}{|c||}{73~MeV} &\multicolumn{2}{|c|}
{74~MeV} \\
\hline
\end{tabular}
\end{center}
\vspace{-4mm}
\caption{\it Typical accepted cross-sections and numbers of events for
         the signal and main backgrounds for the \WWqqqq\ channel for
         two values of c.m. energy, determined from Monte Carlo including
         full detector simulation. Values are given both for basic
         selection cuts, and after demanding a good kinematic fit. 
         The efficiency, purity and expected statistical error on $\Mw$ 
         for an integrated luminosity of 500~pb$^{-1}$ are also given.}
\label{tab:mwdir_eff1}
\end{table}

In order to reconstruct the W mass, a jet-finder is used to force the
selected events to contain four jets. Jets are usually reconstructed
using both tracks and calorimeter information combined to give the
best resolution. The typical jet energy resolution is around 20\%
for a jet energy of 20~GeV, improving to 15\% at an energy of 60~GeV;
over this same energy range the angular resolution improves from 
4$^\circ$ to 1.3$^\circ$ for jets at 90$^\circ$ to the beam direction.
Studies of various jet finders comparing the reconstructed jets with the 
initial quark directions show no major differences among the 
commonly used schemes. The W mass may then be reconstructed by forming the 
invariant mass of pairs of jets. For each event, there are three possible 
combinations; Monte Carlo studies show that the combination where the two 
highest energy jets are combined together is rarely the correct one, and 
combinatorial background can be reduced by discarding this combination. 
The mass resolution can be improved by using beam energy constraints or 
kinematic fits, as described in the next section.

%\noindent{\boldmath  (b) \WWqqlnu}
\subsubsection{\WWqqlnu}

\begin{figure}[htbp]
\vspace{0.1cm}
\centerline{
\epsfig{figure=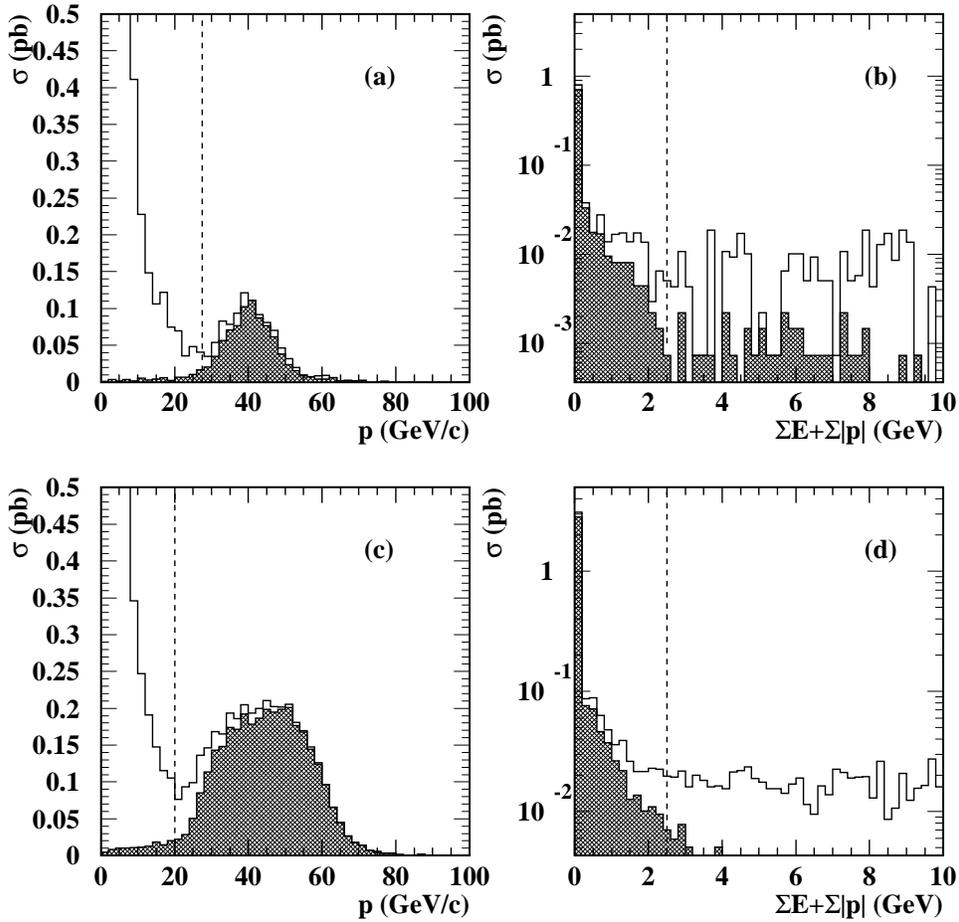,height=14cm,angle=0}
}
\caption{\it (a) Momentum spectrum of particles identified as leptons and
         passing isolation cuts at $161\gev$. 
         (b) scalar sum of charged particle momentum and electromagnetic 
         calorimeter energy in a 200~mrad cone around high momentum
         identified electrons and muons at $161\gev$.
         (c) and (d) as (a) and (b) for $\protect\sqrt{s} = 175\gev$. 
         In each case the solid histogram shows the contribution from 
         leptons in \WWqqlnu\ events, the open histogram the background.
         The dashed lines show possible cuts.}
\label{fig:mwdir_plep}
\end{figure}

The distinguishing feature of \WWqqenu\ and \WWqqmnu\ events is the
presence of a high momentum, isolated lepton. Typical selection cuts
for \WWqqenu\ and \WWqqmnu\ events include:
\begin{itemize}
\item high multiplicity of tracks and calorimeter clusters to remove
      purely leptonic events; the multiplicity of \WWqqlnu\ events
      is lower than that of \WWqqqq\ events, so suitable cuts
      do not remove \Zgqq\ background in this case.
\item the presence of a high momentum, isolated, lepton.
\end{itemize}
\begin{table}[!htbp]
\begin{center}
\begin{tabular}{|l|c|c||c|c|}
\hline
 &\multicolumn{2}{|c||}{$\sqrt{s}$ = 175~GeV} 
 &\multicolumn{2}{|c|}{$\sqrt{s}$ =192~GeV} \\ \hline
 &Accepted  &Events  &Accepted &Events \\
 & cross-section (pb) &for 500pb$^{-1}$ & cross-section (pb) 
 &for 500pb$^{-1}$ \\ \hline
\multicolumn{5}{|l|}{After basic selection cuts:} \\ \hline 
\WWqqlnu\ ($l=e$ or $\mu$) &3.1 &1550 &3.8 &1880 \\
\WWqqtnu &0.2 &100  &0.4 &200  \\
\Zgqq   &0.2 &100  &0.2 &100  \\
\Zee    &0.2 &100  &0.3 &150  \\ \hline
efficiency &\multicolumn{2}{|c||}{77\%} &\multicolumn{2}{|c|}{74\%} \\
purity &\multicolumn{2}{|c||}{83\%} &\multicolumn{2}{|c|}{80\%} \\ \hline
\multicolumn{5}{|l|}{After kinematic fit:} \\ \hline 
\WWqqlnu($l=e$ or $\mu$) &3.0 &1500 &3.4 &1700 \\
\WWqqtnu &0.05 &25  &0.06 &30   \\
\Zgqq   &0.04 &20  &0.05 &25  \\
\Zee     &0.02 &10   &0.05 &25   \\ \hline
efficiency &\multicolumn{2}{|c||}{73\%} &\multicolumn{2}{|c|}{68\%} \\
purity &\multicolumn{2}{|c||}{96\%} &\multicolumn{2}{|c|}{95\%} \\ \hline
$\Delta(\Mw)$ (stat) &\multicolumn{2}{|c||}{72~MeV} &\multicolumn{2}{|c|}
{93~MeV} \\
\hline
\end{tabular}
\end{center}
\vspace{-4mm}
\caption{\it Typical accepted cross-sections and numbers of events for
         the signal and main backgrounds for the \WWqqlnu\ ($l=e$ or $\mu$)
         channel for two values of c.m. energy, determined from Monte Carlo 
         including full detector simulation. Values are given both for basic
         selection cuts, and after demanding a good kinematic fit. 
         The efficiency, purity and expected statistical error on $\Mw$ 
         for an integrated luminosity of 500~pb$^{-1}$ are also given.}
\label{tab:mwdir_eff2}
\end{table}
An isolated lepton can be identified as an electron or muon with fairly
loose, standard experimental cuts with high efficiency ($\sim$95\%),
though only in a limited acceptance, typically $|\cos(\theta)| < 0.93$.
For example electrons can be identified using the match between track 
momentum and energy deposited in the electromagnetic calorimeter,
where the pattern of energy deposition is consistent with an
electromagnetic shower. Muon identification uses matching between a track 
in a central tracking chamber and one in outer muon detectors. 
The lepton momentum spectrum and its degree of isolation, as measured by the 
total scalar sum of charged particle momentum plus electromagnetic energy in 
a 200mrad cone around the lepton, are shown in Fig.~\ref{fig:mwdir_plep}.  
Typical efficiencies, purities, accepted cross-sections and numbers of
events are indicated in Table~\ref{tab:mwdir_eff2}.
\WWqqtnu\ events can be selected as above, but instead of requiring an
identified electron or muon, searching for a narrow, low multiplicity
jet isolated from the other jets.

The W mass can be estimated from these events by simply forming the
invariant mass of the hadronic system, scaling to the beam energy, or
preferably using the full information in a kinematic fit as described below.
In this case, the hadronic system is forced to be two jets, which are
reconstructed as in the \WWqqqq\ case.  The lepton energy resolution is
much better than that for a reconstructed jet, as long as care is taken
to include all the electromagnetic energy associated with an electron.

%%%%%%%%%%%%%%%%%%%%%%%%%%%%%%%%%%%%%%%%%%%%%%%%%%%%%%%%%%%%%%%%

%%%%%% Kinematic Fit Subsection for Direct Reconstruction  %%%%%

\subsection{Constrained fit}
After the event has been reconstructed as a number of jets and a number of
leptons we next turn to the reconstruction of the W mass from these 
four fermions, which are treated as individual objects with measurable 
quantities. We try to reconstruct the best estimator for the W mass on
an event by event basis from the measured quantities and the constraints 
imposed by energy and momentum conservation, and the possible additional
constraint that the masses of the two W's are equal.

Without imposing any constraints the direct reconstruction of the di-jet 
mass gives:
\be
m_{ij} = \sqrt{2E_iE_j(1-\cos\theta_{ij})},
\ee
where $E_i$ and $E_j$ are the jet energies, $\theta_{ij}$ the jet-jet opening
angle and where the jet masses have been neglected. Taking only the errors on 
the jet energies into account, as these are much larger than the
errors on the angle measurements, this then gives:
\be
\frac{\sigma (m_{ij})}{m_{ij}} = 
\sqrt{\left(\frac{\sigma (E_i)}{2 E_i} \right)^2 + 
\left(\frac{\sigma (E_j)}{2 E_j} \right)^2 }
\ee
Typical jet energy measurement errors of 15\% at $45\gev$ lead to 
a relative uncertainty of 10\% on $\Mw$, and give distributions
of reconstructed mass as shown in the top half of Fig.~\ref{mwdir_massplot}. 
To make a precision measurement of $\Mw$ it is necessary to improve
this resolution by making use of the knowledge of the total energy and 
momentum which is given in an \epem\ collider. 

\subsubsection{Rescaling methods}

These methods are especially useful for the analysis of the semi-leptonic 
channels and most have been described previously~\cite{aachen}.
The basic principle is to write the momentum, energy, and equal mass
constraints as functions of the measured fermion momenta and solve for those
variables which have the largest measurement uncertainties, generally
jet energies.

The first step is to include the beam energy constraint, which is equivalent 
to the constraint that the two masses are equal:
\be
E_i + E_j = E_{\rm b}.
\ee
This leads to a determination of the reconstructed $\Mw$:
\be
m_{ij} = \frac{E_{\rm b}}{E_i+E_j}\sqrt{2E_iE_j(1-\cos\theta_{ij})}.
\ee
Assuming that only the errors on the jet energies are important
the error on the mass becomes:
\be
\frac{\sigma (m_{ij})}{m_{ij}} 
\simeq \frac{|E_i-E_j|}{E_i+E_j} 
\sqrt{\left(\frac{\sigma(E_i)}{E_i}\right)^2+
\left(\frac{\sigma(E_j)}{E_j}\right)^2}.
\ee
Together with the other errors which were neglected, this leads to a 
relative uncertainty on the reconstructed mass of typically 5\%. 
This method is especially
suited to the semi-leptonic channel in which the leptonic W decay into 
$\tau \nu_\tau$ does not allow advantage to be taken of the measured lepton 
energy. 

In the case of semi-leptonic decays to an electron or muon, we can include
the parameters of the measured lepton by writing:
\begin{eqnarray} 
\vec{p}_\nu &=& - E_\ell \vec{{\rm e}_\ell} - E_i \vec{{\rm e}_i} 
- E_j \vec{{\rm e}_j} \nonumber \\ 
E_i + E_j &=& E_{\rm b} \nonumber \\ 
E_\ell + E_\nu & = &E_{\rm b},
\label{mwdir_dien}
\end{eqnarray} 
where $\vec{\rm e}$ are unit vectors in the direction of the particles.
These five equations with five unknowns, $\vec{p}_\nu$, $E_i$, and $E_j$, 
can be solved, and yield two distinct solutions.
This ambiguity leads to a problem if the two solutions are close to each 
other. Taking the solution which is closest to the measured jet energies
leads to a relative error on $\Mw$ of about 4\%.

The two exact solutions of Eq.~(\ref{mwdir_dien}) will
give two minima also for the constrained fit in the $\chi^2(\Mw)$ 
distribution when the two solutions are close. 
Monte Carlo studies suggest that about 40\% of the events are afflicted by this 
problem. Current analyses have not yet included this effect in the
determination of $\Mw$ and one might therefore expect an  
improvement in resolution if this effect is taken correctly into account.

\subsubsection{The constrained fit}

%\subsubsection{The constrained fit}
The most effective way to use all the information available in an
event is to perform a constrained kinematic fit. In such a fit, the
measured parameters are varied until a solution is found which satisfies the
constraints imposed and also minimises the $\chi^2$  difference between
the measured and fitted values. Several methods exist to perform this 
minimization. A traditional one solves the problem using
Lagrange multipliers, minimising 
\be{\cal S}(\vec{y},\vec{\lambda})=(\vec{y}-\vec{y_{0}})^{\top}
{\bf V}^{-1}(\vec{y}-\vec{y_{0}})+2\vec{\lambda}\cdot\vec{f}(\vec{y}),
\ee
where ${\bf V}$ is the error matrix, $\vec{y}$ the fitted variables,
$\vec{y_{0}}$ the measured values, $\vec{\lambda}$ Lagrange multipliers,
and $\vec{f}(\vec{y})$ the constraints written as functions which must vanish. 
An equivalent method is to use penalty functions where
terms of the type $\vec{f}^2/\sigma^2$ are added to the $\chi^2$.
The procedure then minimizes the total $\chi^2$ in an iterative way, for 
each step decreasing the $\sigma$ of the penalties. Results are in general
the same but the convergence is slightly slower.

The inputs to the fit are measurements of the energy and angles, of the four 
jets in the hadronic final state or of the two jets and lepton in the 
semi-leptonic final state, together with their error matrix.
Errors on jet parameters can be extracted from data or Monte Carlo,
and may be functions of both energy and position measured in
the detector. In practice most of the jet measurement errors are nearly 
uncorrelated, making the error matrix diagonal. For the jet masses two 
different strategies are used. Either the jets are assumed to be massless and 
the measured jet energy is used as the measured jet momentum, or one includes 
the reconstructed jet masses in the fit.
%\footnote{  
%The $\tau$ lepton can be treated in nearly the same way as jets, except that 
%the errors on the transverse momenta are given by the missing mass and that 
%the $\tau$ mass is used explicitly in the fit. This has not yet been
%studied in detail.
%}.
(Note that the $\tau$ lepton can be treated in nearly 
the same way as jets, except that 
the errors on the transverse momenta are given by the missing mass and that 
the $\tau$ mass is used explicitly in the fit. This has not yet been
studied in detail.)

The fit can be performed using only the constraints of energy and momentum 
conservation, or also including the constraint that the masses of
the two reconstructed W's be equal. In the hadronic channel, this gives a 4C 
or 5C fit respectively. In the semi-leptonic channel, the number of constraints
is reduced by three because the parameters of the neutrino are
unmeasured, resulting in a 1C or 2C fit. 
The equal mass constraint can be included exactly, 
or the width of the W can be taken into account by adding a term to
the $\chi^2$ proportional to the difference in mass of the two 
W's. In practice, because the mass resolution is larger than the
W width, both methods give almost the same results. If an equal mass
constraint is not applied, the reconstructed masses of the two di-fermion
systems are strongly anticorrelated. Thus only the average invariant
mass can usefully be extracted per event. The inclusion of an equal
mass constraint is preferred over a fit using only energy and
momentum conservation because it gives improved mass resolution and
superior background rejection.  

In the 4-jet channel we do not know which jets are to be combined to produce 
the heavy particles we are interested in reconstructing. Therefore the 
constrained fit is performed for all three possible combinations.
In the case with no equal mass constraint there are three
different mass solutions with the same $\chi^2$. When an equal mass 
constraint is imposed the three different solutions will in general
have different $\chi^2$ and one can therefore use this information
to distinguish between the solutions. The procedure employed by most 
analyses is to first define a mass window, wide enough not to bias the
mass measurement, and if more than one solution exists inside this window 
choose the solution with the lowest $\chi^2$. This procedure is however not 
perfect and one is left with a fraction of wrong combinations as shown in 
Fig.~\ref{mwdir_massplot}.

\subsubsection{Results of the fit}

\begin{figure}[p]
\begin{center}
\mbox{\epsfig{file=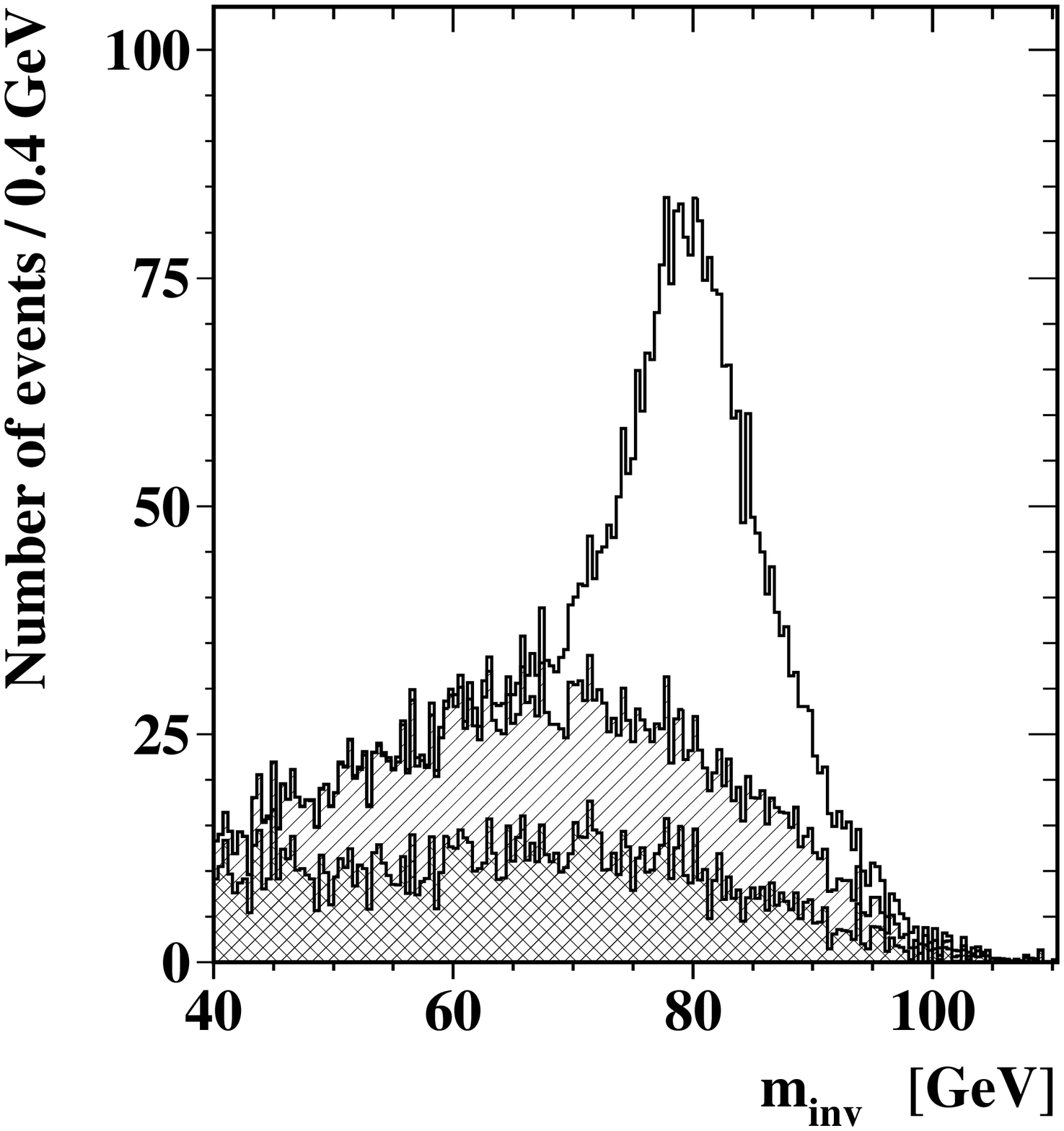,height=8cm}}
\hfill
\mbox{\epsfig{file=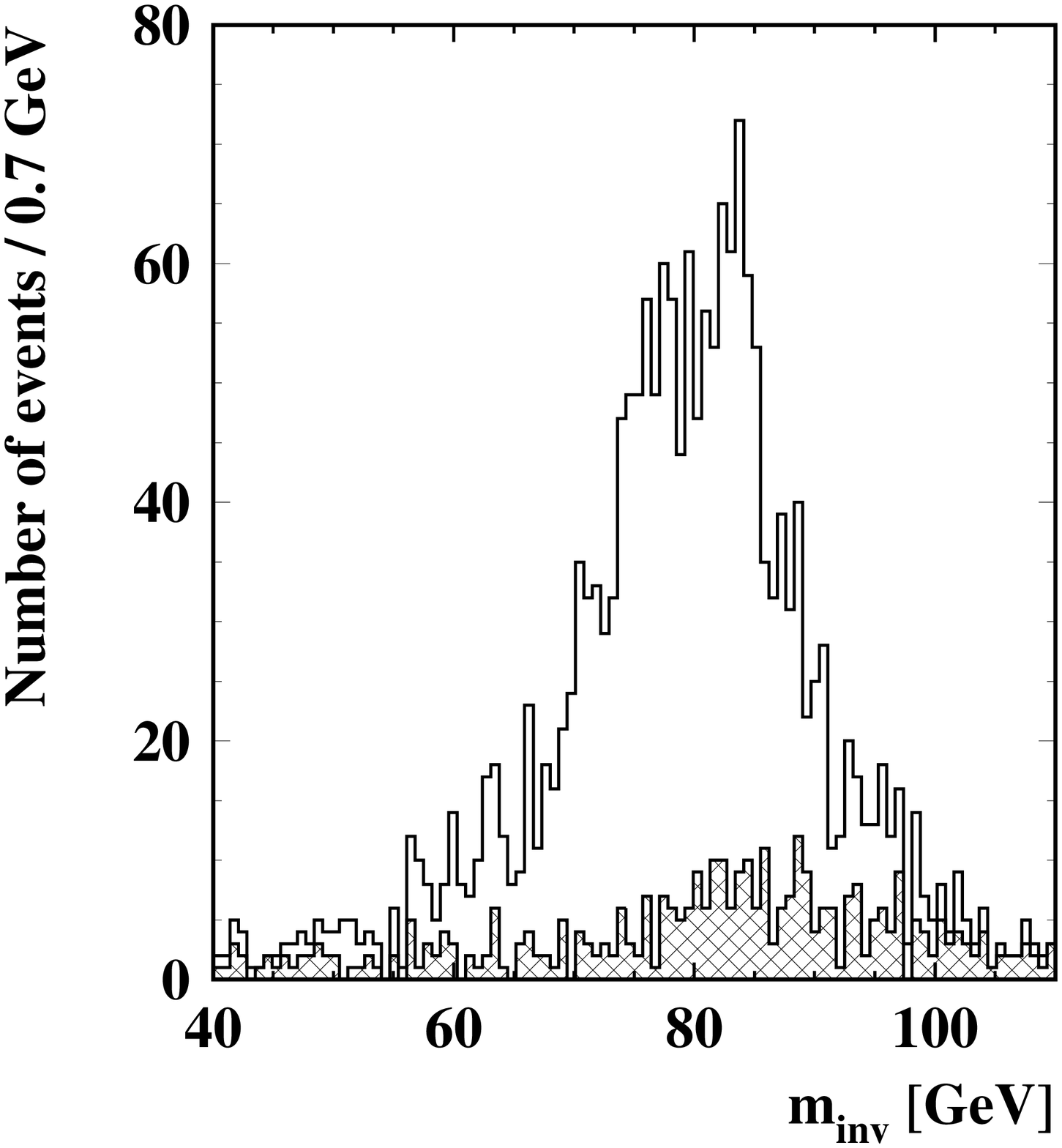,height=8cm}} \\
\mbox{\epsfig{file=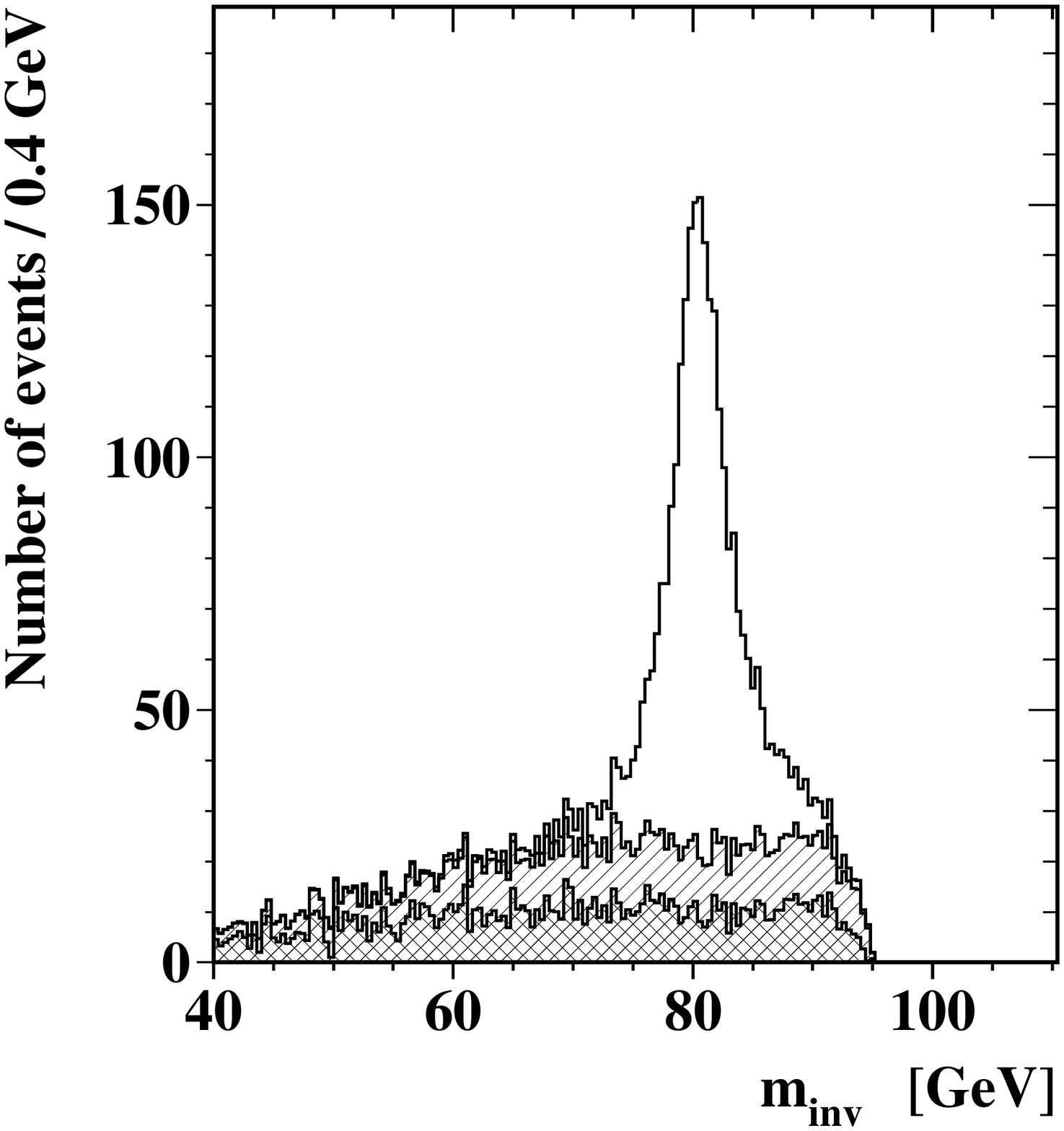,height=8cm}} 
\hfill
\mbox{\epsfig{file=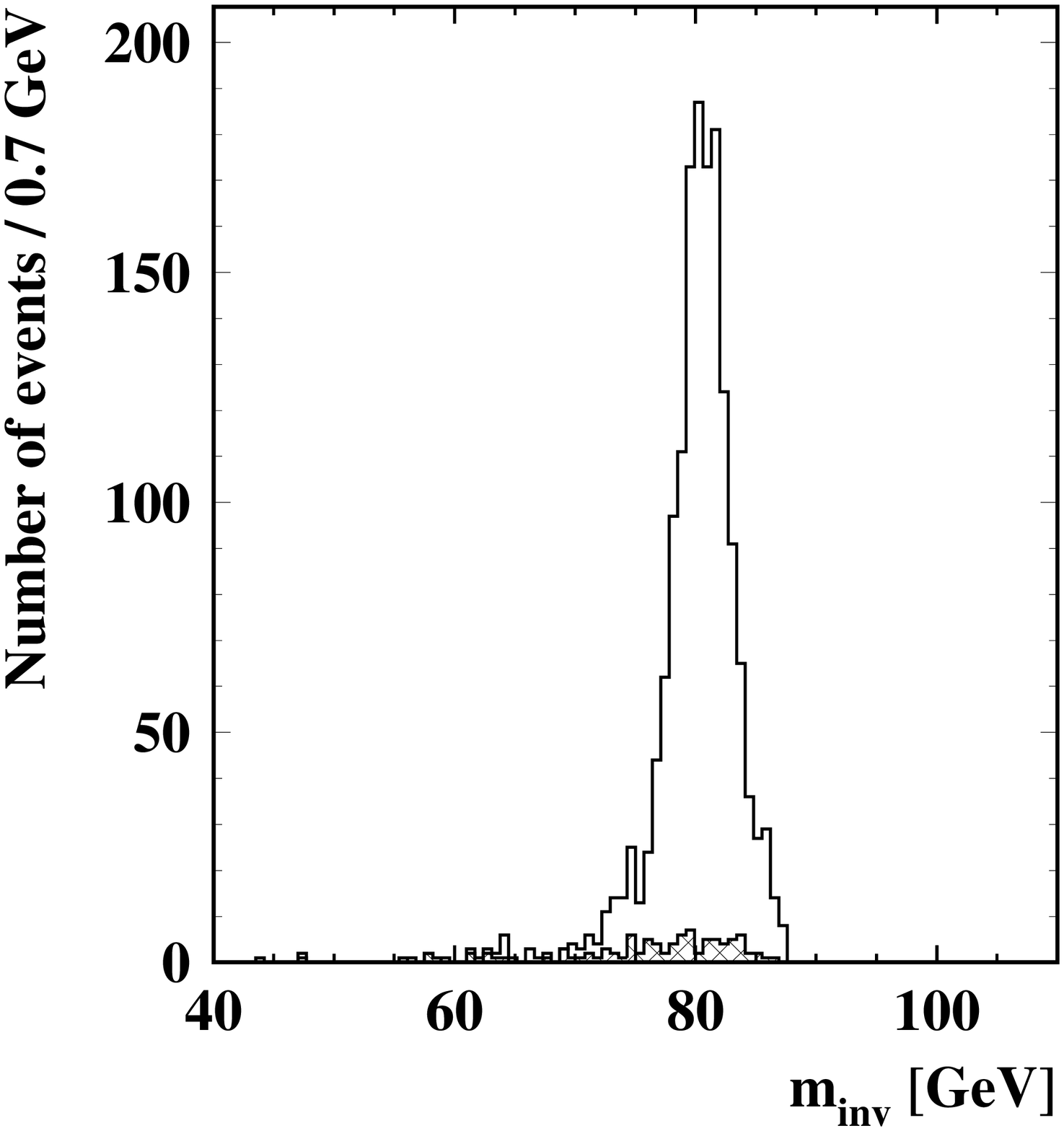,height=8cm}} \\ 
\end{center}
\caption[]{\it Distributions of the invariant mass for selected events
  for the \WWqqqq\ channel at $\sqrt{s} = 192\gev$ (left) and the \WWqqlnu\
  channel at $\sqrt{s} = 175\gev$ (right). The top plots are before the kinematic 
  fit, the bottom ones after a five-constraint (\WWqqqq)
  or two-constraint  (\WWqqlnu) kinematic fit. 
  For the \WWqqqq\ channel, the open histograms show the correctly found 
  jet-jet combinations, the singly hatched areas correspond to the 
  incorrectly found jet-jet combinations and the cross-hatched parts 
  indicate the background from other sources (mainly $q\bar q(\gamma)$).
  For the \WWqqlnu\ channel the open histograms show signal events,
  the cross-hatched areas background including the \WWqqtnu\ contribution.
  The statistics correspond to an integrated luminosity of 500~pb$^{-1}$.}
\label{mwdir_massplot}
\end{figure}

In Fig.~\ref{mwdir_massplot} we show invariant mass distributions
before and after the fit, in the latter case taking only those events
which give a good fit. Before the fit, the mass distributions are
very broad. After the fit the mass resolutions obtained are typically 
3.5\% for the \WWqqlnu\ channel and around 2.5\% for the 4-jet channel.
It is also clear from this figure that the $\chi^2$ of the constrained fit 
can be used to eliminate possible background which does not comply with the 
$\WW$ hypothesis: the level of background is much lower in both
channels after demanding a good fit.
However, a fraction of $\WW$ events, varying from about 10\% to 30\% 
depending on the channel, also fails to give a good fit. These events are
discussed below. Typical values of efficiency and purity after the
fit are given in Tables~\ref{tab:mwdir_eff1} and~\ref{tab:mwdir_eff2}.

From the constrained fit we can calculate the error on the fitted mass. 
This error is highly correlated to the actual value of the fitted mass, so 
selecting events with a simple cut on this quantity would seriously bias the
measurement. The reason for this effect is related to the kinematics of the 
events. When the mass approaches the kinematic limit the precision on the 
sum of the unknown masses will be better and better while the precision on 
the difference between the masses will deteriorate. This leads to a mass 
resolution that broadens with increasing energy. Going from $175\gev$ to 
$192\gev$ typically increases the measurement errors by 25\% on an event by 
event basis. The effect of this on the final statistical error on $\Mw$
is diluted by the fixed $\Gamma_{\rm W}$ and compensated by an increase of 
the $\WW$ cross-section.

The constrained fit assumes that the errors on the measured quantities
are Gaussian and uncorrelated between jets. Several effects lead to 
non-Gaussian errors and correlations. Each of these will lead to tails in 
the distributions and hence to a peak at low probability for the fitted 
$\chi^2$. The most important is gluon radiation, but also overlapping
jets, initial state radiation, $\Gamma_{\rm W}$, and acceptance effects 
play a r\^ole. 
The hard gluon radiation is of course in direct disagreement with the 
treatment of jets as independent objects. Even rather soft gluon radiation 
lead to jets being broadened in a specific direction, giving correlations 
that are not included in the current implementations of the constrained fit.
Studies have been performed to try to recover some of the 4-jet events
which fail to give a good fit by treating them as 5-jet events. 
However, although some fraction of these events then give a good fit, the
combinatorial problem is severe, and it appears that their inclusion
has little effect on the ultimate mass resolution.  

\subsubsection{Inclusion of initial state radiation}

As will be seen below, energy lost in initial state radiation biases
the fitted mass if it is not included in the fit.
Initial state radiation can be included in the constrained fit using
the following procedure.
We know that there is a large probability that the photon 
is produced close to the collision axis and hence not detected. 
We can therefore as a good assumption take the momentum to be collinear with
the $z$-axis. We also know to a high precision the expected distribution
of the photon energy. In a simple approximation this reads:
\be
p(x) dx = \beta x^{\beta-1} dx,
\ee
where $x=E_\gamma/\sqrt{s}$ and $\beta$  is a parameter which is 
smaller than unity and depends on $\sqrt{s}$. 
Since this is non-Gaussian we cannot take this term directly
into the $\chi^2$ expression. Instead we introduce the likelihood concept
and use the standard -2 ln(likelihood) as the term to add to the $\chi^2$.
If we just use the probability as the likelihood this approach will 
not work since the distribution has a pole for $x \rightarrow 0$. 
Instead one can choose to use the confidence limit as an estimator for the 
likelihood:
\be
C(x) = \int_x^{\infty} p(y) dy 
\ee
With this formulation the constrained fit works but its 
implementation is rather difficult since one has to divide the fit into
two parts: one where one assumes the photon goes in the forward
direction and one where one assumes the opposite. When the fitted $E_\gamma$
approaches zero the first order derivative of $\chi^2$ will still diverge,
but this only means that the fit prefers the zero solution, since the 
measurement does not have sufficient resolution to distinguish 
between no $\gamma_{\rm ISR}$ and a $\gamma_{\rm ISR}$ with a small energy.
Monte Carlo studies show that photons below about $3\gev$ cannot be resolved
and when photons are above typically $8\gev$ the W bosons can not be on 
mass-shell. The final improvement in the $\Mw$
measurement is therefore limited.

%%%%%%%%%%%%%%%%%%%%%%%%%%%%%%%%%%%%%%%%%%%%%%%%%%%%%%%%%%%%%%%%

%%%%%% Mass Fit Subsection for Direct Reconstruction       %%%%%

\subsection{Determination of the mass and width of the W}
\label{sec:mwdir_wfit}

In this section we describe several strategies for extracting 
 $\MW$ from distributions of reconstructed 
invariant masses such as those show in Fig.~\ref{mwdir_massplot}. 
As we aim for a precision measurement of $\MW$ with 
sub-permille accuracy, this is a non-trivial task, because we 
have to control any systematic effect to an accuracy of a few MeV. The
total width of the W boson, $\GW$, may either be extracted 
simultaneously with the mass, or the functional dependence 
$\GW=\GW(\MW)$ of the Standard Model may be imposed as a constraint
for increased accuracy on $\MW$.
The methods to analyse the data in terms of $\MW$ and $\GW$ fall into 
four groups:
(1) Monte Carlo calibration of simple function;
(2) (De-) Convolution of underlying physics function;
(3) Monte Carlo interpolation;
(4) Reweighting of Monte Carlo events.

In general, all methods make use of Monte Carlo 
event generators~\cite{WWGROUP} and detector simulation 
to determine 
the effects of the detector such as resolution. Thus any method has 
to be checked for possible systematic biases introduced by using 
Monte Carlo event samples generated with certain input values for 
$\MW$ and $\GW$. In addition, systematic errors may arise due to 
deficiencies in the Monte Carlo simulation describing the detector 
and/or the data. Other systematic errors arise from the technical
implementation of the fitting methods, such as fit range, bin width in
case of binned data, choice of functions to describe signal and 
background etc.

%\noindent{\bf (a) Monte Carlo Calibration}
\subsubsection{Monte Carlo calibration}

To fit the invariant mass distribution, this method uses a simple, 
ad-hoc function, e.g., a double Gaussian or a Breit-Wigner convoluted
with a Gaussian, to describe the signal peak and another simple
function to describe the background. One of the fit parameters is 
used as an estimator, $M$, for the W mass, e.g., 
the mean of the Gaussian or the Breit-Wigner. The same fitting 
procedure is applied to both data and Monte Carlo events. Since for 
the latter the input W mass, $\MW^{\mathrm{MC}}$, is known, 
the Monte Carlo result is used to evaluate the bias $\Delta$ of this 
method, $\Delta \equiv M^{\mathrm{MC}}-\MW^{\mathrm{MC}}$, where this bias may 
depend on the final state analysed. The mass of the W measured 
in the data is now simply given by the estimator $M^{\mathrm{data}}$ 
derived from fitting the data distribution, corrected for the bias 
$\Delta$ evaluated from Monte Carlo events: 
$\MW=M^{\mathrm{data}}-\Delta$. 

This procedure automatically takes into account all corrections for all
biases as long as they are implemented in the Monte Carlo simulation,
such as initial-state radiation, background contributions, detector 
resolutions and efficiencies, selection cuts etc. The knowledge of 
how well the Monte Carlo describes the underlying physics and the 
detector response enters in the systematic error on the bias $\Delta$. 

The fundamental drawback of this method is that the simple function 
used in the fit is not unique. Depending on its choice, even different 
statistical errors on $\MW$ can be obtained. In addition, the estimate of 
the bias correction 
$\Delta$ depends itself to some extent on the Monte Carlo parameters
$\MW^{\mathrm{MC}}$ and $\GW^{\mathrm{MC}}$, or on the centre-of-mass 
energy, $\sqrt{s}$, 
$\Delta=\Delta(s;\MW^{\mathrm{MC}},\GW^{\mathrm{MC}})$. 
Such a dependence, however, can be corrected for by iteration. 

%\noindent{\bf (b) Convolution}
\subsubsection{Convolution}

The drawbacks listed above are alleviated in the convolution method. 
Here, the correct function, i.e., the underlying physics function, 
is used as a fitting function. This function is simply the 
differential cross-section in the two invariant masses (denoted by 
$m_1,m_2$). Note that this function is not a simple Breit-Wigner 
distribution in $m_1$ and $m_2$ due to phase space effects and
radiative corrections. Several analytical codes exist (e.g., GENTLE 
etc.~\cite{WWGROUP}), which calculate the differential cross-section
including QED corrections, 
$\frac{\dd\sigma(s;\MW,\GW)}{\d m_1 \d m_2}$, as a function of the 
centre-of-mass energy, $\sqrt{s}$, and the Breit-Wigner mass and 
total width of the W boson, $\MW$ and $\GW$. 

The effects of the detector are included by convolution. The 
prediction for the distribution of the reconstructed invariant masses 
(denoted by $\overline{m}_1,\overline{m}_2$) is thus given by:
\begin{equation}
\frac{\dd\sigma(s;\MW,\GW)}{\d\overline{m}_1~\d\overline{m}_2} ~ = ~
\int \d m_1 \int \d m_2~G(s;\overline{m}_1,\overline{m}_2,m_1,m_2)\cdot
\frac{\dd\sigma(s;\MW,\GW)}{\d m_1~\d m_2}\,.
\end{equation}
The transfer or Green's function $G$, which depends on the final
state analysed, can be interpreted as the probability of reconstructing
the pair of invariant masses $(\overline{m}_1,\overline{m}_2)$ given 
the event contained the pair of true invariant masses $(m_1,m_2)$. 
Several simplifications for $G$ are possible, down to a 1-dimensional
resolution function 
$G=G\left([{\overline{m}_1+\overline{m}_2}]/{2}-[{m_1+m_2}]/{2}\right)$.

The actual fitting of invariant mass distributions can be performed 
either on the measured or on unfolded distributions. In the former 
case, the underlying physics function, 
$\frac{\dd\sigma(s;\MW,\GW)}{\d m_1 \d m_2}$,
is convoluted with $G$, and the result is fitted to the data with $\MW$ 
and $\GW$ as fit parameters. In the latter case, the measured 
distribution,
$\frac{\dd\sigma(s;\MW,\GW)}{\d \overline{m}_1 \d \overline{m}_2}$,
is first unfolded for detector effects, by applying the ``inverse'' of
$G$. The underlying physics function, 
$\frac{\dd\sigma(s;\MW,\GW)}{\d m_1 \d m_2}$, can now  be 
fitted directly to the unfolded distribution. From a purely mathematical point 
of view, both methods are equivalent.
In practice, however, $G$ is determined only
up to a certain statistical accuracy. Since folding is an intrinsically
stable procedure in contrast to unfolding, which is unstable, the 
former method is preferred.  Reference~\cite{unfolding} gives more 
details on the features of unfolding procedures. Backgrounds are
described by simple functions, derived from data and from Monte Carlo
simulations, which are added to the signal.

This method allows cuts on generated invariant masses ($m_1,m_2$) and
ISR energy loss, since these are the only variables used in most
semianalytical codes. In addition, cuts on the reconstructed invariant
masses ($\overline{m}_1,\overline{m}_2$) are possible. Since cuts 
on other variables cannot be applied, it must be checked whether 
selection cuts bias the invariant mass distribution, for example by 
testing the method with Monte Carlo events (cf. method (1)).

%\noindent{\bf (c) Other Monte Carlo based Methods}
\subsubsection{Other Monte Carlo based methods}

The problems of the previous two methods can be solved by a 
Monte Carlo interpolation technique. Several samples of Monte Carlo 
events corresponding to different input values of $\MW^{\mathrm{MC}}$ 
and $\GW^{\mathrm{MC}}$ are simulated, e.g., in a grid around the
current central values for $(\MW,\GW)$ extending a few times the total
(expected) error in all directions. The Monte Carlo samples of the 
accepted events are compared to the accepted data events, thereby 
taking the influence of event selection cuts into account. Backgrounds
are included by adding the corresponding Monte Carlo events. The 
compatibility of the invariant mass distributions is calculated, e.g.,
in terms of a $\chi^2$ quantity. Interpolation of the $\chi^2$ within 
the generated ($\MW^{\mathrm{MC}}$, $\GW^{\mathrm{MC}}$) grid allows 
to find the values $\MW$ and $\GW$ which minimise the $\chi^2$. Like 
method~(1), this method corrects automatically all possible biases due
to all effects considered in the Monte Carlo simulation. The only
drawback is that a rather large amount of Monte Carlo events must be 
simulated.

This problem, however, can be solved by a reweighting procedure. In
that case only one sample of Monte Carlo events is needed, which has 
been generated with fixed values $\MW^{\mathrm{MC }}$ and 
$\GW^{\mathrm{MC }}$. Event-by-event reweighting in the generated
invariant masses ($m_1,m_2$) is performed to construct the prediction 
for the invariant mass distributions corresponding to values 
$\MW^{\mathrm{fit}}$ and $\GW^{\mathrm{fit}}$. These distributions 
are then fitted to the data distributions. 
Since individual Monte Carlo events are reweighted, it is straight 
forward to implement the effects of selection cuts. Moreover, using 
a Monte Carlo generator also in the calculation of the event weights, 
it is even possible to extend further the set of variables on which 
the event weights depend to include any kinematic variable describing 
the four-fermion final state, such as the reconstructed fermion 
energies and angles. 

\subsubsection{Expected statistical error on $\Mw$}

So far the experiments have studied methods (1), (2) and 
(3)~\cite{ALEPH-MINV-FITS,DELPHI-MINV-FITS,L3-MINV-FITS,OPAL-MINV-FITS}. 
For the statistical errors on $\MW$ quoted in Tables~\ref{tab:mwdir_eff1} 
and~\ref{tab:mwdir_eff2}, the experiments have mainly used method
(1), which is adequate for this purpose. It should be noted that the 
more involved analyses (2), (3) and (4) do not aim for a reduction in 
the statistical error on $\MW$. This error is fixed by the number of 
selected events, the natural width of the W boson and the detector 
resolution in invariant masses. Instead the advantage of these methods
lies in the fact that they allow a thorough investigation of the various 
systematic biases arising in the determination of the W  mass. 
It is possible to disentangle systematic effects due to background 
diagrams, initial-state radiation, event selection, detector 
calibration and resolution etc. Thus, in order to have a better
control of systematic effects, it is envisaged that the final analyses
will 
 use the more involved strategies (2), (3), (4) or a combination thereof.

As shown in Table~\ref{tab:mwdir_eff1}, the statistical error on $\Mw$
from the \WWqqqq\ channel is expected to be about $73\mev$, roughly
independent of $\sqrt{s}$, for an integrated luminosity of $500\pb^{-1}$.
For the \WWqqlnu\ channel a similar value is expected at $175\gev$,
but at higher energies the worsening resolution on $\Mw$ causes this
value to increase somewhat.

%%%%%%%%%%%%%%%%%%%%%%%%%%%%%%%%%%%%%%%%%%%%%%%%%%%%%%%%%%%%%%%%

%%%%%% Systematic Errors Subsection for Direct Reconstruction %%%%%

\subsection{Systematic errors}
\label{sec:sys_error}
Three classes of systematic error have been envisaged: errors coming from the
accelerator, from the knowledge of the underlying physics phenomena, and from
the detector. The last two classes are sometimes related as some physics
parameters may affect the detector response.
In contrast with the threshold cross-section method, effects which distort
the mass distribution must be considered here.
Most of the following error
estimates have been obtained using simulations. The models used will be checked
against LEP2 data when available and represent today's state of the art.

%\noindent{\bf (a) Error from the LEP beam}
\subsubsection{Error from the LEP beam}

The direct mass reconstruction method relies on constraining or rescaling
the energies of the reconstructed W's to the beam energy.
The error on the beam energy is foreseen to be less than $12\mev$
(see Section~\ref{sec:lep2wmachine}), translating to
$12\mev\ \times  \Mw / E_{beam}$ on the mass.
The beam energy spread will be
$\approx 40\mev \times s / (91\gev)^2$ ~\cite{lep2wmachine} with a negligible
effect on the statistical error.
The operation with bunch trains should not significantly affect the beam 
energy.
Experience gained at LEP1 will guarantee a very good follow up of the behaviour
of the beam even at short time scales.

\subsubsection{Error from the theoretical description}
%\subsubsection{Error from phenomenology}

The most important distortion of 
the mass distribution comes from initial state
radiation (ISR). The average energy carried by radiated ISR photons is
$\langle E_{ISR}\rangle = 1.2\gev$ per event at $\sqrt{s}=175\gev$
and rises linearly up to $2.2\gev$ at $\sqrt{s}=200\gev$.
From the beam energy constraint, one expects an average mass shift of the
order of $\langle E_{ISR}\rangle
  \Mw /\sqrt{s}$, i.e. about $500\mev$ at $175\gev$.
A fit to the mass distribution gives more weight to the peak
yielding a shift of only $200\mev$ at this energy.
As a direct measurement of the ISR spectrum seems impossible to an accuracy
relevant for the W mass measurement, we will rely on the theoretical
calculations as described in Section~\ref{sec:theory_isr}
 which present an error of
$10\mev$ on $\langle E_{ISR}\rangle$.
 Together with uncertainties on the shape of the
spectrum, this translates into an error smaller than $10\mev$ on the W mass.

Background events contained in the final sample will also distort the mass
distribution (Fig.~\ref{mwdir_massplot}). These backgrounds can be
monitored
with the data themselves under a dedicated analysis. The most important source
of background comes from \Zgqq\ events which has a
flat mass distribution in the fit area. The effects of an error on the level or
in the shape of
this background have been estimated by changing its production cross-section by
20\% and by
shifting the background mass distribution by $0.5\gev$ resulting in
an error of $12\mev$ for the \WWqqqq\ channel and $6\mev$ for the
\WWqqlnu\ channel. The background four jet rate as measured at LEP1 is not
perfectly reproduced by the simulations~\cite{qcd_fourjet}
 and will have to be 
better determined.

The fragmentation model for W events is the same as was fitted to
Z events at LEP1. To estimate errors from this model, each of its  parameters
has been varied and the error is derived
according to a 1 standard deviation from the best fit value.
The total error is $16\mev$, which should be taken as an upper
limit as the parameters have been varied separately without taking into account
correlations.

Colour reconnection and Bose-Einstein correlations may seriously affect the
mass distribution in the \WWqqqq\ channel. These are discussed in detail in
Section~\ref{sec:colrec}.

%\noindent{\bf (c) Error from the detector}
\subsubsection{Error from the detector}

Effects of miscalibration appear to be very small. This is
due to the use of the beam energy which gives a precise scale. For instance,
a conservative shift of 2\% in the momentum of charged tracks, and
miscalibration of 5\%
for electromagnetic and hadronic calorimeters produces a shift of
$10\mev$ on the W mass.

During the LEP2 era, it is foreseen to run for short periods at $\sqrt{s}=\Mz$
in order to calibrate and determine the efficiencies of the detectors with high
statistics. Genuine LEP2 events like $\epem\rightarrow\Zzero\gamma$ will also
be used to monitor the hadronic recoil mass to the photon.
Two Z events produced at LEP1 can be mixed and boosted to a LEP2 energy to
study the adequacy of the mass reconstruction in comparison with the
simulations.

In the \WWqqqq\ channel, the wrong assignment of a jet to a W distorts the mass
distribution giving rise to a low mass tail. The amount of such misassignments
depends on the algorithms used. By changing parameters in such algorithms
we estimate an error of up to $5\mev$.

Several methods to extract the W mass from the measurements have been studied
(see Section~\ref{sec:mwdir_wfit}). If a simple function is
fitted to the measured mass distribution, errors from the choice of the mass
fit range are to be expected, and are
typically $10\mev$. A convolution of the physics function with (or a
deconvolution from) detector effects should largely reduce such errors.
However this convolution depends on the accuracy of the simulation of the event
structure, for instance on the fragmentation model used (see above).

In the present simulation studies, limited statistics have been generated.
The statistical error on the mass shift with 500k events
simulated is of the order of $10\mev$.

\begin{table}[htbp]
\begin{center}
\begin{tabular}{|l|c|c|}
\hline
Source             & \WWqqqq & \WWqqlnu \\
\hline
$E_{beam}$         & 12 & 12 \\
ISR                & 10 & 10 \\
fragmentation      & 16 & 16 \\
backgrounds        & 12 &  6 \\
calibration        & 10 & 10 \\
MC statistics      & 10 & 10 \\
mass fit           & 10 & 10 \\
jet assignment     &  5 &  - \\
interconnection    &  ? &  - \\  
\hline
{\bf total }       &{\bf 31} &{\bf 29 } \\
\hline
\end{tabular}
\end{center}
\vspace{-4mm}
\caption{\it Estimated systematic errors on $\Mw$ per experiment in$\mev$.
Colour reconnection and Bose-Einstein effects are not included in the total.}
\label{mwdir_systab}
\end{table}

%\noindent{\bf Common errors}
%\noindent{\bf (d) Common errors}
\subsubsection{Common errors}

The effects mentioned above are summarized in Table~\ref{mwdir_systab}
and total
to about $30\mev$ for each channel at $\sqrt{s} = 175\gev$.
Although errors from the beam energy, ISR and background will vary with energy,
we do not expect large variations over the 165 to $190\gev$ range.

Errors on the beam energy and ISR are common to all experiments.
Errors from fragmentation and background are partially common
as far as the physical parameters are concerned, but a particular analysis
will retain a level of background or be sensitive to fragmentation
tails in a different way from another analysis, hence with a slightly different
error. If these errors were considered common to the four
experiments, they would amount to a total of $25\mev$
($23\mev$)  in the \WWqqqq\ (\WWqqlnu) channel.

%%%%%%%%%%%%%%%%%%%%%%%%%%%%%%%%%%%%%%%%%%%%%%%%%%%%%%%%%%%%%%%%

%%%%%% Summary Subsection for Direct Reconstruction %%%%%

\subsection{Summary}
In Table~\ref{mwdir_sumtab} we show the total error on $\Mw$ which
might be expected by combining four experiments each with an integrated 
luminosity of $500\pb^{-1}$ at $\sqrt{s} = 175\gev$. As discussed in the
previous section, systematic errors from the beam energy measurement
and initial state radiation, fragmentation and backgrounds are considered 
common to both channels and to all experiments. The total expected error is 
around $34\mev$. with roughly equal contributions from statistics and
systematics. This value is expected to rise, but only very slowly, with 
$\sqrt{s}$
because of the worsening mass resolution.

\begin{table}[htbp]
\begin{center}
\begin{tabular}{|l|c|c|c|}
\hline
Source             & \WWqqqq & \WWqqlnu  &Combined\\
\hline
Statistical        & 36      & 36         & 25 \\
Common systematic  & 25      & 23         & 23 \\
Uncorr. systematic &  9      &  9         &  6 \\
\hline
{\bf total }       &{\bf 45} &{\bf 44 }   &{\bf 34}\\
\hline
\end{tabular}
\end{center}
\vspace{-4mm}
\caption{\it Estimated total error on $\Mw$, in$\mev$, which could be
obtained by combining four experiments each with an integrated
luminosity of $500\pb^{-1}$ at $\protect\sqrt{s} = 175\gev$.
Colour reconnection and Bose-Einstein effects are not included in the 
systematic error.}
\label{mwdir_sumtab}
\end{table}

%\input s4_new.tex
%%%%%%%%%%%%%%%%%%%%%%%%%%%%%%%%%%%%%%%%%%%%%%%%%%%%%%%%%%%%%%%%%%%%%
%
%                Draft Interconnection Panel of W Mass Group
%                ===== =============== ===== == = ==== =====
%
%                Latest update : 12 December 1995
%
%%%%%%%%%%%%%%%%%%%%%%%%%%%%%%%%%%%%%%%%%%%%%%%%%%%%%%%%%%%%%%%%%%%%%
\section{Interconnection Effects\protect\footnotemark[4]}
\footnotetext[4]{ prepared by V.A.~Khoze, L.~L\"onnblad, 
R.~M{\o}ller, T.~Sj\"ostrand, \v{S}.~Todorova and N.K.~Watson.}
\label{sec:colrec}
\subsection{Introduction}

The success of the precision measurements of the W boson mass
\MWImw\ strongly relies on accurate theoretical knowledge of the
dynamics of the production and decay stages in 
$\MWIe^+ \MWIe^- \rightarrow \MWIw^+ \MWIw^- \rightarrow 4$ fermions.  
Owing to the large W width, \MWIgw, these stages are not independent but 
may be interconnected by QCD (and electroweak) interference effects,
which must be kept under theoretical control.  Interconnection
phenomena may obscure the separate identities of the two W
bosons, so that the final state may no longer be considered as a
superposition of two separate W decays \cite{MWIGPZ,MWISK}. Thus the
``direct reconstruction'' method of measuring \MWImw\ at LEP2
using the hadronic $(\MWIq_1 \MWIqbar_2 \, \MWIq_3 \MWIqbar_4)$
channel has an important caveat --- the colour reconnection
effects \cite{MWIGPZ,MWISK,MWIGH} --- which may distort the mass 
determination \cite{MWISK}.  Another delicate question is whether 
Bose--Einstein effects could induce a further uncertainty in the mass
determination \cite{MWISK,MWILS}.

The hadronic effects mentioned above are all well-known from other 
processes. Interferences in the production and decay of unstable objects 
are observed e.g. in multipion final states at low invariant masses 
\cite{MWICB}. QCD interconnection could be exemplified by J/$\psi$ 
production in B decay: in the weak decay 
$\mbox{b} \to \mbox{c} \mbox{W}^- \to \mbox{c} \, \MWIbar{c} \mbox{s}$
the $\mbox{c} \MWIbar{c} \to \mbox{J}/\psi$ formation requires a
``cross-talk'' between the $\MWIbar{c}$$+$s and the c$ + $spectator colour 
singlets. Bose--Einstein effects are readily visible in high-energy
multiparticle production, including LEP1. In view of the precedents,
the working hypothesis must be that interconnection effects are at play
also for W$^+$W$^-$ events.

The space--time picture of hadroproduction in hadronic $\MWIw^+ \MWIw^-$
decays plays a very important r\^ole in understanding the physics of
QCD interference phenomena \cite{MWISK,MWIKOS}.  Consider,
for instance, a centre of mass energy of 175~GeV where the typical
%% NKW. for instance, a typical c.m.\ energy of 175 GeV.  Then the mean
separation of the two decay vertices in space and in time is of the
order of 0.1 fm.  A gluon with an energy $\omega \gg \MWIgw$
therefore has a wavelength much smaller than the separation between
the $\MWIw^+$ and $\MWIw^-$ decay vertices, and is emitted almost
incoherently either by the $\MWIq_1 \MWIqbar_2 $ system from one W
or by the $\MWIq_3 \MWIqbar_4$ one from the other.  Only fairly soft 
gluons, $\omega~ \raisebox{-0.8mm}{$\stackrel{<}{\sim}$} ~\Gamma_W$, 
feel the joint action of all four quark colour
charges.  On the other hand, the typical distance scale of
hadronization is about 1~fm, i.e.\ much larger than the decay
vertex separation. Similarly, observed Bose--Einstein radii are of this
%% NKW. vertex separation. Also observed Bose--Einstein radii are of this
order. As a result, the hadronization phase may
induce sizeable interference effects.

As one could anticipate from the above, the perturbative QCD 
interconnection effects
appear to be very small \cite{MWISK}. In examining the
size of the non-perturbative W mass distortions one has to rely on  
%% NKW. size of the nonperturbative W mass distortions one has to rely on  
existing QCD Monte
Carlo models rather than on exact calculations.  These models
have done a very good job in describing a large part of the
experimental data at the Z$^0$ resonance, so it seems plausible
that they (after the appropriate extensions) could provide a
reliable estimate of the {\em magnitude\/} of the 
interconnection-induced shift in the W mass.  However, one has to bear 
in mind that there is a true limit to our current understanding of
the physics of hadronization \cite{MWISK,MWILL}, so the actual 
{\em value\/} of the shift may be beyond our current reach.

One of the achievements of the activity of our working group is
that the colour reconnection physics has been tested in several
approaches \cite{MWISK,MWIGH,MWILL,MWIBW, MWIST}. The Bose--Einstein 
studies are somewhat lagging behind, but progress is visible
also here \cite{MWILS,MWILStwo,MWIBM}. 

There is another challenging reason to study the phenomenon of
colour recoupling in hadronic $\MWIw^+ \MWIw^-$ events:
it could provide a new laboratory for probing non-perturbative 
QCD dynamics \cite{MWIGPZ}.  The very fact that different
assumptions about the confinement forces may give different
predictions for various final-state characteristics means that it
might be possible to learn from experiment about the structure of 
the QCD vacuum \cite{MWIGPZ,MWISK,MWIGH}.

Finally, note that there are other effects --- this time originating
in purely QED radiative phenomena --- which, in principle,
prevent the final state from being treated as two separate W decays.
For instance, final-state QED interactions, for the threshold region
exemplified by the Coulomb forces between two unstable W bosons,
induce non-factorizable corrections to the final-state mass 
distributions in $\MWIe^+ \MWIe^- \rightarrow \MWIw^+ \MWIw^- 
\rightarrow 4$ fermions \cite{MWIKSt,MWIKSj,MWIMY}.  Of course, there 
is no reason why all such QED effects cannot, in principle, be
computed to arbitrary accuracy, and taken into account in the
mass determination.  However, at the moment no complete formulae
are available which could be relevant for the whole LEP2 energy
range. This topic certainly needs further detailed studies and a
comprehensive activity here is in progress. For further comments
see  Section~\ref{sec:mw-intro}.

\subsection{Colour reconnection}

Several quite different colour reconnection models have been presented 
%% NKW. Several quite different colour-reconnection models have been presented
by now. A crude survey of features is found in Table \ref{MWImodeltab}.
The listing is by no means complete; several simpler toy models have
also been proposed \cite{MWISK}. The three main philosophies are the
following:
\begin{itemize}
\item 
A colour-confinement string is created in each W decay, spanned
between the decay-product partons. These strings expand out from 
the respective decay vertex and eventually fragment to hadrons. 
Before that time, a reconnection can occur by a space--time encounter 
between sections of the two strings. In one extreme (QCD vacuum structure
analogous to a type I superconductor) the reconnection probability
is related to the integrated space--time overlap of the extended
strings, in another (likewise, with type II superconductor) to the crossing 
%% NKW. strings, in another (ditto with type II superconductor) 
%% NKW. to the crossing
of two string cores (vortex lines). Many subvariants are conceivable.
The models in \cite{MWIST} represent further refinements of the ones
presented in \cite{MWISK}. 
\item Perturbative QCD prefers configurations with minimal string length
in Z$^0$ decays. Here length is defined in terms of the $\lambda$ measure,
which may be viewed as the rapidity range along the string:
$\lambda \approx \sum_i \ln (m_i^2/m_{\rho}^2)$, where $m_i$ is the 
invariant mass of the $i$'th string piece and $m_{\rho}$ sets a
typical hadronic mass scale. 
It is plausible that, when the partons of a W pair are separating and 
strings formed between them, such configurations are favoured which 
correspond to a reduced total string length. Therefore a reasonable 
criterion is to allow $\lambda$-reducing reconnections at the scale of 
\MWIgw, within the limits of what is given by colour algebra factors. 
%reconnections after each step of the parton-shower evolution, within the 
%It is plausible that also reconnections 
%between strings are favoured when the total length is decreased in the 
%process. Therefore a reasonable criterion is to allow $\lambda$-reducing 
%reconnections after each step of the parton-shower evolution, within the 
%limits of what is given by colour algebra factors. 
The models in
\cite{MWIGH} and \cite{MWILL} are variants on this theme.
\item In a cluster approach to fragmentation, clusters are
formed by the recombination of a quark from one gluon branching with
an antiquark from an ``adjacent'' branching (alternatively primary quark 
flavours or diquarks). Normally ``adjacent'' is defined in terms of
the shower history, but another reasonable criterion is the space--time
size of the formed clusters. Therefore reconnection relative to the
ordinary picture could be allowed anytime the sum of the squared sizes 
of the formed clusters is reduced. Restrictions are imposed, so that
e.g. the quark and antiquark from a gluon splitting cannot form 
a cluster together. This approach is found in \cite{MWIBW}.
\end{itemize}
The models need not be viewed as mutually contradictory. Rather,
each may represent some aspect of the true nature of the process.

\begin{table}[tbp]
\begin{center}
\begin{tabular}{|c|c|c|c|c|c|} 
\hline
authors & Khoze & Todorova & Gustafson & L\"onnblad & Webber \\
        & Sj\"ostrand & & H\"akkinen & & \\
reference & \cite{MWISK} & \cite{MWIST} & \cite{MWIGH,MWIGHtwo} & 
          \cite{MWILL} & \cite{MWIBW} \\
\hline
based on & \multicolumn{2}{|c|}{{\sc Pythia}} &
          \multicolumn{2}{|c|}{{\sc Ariadne}} & {\small HERWIG} \\
\hline
reconnection & \multicolumn{2}{|c|}{space--time overlap (I)} & 
          \multicolumn{2}{|c|}{string length} & cluster space--time \\
criterion & \multicolumn{2}{|c|}{or crossing (II) of strings} &
          \multicolumn{2}{|c|}{reduced} & sizes reduced \\
\hline
reconnection & \multicolumn{2}{|c|}{I: free parameter} & free & 
partly & free \\ 
probability &  \multicolumn{2}{|c|}{II: partly predicted} & parameter & 
        predicted & parameter \\
\hline
model of & yes & yes & no & yes &yes \\
all events & & & & & \\
\hline
\multicolumn{6}{|l|}{space--time picture implemented for \hfill
(--- = not applicable)} \\
W vertices & yes & yes & no & no & yes \\
parton shower & no & yes & --- & --- & yes \\
fragmentation & yes & yes & --- & --- & --- \\
\hline
multiple & no & yes & yes & yes & yes \\
reconnections & & & & & \\
\hline
reconnection & no & yes & yes & yes & yes \\
inside W/Z & & & & & \\
\hline
change of event & \multicolumn{2}{|c|}{almost} & small but & 
visible, needs & large, needs \\
properties & \multicolumn{2}{|c|}{invisible} & visible & 
retuning & retuning \\
\hline
\end{tabular}
\caption{Survey of colour reconnection models. The information should
be taken as indicative; only the original literature gives the
ideological motivation behind the choices.}
\label{MWImodeltab}
\end{center}
\end{table}

The main philosophies can be varied in a multitude of ways ---
Table \ref{MWImodeltab} is to be viewed as an appetizer to a proper
appreciation of this. Program details can be found in the
QCD event generators section and in the original literature.
The overall reconnection probability depends on free
parameters in all the models, since we do not understand the 
non-perturbative dynamics; colour suppression factors like $1/N_C^2$ 
%% NKW. nonperturbative dynamics; colour suppression factors like $1/N_C^2$ 
may be present, as in the perturbative phase, but are multiplied by 
unknown functions of the space--time and momentum variables.
Furthermore, the models can only be incomplete representations
of the physics, e.g., the effects of ``negative antennae/dipoles'' 
\cite{MWISK} and of virtual-gluon corrections remain to be addressed. 

Several studies have been performed to estimate the effect (bias) of
colour reconnections on the W mass measurement, and to evaluate the 
possibility of determining experimentally whether such phenomena are 
taking place in the W$^+$W$^-$ system. Reconnection effects could also 
give visible signals at LEP1. To allow unbiased comparisons, each
reconnection model should be retuned to general event-shape data
with the same care as for the ``no reconnection'' scenarios. This 
process is now under way. The {\sc Ariadne} studies make use of a
retuning of the model with reconnection effects, by its author
\cite{MWILL,MWILLparams}.

In order to set limits on possible W mass shifts, the experimental 
sensitivity with which reconnection may be observed has to be evaluated 
for each model (as a function of the model parameters, where appropriate). 
If no effects are observed in the LEP2 data, this sets a limit on the
reconnection probability, which can be turned into an estimate of the 
systematic uncertainty in the W mass. To complicate matters, different 
models will in general predict different mass shifts for the same
probability of reconnection. This complication will also appear if 
reconnections are observed in the data and one would like to correct
the observed W mass accordingly. 

Predictions for the systematic error in W mass measurement are presented 
in Table~\ref{MWIreconntab}. It is important to understand that the value 
of the W mass shift depends {\em very\/} strongly on the definition of this
shift, as well as on the method used to reconstruct the jets, the fitting
function used, and the tuning of the models. These factors need to be
controlled strictly to allow meaningful comparisons. Three estimates of the
mass shift have been studied:
%% NKW. \begin{MWIitemize}
\begin{itemize}
\item {\bf Averaging}~~ Some Monte Carlo authors \cite{MWISK,MWILL} form a
  distribution (difference between reconstructed and generated W mass)
  event-by-event, and use the difference in the means of two such
  distributions (with and without reconnection) as an estimator of the
  systematic shift.
\item {\bf Fitting}~~ Two mass distributions are formed, one with and one without
  reconnection, and the W mass is obtained by fitting each separately. The
  function used was a Breit-Wigner, modulated by phase space factors.  The
  difference between the fit results gives the estimated mass shift. The
  definition of jets and correspondence between jets and W's followed method
  3 of \cite{MWISK}.
\item {\bf Detector}~~ The analysis was performed as if using real data.  Mass
  distributions are obtained after the events (with initial state 
  radiation) have been passed
  through a LEP detector simulation program, thereby including typical
  experimental acceptance and quality cuts. The W mass is determined for each
  event using a 5 constraint kinematic fit (see, e.g.,
  Section~\ref{sec:mwdir}).  The shift is taken as the difference between
  two fitted mass distributions as above, although the definition of jets,
  the association between jets and W's and the fitting function were
  different. The experimental results presented here were obtained using a
  simulation of the OPAL experiment~\cite{MWIGPW}.
\end{itemize}
%% NKW. \end{MWIitemize}
Table~\ref{MWIreconntab} contains an (incomplete) comparison of the shifts
observed using all three definitions of the shift in the W mass, where the
errors are from finite Monte Carlo statistics.  As indicated, the three are
not expected to agree. It is noted that there is an upwards shift after
detector simulation. The study of \cite{MWIGPW}, including the same kinematic
fitting, jet and W reconstruction, was repeated using events without detector
simulation.  Mass shifts which are 1--2$\sigma$ (statistical) lower than the
detector simulation fits were observed; they were stable against changes in
fit range, and cuts on $P_t>150$~MeV/c and kinematic fit probability 
($>$1\%).
They are also consistent with the shifts in the `fitting' column in
Table~\ref{MWIreconntab}.

%% NKW 951208. Changed, as the 'fitted' results have now become larger
%%             than the repeated OPAL analysis without detector simulation.
If the shift is genuinely caused by specific experimental cuts, it may be
possible to redesign those cuts to reduce $\Delta$\MWImw\ to the level
seen without detector simulation. If so, we might also construct a related
observable for the detection of reconnection.

\begin{table}[tbp]
\begin{center}
\begin{tabular}{|c|c|c|c|c|c|c|} \hline
 Model & Reconn. & \multicolumn{3}{|c|}{$\Delta$\MWImw\ 
   (MeV)} &
         \multicolumn{2}{|c|}{Sensitivity ($\sigma$)} \\ 
\cline{3-7}
       & prob.(\%) & averaging & fitting & detector & Central & Interjet \\ 
\hline \hline
\cite{MWISK} Type I,
        $\rho$=0.9 & 46.7 & 27$\pm$14 & 61$\pm$20 & 130$\pm$40 &  2.4 & 1.9 \\
\hline
 \multicolumn{1}{|r|}{$\rho$=0.6~~}
               & 36.5 & 22$\pm$14 & 68$\pm$20 &  80$\pm$40 &  2.0 & 1.5 \\ 
\hline
 \multicolumn{1}{|r|}{$\rho$=0.3~~}
                   & 22.4 & 10$\pm$14 & 35$\pm$20&  60$\pm$40 &  1.4 & 1.2 \\ 
\hline
 \multicolumn{1}{|r|}{Type II$'$~~}  
         & 34.2 & $-$22$\pm$14 & $-$25$\pm$19 &  50$\pm$40 &  0.8 & 0.8 \\ 
\hline \hline
\cite{MWIST} Type I, $\rho$=1. & 22 & 0$\pm$14 & 33$\pm$18  & -- & -- & -- \\ 
\hline
 \multicolumn{1}{|r|}{Type I, $\rho$=100.} & 64 & 17$\pm$14 & 47$\pm$18  
& -- & -- & -- \\ 
\hline
 \multicolumn{1}{|r|}{Type II, $d$=0.~fm} & 8 & $-$18$\pm$14  & $-$1$\pm$18 
& -- & -- & -- \\ 
\hline
 \multicolumn{1}{|r|}{Type II, $d$=0.01~fm} & 34 & $-$2$\pm$14 & 32$\pm$18 
& -- & -- & -- \\ 
\hline \hline
 \cite{MWIGH}     & 100  & 70$\pm$20 & -- & 150$\pm$40 & 6.5 & 6.8 \\ 
\hline \hline
 \cite{MWILL}     & 52 & 30$\pm$10  & 58 $\pm$ 21 & 154$\pm$36 & -- & -- \\ 
\hline
\cite{MWILL} retuned(\cite{MWIHWtwo})  & 46 & $-$6$\pm$14  & 17 $\pm$ 19  
& -- & -- & -- \\ 
\hline \hline
 \cite{MWIBW}     & 8 & -- & 11$\pm$10 & -- & -- & -- \\ 
\hline
\end{tabular}

\protect\caption{Summary of effects on \MWImw\ at 175 GeV. 
See the text for details. The $\rho$
parameter in Type I models relates the string overlap integral to the 
reconnection probability. The $d$ parameter gives the vortex line 
core diameter in Type II models. The numbers for \protect\cite{MWIGH}
can be rescaled by the reconnection probability, which here is a 
free parameter. The sensitivities in the last two columns include 
detector simulation \protect\cite{MWIGPW}. Several of the numbers in 
the table are preliminary and further studies are underway to 
understand them.}
\label{MWIreconntab}
\end{center}
\end{table}

Two related experimental methods of detecting the presence of reconnection 
phenomena at LEP2 have been studied, including OPAL simulation 
\cite{MWIGPW}: multiplicity in a central rapidity bin~\cite{MWIGH}
and interjet multiplicity~\cite{MWIGPW}.

%The rapidity distribution at $\sqrt{s}$ = 175~GeV is shown in
%figure~\ref{MWIrapidityfig} for normal {\sc Pythia} events and for a
%sample of events generated with an independent implementation of the GH
%model~\cite{MWIGPW}.  The rapidity, $y$, defined with respect to the
%thrust axis, is measured for events in which the thrust, $T$, is greater
%than 0.76 and also for $T>0.92$.  The dashed line is the spectrum obtained
%with the fully reconnected event sample, the solid histogram corresponds to
%normal {\sc Pythia} events. In this implementation of the GH model, these
%distributions are not particularly sensitive to reconnection effects
%\cite{MWIGPW}.
%% Here the full angular distributions of W decays are used, while the
%% original GH studies were performed with isotropic W decays; presumably 
%% this is at the origin of differences. 
%Other recent studies \cite{MWIGHthree} confirm that effects presumably
%are smaller than original studies indicated.   

%\begin{figure}[tb]
%\vspace*{-1cm}
%\begin{center}
%\epsfig{file=MWIrapidityfig.eps,height=12cm}
%\caption{Rapidity distributions at 175 GeV (\`{a} la GH model).}
%\label{MWIrapidityfig}t
%\end{center}
%\end{figure}

In ref.~\cite{MWIGH} a significant decrease was found in the central
rapidity region for events with a thrust cut $T > 0.76$. (A larger
difference but poorer statistical significance was obtained for more
restrictive thrust 
cuts.) An independent, {\sc Pythia}-based implementation of the GH
model \cite{MWIGPW}
showed a much reduced signal. It was here pointed out that one reason 
for the difference is the effect of W polarization, which was neglected
in ref.~\cite{MWIGH}. Recent studies \cite{MWIGHthree} confirm that
the polarization is important; it reduces the signal by roughly a 
factor of 2 compared to the original results. However, although
significantly reduced, a clear signal for reconnection effects is
still visible in \cite{MWIGHthree}; this appears to be larger than 
the result obtained in ref.~\cite{MWIGPW}.   

Following the prescription of \cite{MWIGH}, the central multiplicity is
defined as the number of particles, $n$, observed within a given rapidity
interval. In order to quantify the level at which colour reconnection could
be observed, the sensitivity at a given multiplicity, $n$, is defined as
 \cite{MWIGPW}:
\begin{equation}
    \left[   \sum_{i=1}^{n}  N_{\mathrm{rec}}(i)
 -      \sum_{i=1}^{n}  N_{\mathrm{norm}}(i)     \right]
 /  \sqrt{ 1  +   \sum_{i=1}^{n} N_{\mathrm{norm}}(i)  }
 \label{MWIsensiteq}
\end{equation}
where $N_{\mathrm{rec}}(i)$ and $N_{\mathrm{norm}}(i)$ are the number of
events in the reconnected and normal event samples having a central
multiplicity $i$, respectively.  The sensitivities given in
Table~\ref{MWIreconntab} are obtained assuming that 3\,000 four quark
events have been observed, that is about what is expected from
the nominal luminosity of 500~pb$^{-1}$, by scaling the sensitivities from
samples of 50\,000 generated events by $\sqrt{3\,000/50\,000}$. 
If the results of the four experiments are combined, the statistical
significance will double. With such statistics a
tentative conclusion from the models so far studied is that reconnection
effects could be observed down to 25--30\%, corresponding to a
$3\sigma$ effect. A smaller integrated luminosity is not very encouraging 
at an energy of 175~GeV. It should also be remembered that the significance
level is based on an assumed perfect knowledge on the no-reconnection
multiplicity distribution; an entirely experimental procedure based on a
comparison with mixed leptonic--hadronic events would have a
smaller sensitivity.

A reconnection probability of 30\% could induce a systematic effect 
of about 50~MeV on the W mass measurement using the four-jet method, 
which is of the same order as the total measurement error expected in this 
mode for all four LEP experiments combined. This mass shift is found to 
depend upon the model used and may be larger than estimated above.
The uncertainty is further increased by other reconnection sources,
including perturbative interconnection, interplay between perturbative 
and non-perturbative effects \cite{MWISK}, virtual effects, and so on.
%% NKW. and nonperturbative effects \cite{MWISK}, virtual effects, and so on.
The 50~MeV number above is therefore to be viewed as a realistic
uncertainty.  

It is also interesting to study event topologies at LEP1.  For instance, a
qgg$\MWIqbar$ event is normally expected to have the partons ordered along a
single string, but an alternative subdivision into a q$\MWIqbar$ and a gg
singlet is possible. Such ``reconnection'' phenomena (though not quite of the
same character as the ones we worry about for the W mass issue) could give
rapidity gaps in events where the quark and antiquark are tagged to lie in
the same hemisphere with respect to the thrust axis \cite{MWIGHtwo,MWIGPW}.
Generic event shapes presumably cannot lead to any definite conclusions; in
part, the information on event shapes is used to tune generators in the first
place. If a properly tuned generator with ``reconnection'' included could
perform as well as the conventional ones, it would provide indications that a
given approach is not unreasonable. Models including ``reconnection'' which
do not, even with appropriate retuning, describe the data are less useful.  A
recent retuning of {\sc Ariadne} with ``reconnection'' by DELPHI \cite{MWIHW}
shows that the quality of the description of LEP1 data can even be improved.
%% NKW. This next clause seems to overestate the importance of the tuned
%%      parameters, without adding information to the discussion.
%%
%% This could open the way towards more incisive tests;
However, it may be entirely fortuitous.

\subsection{Bose--Einstein Effects}

Since the hadronization regions of the $\MWIw^+$ and $\MWIw^-$ overlap, 
it is natural to assume that some coherence effects are present between
identical low-momentum bosons stemming from different W's due to
Bose--Einstein (BE) correlations. How much
such effects would influence the W mass measurement is difficult
to answer. Intuitively, since the BE effect favours production
of identical bosons close in phase-space, one would expect the
softest particles from each W to be ``dragged'' closer to each other.
This reduces the momentum of the W's and thus
increases the measured W
mass. However, most of our understanding of the details of
multi-particle production comes from probabilistic hadronization
models where BE correlations are absent. Only a few attempts have
been made to include such effects and to investigate their influence
on the W mass measurement.

One such attempt \cite{MWILS} used an algorithm ({\tt LUBOEI})
implemented in the {\sc Jetset}. This algorithm is based on the 
assumptions that BE effects are local in phase space and do not alter
the event multiplicity. (Notice that any effects on the multiplicity
are already, at least partially, accounted for by the tuning of the
generators to data.) A re-weighting of events with a BE factor 
is then equivalent, in some approximation, to letting the momenta
of the bosons produced in the hadronization be shifted somewhat, e.g. 
to reproduce the difference in two-particle correlation functions
expected for a source with Gaussian distribution in space--time,
\begin{equation}
  \frac{C_{\MWIindx{BE}}(Q)}{C_{\MWIindx{noBE}}(Q)} = 
        1 + \lambda \exp(-Q^2 R^2) ~.
  \label{MWIf2}
\end{equation}
Here $R$ is the source radius and $\lambda$ is an incoherence 
parameter. $C_{\MWIindx{BE}}(Q)$ and $C_{\MWIindx{noBE}}(Q)$ are 
the two-particle correlations as functions of relative 
four-momenta in a world with and one without BE effects, respectively.
This is a well-defined generator procedure but,
since it is not possible to switch off BE in the data, the main
challenge of experimental BE studies is to define an appropriate
no-BE reference sample.  With this algorithm, the
observed correlations at LEP are reproduced using $\lambda\approx 1$
and $R\approx 0.5$ fm \cite{MWIBEL}. One problem with this
algorithm is that it does not inherently conserve energy and
momentum. This has to be done separately by rescaling all momenta in
an event. Therefore, when this algorithm is used on $\MWIw^+\MWIw^-$ 
events,
identical bosons within each W are shifted closer to each other in
phase space resulting in an artificial negative shift in the measured
W mass even if there is no cross-talk between the W's. In Ref.\
\cite{MWILS} this artificial shift was corrected for and an
estimate for the true shift in the measured W mass was obtained. The
result was $\langle\Delta M_W\rangle\approx 100$ MeV for 170 GeV
center of mass energy. This shift increases with energy and decreasing
source radius.

A variation of this algorithm  has been investigated \cite{MWILStwo},
where not only the momenta of identical particles are shifted closer
to each other, but also non-identical particles are shifted away from
each other, so that the ratio of correlation functions of identical
and non-identical particles is the same as with the original
algorithm. In this way, the
energy--momentum non-conservation can be made much smaller, minimizing
possible dependence on the correction procedure. The shift in the W
mass is in this case smaller, around 30 MeV. But modifying the
algorithm slightly, not allowing the shifting of non-identical bosons
from different W's, again gives a mass shift around 100 MeV.

In both these cases the shift was calculated on the generator level,
assuming that the origin of each particle was known. It is clear that
an experimentally feasible reconstruction algorithm may be more or
less sensitive to these BE effects.

Another attempt to estimate the influence of BE correlations on the
W mass measurement was based on a simplified toy model of W
decays, in which only decays into two or three particles were
considered \cite{MWIBM}. Each of the two W's was treated as a
scalar, decaying into a heavy particle and one or two pions, according
to a relativistic Breit-Wigner. The decay amplitudes can thus be
written down explicitly, with and without symmetrization with respect
to exchange of the pions. This allowed the generation of such 
low-multiplicity events, each with two weights corresponding to the
symmetrized and non-symmetrized cases, respectively, and thus the study
of the effect on the reconstructed W mass. The mass of the heavy
particle in the model was chosen to correspond roughly to the total
mass of the hadronic decay products of a W boson with one or two
pions removed, i.e.\ 75--79 GeV.

The model calculations confirm that substantial shifts of the W
mass due to the BE symmetrization can occur, and also
suggests that the magnitude of the effect can depend significantly
on the precise way in which the W mass is reconstructed from the
data. In the particular case studied, the peak position of the mass
distribution was less sensitive to the BE effect than the
average mass. The interpretation was hampered by severe threshold 
effects, however, and it is in any case not clear how to draw
quantitative conclusions valid for the multi-particle decays
that form the bulk of real hadronic W decays.

At our present level of understanding, effects of BE correlations on
the W-mass measurement of the order of 100 MeV cannot be
excluded, though this number may be viewed as a ``worst case'' estimate.
Much more work is needed to develop realistic models and to
study the sensitivity of different reconstruction algorithms. It may
also be possible to find signals, e.g.\ by studying the two-particle
correlation function using only pairs where the particles come from 
different reconstructed W's, that could enable us to deduce from data 
the magnitude of the effect.

\subsection{Summary and Outlook}

Several activities are under way, especially on the experimental front,
so it would be premature to draw any definite conclusions. However,
it is not  excluded that interconnection and Bose--Einstein
effects could each contribute a 50~MeV mass shift in the hadronic
$\MWIw^+ \MWIw^-$ decay channel. Most interconnection models predict
a positive mass shift, but we know of no general argument why it would
have to be so.
%% NKW. Goes without saying.
%% TS. Changed back!
Furthermore,  not all possible interconnection
effects have been studied so far.
In contrast, a positive mass shift from BE effects appears plausible on
general grounds.

In spite of the bias towards positive shifts, the conservative approach 
would be to quote a symmetric ``theory uncertainty'' of the order of 
$\pm100$~MeV. This is to be viewed as some sort of theorist's $1 \sigma$ 
error, that is, larger numbers can not be excluded but are less likely.
The value above refers to studies at 175~GeV. The energy dependence
has not yet been well studied, but indications are that the uncertainty
is not smaller even at the maximum LEP2 energy. We note here that
the interconnection phenomena are not expected to die out rapidly with 
increasing energy \cite{MWISK}, since the W resonance is so wide that 
the two W decays appear ``on top of each other'', in terms of hadronic 
distance scales, over the full LEP2 energy range.

Interconnection effects might be found in the data itself, at LEP2 or 
even by a careful analysis of LEP1 data. If so, it would be {\em very\/} 
interesting. However, a non-discovery would not prove non-existence:
so far, we have failed to find realistic signals for some of the current 
approaches. We could only hope to exclude some extreme scenarios 
with large values of the reconnection probability. Conversely, 
at our present level of understanding, a 
discovery would not guarantee a unique recipe for how to ``correct''
the data. This is illustrated by BE effects, where it should be
fairly straightforward to observe cross-talk at LEP2 in principle,
but where statistics presumably will not allow a sufficiently precise 
investigation for the effects on the measured W mass to be well defined.

At the end of this workshop, another model of interconnections appeared
\cite{MWIEG}, with which its authors find mass shifts of up to several hundred
MeV, i.e.  more than observed in previous studies. This exemplifies the need
for continued theoretical and experimental activity in the field to better
understand the issues. However, in view of the well-known complexity of
hadronic final states, it would be unrealistic to expect these studies to
give any simple answers. Monte Carlo models are all we have so far, and may be
the best we can get in the foreseeable future. Therefore the main conclusion
is clear: {\em one cannot blindly average W mass results in the hadronic}
$\MWIw^+ \MWIw^-$ {\em channel with numbers obtained elsewhere.}

%\input s5_new.tex
            
%%%%%%%%%%%%%%%%%%%%%%%%%%%%%%%%%%%%%%%%%%%%%%%%%%%%%%%%%%%%%%%%%%%%%%%%%%
%   please note the following conventions and macros:                    %
%                                                                        %
%   Fig.~\ref{}, Table~\ref{}, Section~\ref{} and Ref.~\cite{}.          %
%                                                                        %
%   for masses please use $\Mw$, $\Mh$, $\Mt$, etc.,                     %
%   and if you wish, units are provided by e.g. $25\mev$, $161\gev$,     %
%   $100\pb^{-1}$ which puts the units in roman and adds a space after   %
%   the number.                                                          %
%                                                                        %
%%%%%%%%%%%%%%%%%%%%%%%%%%%%%%%%%%%%%%%%%%%%%%%%%%%%%%%%%%%%%%%%%%%%%%%%%%

\section{Conclusions}

\begin{figure}[htb!]
  \centerline{\epsfysize12cm\epsffile{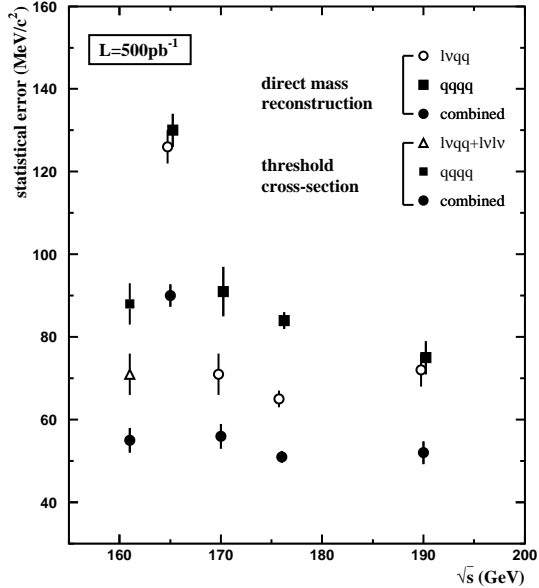}}
  \caption{\it 
Collider energy dependence of the statistical errors.
          }
  \label{fig:energy_dep}
\end{figure}

The goal of $\Delta\Mw < O(50)\gev$ appears to be achievable at LEP2.
The `worst-case' scenario, in which only the $\qq l\nu$ channel
is used in the direct reconstruction method because of colour
reconnection in the $\qq\qq$ channel (see below), gives a total
estimated error of $\Delta\Mw = 44\mev$, assuming four
experiments each collecting $500\pb^{-1}$ at $\sqrt{s} = 175\gev$.

Two methods have been studied in detail. A precise measurement
of the $ \epem\to\WW$ {\bf threshold cross section} can be compared to
the theoretical prediction as a function of $\Mw$, see Fig.~\ref{fig:xsec_xsec}.
The maximum statistical power is obtained by running the collider
at an optimal energy $(\sqrt{s})^{\,\mathrm{opt}} \simeq 
2\Mw + 0.5\gev \simeq 161\gev$. All three decay channels ($\qq\qq$, 
$\qq l \nu$ and $l\nu l\nu$) can be used, and the total estimated
errors are listed in Table~\ref{threshold_errors_summary}. 
Some marginal improvement in the systematic
error can be expected, but ultimately the measurement is statistics limited.
An integrated luminosity of {\it at least} $50\pb^{-1}$ per experiment
is required to obtain a precision comparable to that expected
from the combined Fermilab Tevatron experiments. The total 
error using the threshold cross-section method
 for four experiments each with $50\pb^{-1}$ is estimated to
be $108\mev$. 

One very attractive feature of the threshold measurement is that 
it appears to fit in very well with the anticipated LEP2 machine 
schedule, viz. a run of   at least  $25\pb^{-1}$ 
luminosity  at $\sqrt{s} = 161\gev$
in 1996 followed by  the bulk of the luminosity at higher energy.

The {\bf  direct reconstruction method}, in which $\Mw$
is reconstructed from the invariant mass of the W$^\pm$ decay products,
appears to offer the highest precision.
The statistical and systematic errors are summarised in 
Table~\ref{mwdir_sumtab}.
More work is needed to investigate different fitting procedures,
different fragmentation parameters etc., 
and this may lead to small but important reductions in the errors.
The major outstanding problem concerns the issues of
{\it colour reconnection}, i.e. non-perturbative strong interactions
between the decay products of two hadronically decaying W bosons which
may significantly distort the invariant mass distributions.
This question is currently receiving much attention, but at this time it
is impossible to say whether the problem can be completely overcome.
At best, one may eventually hope to demonstrate that the effect 
on the reconstructed $\Mw$ value is either small or under 
theoretical control, so that the results
from the $\qq\qq$ and $\qq l\nu$  channels can be combined.
At worst,  the precise magnitude of the colour reconnection  effect
may remain unknown, with different models giving different mass shifts
larger or comparable to the remaining systematic and statistical
errors. We have therefore chosen to present results both 
for the $\qq l\nu$ channel only (where colour reconnection is absent)
and  for the two channels combined. The estimated total  errors are 
$44\mev$ and $34\mev$ respectively, for $500 \pb^{-1}$ luminosity
for each experiment at $\sqrt{s}=175\gev $.
 
The collider energy dependence of the direct reconstruction measurement
has also been studied. The results are summarised in
Fig.~\ref{fig:energy_dep}. There is clearly a `blind window' between
$161\gev$ and $170\gev$ where {\it a precision measurement 
of $\Mw$ is impossible}. In this window the threshold cross section method
loses sensitivity to $\Mw$, and the direct reconstruction method has
insufficient statistics.

Increasing the 
LEP2  energy from 175~GeV to 192~GeV has an
essentially neutral effect on the measurement of $\Mw$.
Most of the anticipated errors, in
 both the $\qq\qq$ and $\qq l\nu$ channels, 
  appear to be approximately energy 
independent.\footnote[5]{Note that the statistical errors
in Fig.~\ref{fig:energy_dep} show less energy dependence
for the $\qq l\nu$ channel than those in Table~\ref{tab:mwdir_eff2}.
The former correspond to an average of the ALEPH and OPAL studies
while the latter correspond to the OPAL study only.}
Two  important energy-dependent systematic errors, which are common
to all experiments and both decay channels, are from
initial state radiation and  from beam energy calibration.
Both are expected to increase slightly with energy, 
although the overall effect 
is marginal.  
%If colour reconnection ultimately rules out the $\qq\qq$ channel,
%then the slight ($10 - 15\%$) increase with energy of the 
%statistical error from the $\qql\nu$ channel  might also argue 
%in favour of a lower energy.

%\input sectionref.tex
%
% This is the standard for references
%

%%%%%%%%%%%%%%%%%%%%%%%%%%%%%%%%%%%%%%%%%%%%%%%%%%%%%%%%%%%%%%%%

\end{document}